\documentclass[twoside,12pt]{article} \usepackage{epsfig}
\usepackage{graphics}
\usepackage{mathrsfs}
\usepackage[percent]{overpic}
\usepackage{amsmath}
\usepackage{setspace}
\usepackage[margin=2.5cm]{geometry}

\def\ss{\mbox{\boldmath $\sigma$}}
\newcommand{\be}{\begin{equation}}
\newcommand{\ee}{\end{equation}}
\newcommand{\bea}{\begin{eqnarray}}
\newcommand{\eea}{\end{eqnarray}}

\begin{document}

\title{Recent developments in radioactive charged-particle emissions
and related phenomena}
\author{Chong\ Qi, Roberto Liotta, Ramon Wyss\\
\\
Department of Physics, Royal Institute of Technology (KTH), \\SE-10691 Stockholm, Sweden}

\maketitle
\begin{abstract}

The advent and intensive  use of new detector technologies as well as
radioactive ion beam facilities have opened up
possibilities to investigate alpha, proton and cluster decays of highly
unstable nuclei. This article provides a review of the current status of our 
understanding of clustering and the corresponding radioactive particle
decay process in atomic nuclei. We put
alpha decay in the context of charged-particle emissions which also include one- and two-proton emissions as well as heavy
cluster decay. The experimental as
well as the theoretical advances achieved recently in these fields are
presented.  Emphasis is
given to the recent discoveries of charged-particle decays from
proton-rich nuclei around the proton drip line. Those decay measurements
have shown to provide an important
probe for studying the structure of the nuclei involved. Developments on the theoretical
side in
nuclear many-body theories and supercomputing facilities have also made
substantial progress, enabling one to study
the nuclear clusterization and decays within a microscopic and consistent
framework. We report on properties induced by the
nuclear interaction acting in the nuclear medium, like the pairing
interaction, which have been
uncovered by studying the microscopic structure of clusters. The
competition between cluster formations as compared to the corresponding
alpha-particle formation are included.  In the review we also describe the search for super-heavy
nuclei connected by chains of alpha and other radioactive particle decays.

\end{abstract}
\tableofcontents

\section{Introduction}
The process leading to the emission of alpha-particles from nuclei is a
subject that has been studied since the beginning of modern physics at the
end of the 19th Century \cite{Ruther99,Ruther01}. However nearly three decades 
had to pass before Gamow  could explain how an alpha-particle 
can overcome the Coulomb and
centrifugal barriers that trap it inside the nucleus \cite{Gamow1928}. That was 
a great breakthrough which can be viewed as a cornerstone of the probabilistic
interpretation of quantum mechanics.
Gamow explained the decay as the penetration of an already formed alpha
particle through the Coulomb and centrifugal barriers. To obtain the proper 
units, Gamow also introduced the concept of ``assault frequency" which is an 
effective quantity that, due to the Pauli principle, does not carry any 
quantum mechanics validity. This theory has been extremely 
successful in explaining relative decay widths, but could not describe
absolute decay widths. Yet the calculation of the penetrability is relatively easy and 
therefore the theory was applied in many situations, trying to get the absolute 
decay widths by adjusting effective parameters, such as the assault frequency, to 
fit the corresponding decay width. These effective theories are very useful 
because they are easy to apply. However, a proper calculation of the decay 
process needs to address the clustering of the nucleons on the surface of the 
mother nucleus and the following penetrability of the cluster thus formed 
through the Coulomb and centrifugal barriers. The evaluation of the 
cluster formation 
probability is a  challenging undertaking because a proper description of the 
cluster in terms of its components requires a microscopic framework that is 
highly complex. This is the reason why effective approaches are used when 
dealing with clusterization.

Yet, one has to surmise the lack of a firm theoretical foundation for the 
effective quantities thus introduced. This would require a microscopic
treatment of the decay process which includes the degrees of freedom of all
nucleons involved in the decay. Microscopic theories 
are not always famous for their plausibility or
accurate predictions. Phenomenological approaches often surpass them in
both respects by their simplicity and aptitude, owing largely to their
tendency to wrap unknown aspects and even inconsistent ingredients into
adjustable parameters.  One may wonder how can effective theories, in spite of their obvious
shortcomings,  be so successful in describing
alpha decay. This is due to the efforts of many researches through years of 
adjustments to new experimental
data which resulted in methods or, perhaps more proper, effective formulae 
which reproduce reasonable well alpha-decay data. In really fundamental microscopic theories, on the
contrary, no ad hoc assumptions must be invoked, and the number of free
parameters is to be reduced to a minimum, preferably to none. One cannot
really understand the underlying mechanism of decay processes without
describing them by parameter-free microscopic theories, free of ad hoc
elements. To approach this ideal, one has to use dynamical theories in
which the states involved are constructed by some (approximate) solution of
a model Schr\"odinger equation.

\begin{figure}[tb]
\epsfysize=9.0cm
\begin{center}
\begin{minipage}[t]{18 cm}
\epsfig{file=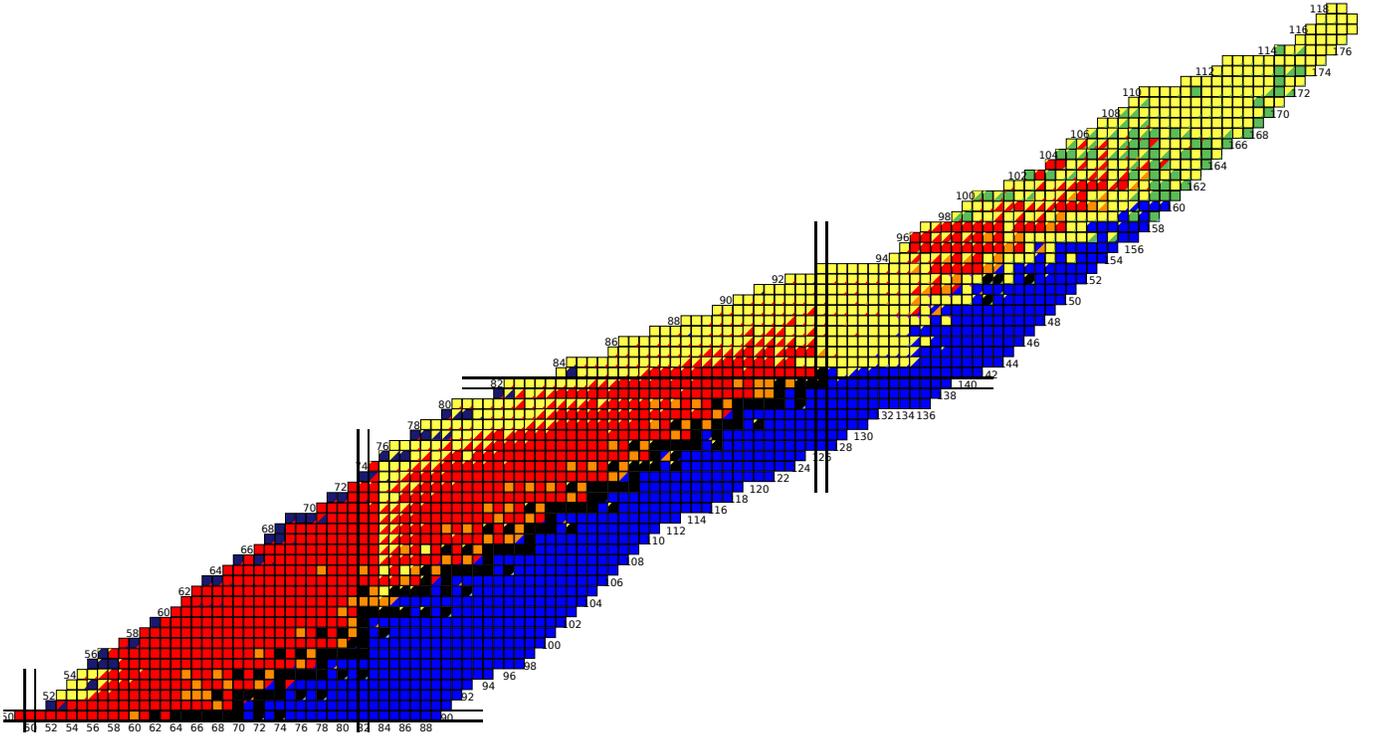,width=18cm}
\end{minipage}
\begin{minipage}[t]{16.5 cm}
\caption{Ground-state decay modes of nuclei with proton number 
$Z\geq50$ extracted from Nubase2016 \cite{1674-1137-41-3-030001} which includes $\alpha$ decay (yellow), $\beta^+$ (red), $\beta^-$ (green), electron capture (dark orange), fission (green). The more rare cases of proton decays and cluster decays are marked with dark blue and purple colors, 
respectively. We use triangles to indicate secondary decay modes. The solid lines correspond to the proton shell 
closures $Z=50$, 82 and neutron shell-closures $N=50$, 82 and 126.
\label{fig1all}}
\end{minipage}
\end{center}
\end{figure}

The study of
particle radioactivity have been a primordial interest in nuclear physics.   
These developments are going on even at present. In this review we will present  
developments that took place through the many years.
Indeed, nuclear physics is undergoing a renaissance from both experimental and 
theoretical physics point of views with the availability of intense 
radioactive ion beams and new detecting and supercomputing technologies. The new nuclear facilities 
have opened up  possibilities to 
investigate highly unstable nuclei.  One of the most important aspects in these searches is the 
possibility of discovering so far unknown superheavy nuclei. Here alpha emission 
is one of the dominant forms of decay.   
This decay occurs most often in 
massive nuclei that have relatively large proton to neutron ratios, where it can reduce the 
ratio of protons to neutrons in the parent nucleus, bringing it to a more stable 
configuration in the daughter nucleus. Almost all observed proton-rich or 
neutron-deficient nuclei
starting from mass number $A \sim 150$ decay through alpha-emission channels, as shown 
in Fig. \ref{fig1all} where we plotted the observed ground state decay modes for nuclei with $Z\ge 50$. 
The emission of heavy clusters, which is also observed in heavy nuclei,
is closely related to spontaneous fission 
\cite{0034-4885-62-4-001,RevModPhys.85.1541,0034-4885-79-11-116301}. In fact
fission and $\alpha$ emission may compete with similar probabilities  
as decay modes in heavy and superheavy nuclei.

Besides the decay of clusters from nuclei, we will also report studies on 
proton and two-proton radioactivity. We use the term ``proton radioactivity" to describe  
the process leading to the 
decay of a nucleus by emitting a proton. It may seem more natural to call
this process ``proton decay", as one calls ``alpha decay" the emission of an
alpha particle from a nucleus. The problem with the name "proton decay" is
that it may be confused with the hypothetical decay of a proton into lighter
particles, such as a neutral pion and a positron. There is no evidence at
present that proton decay in this sense exists.  
In this review we will use indiscriminately the terms ``proton 
radioactivity", ``proton emission" and ``proton decay" to describe the emission 
of a proton from a nucleus. The name used for the
proton decay process should not be confusing here, since we refer to nuclear
decay only.

Our aim in this review is to present in a clear and pedagogical fashion the
most relevant investigations that have been performed during the last couple
of decades by outstanding researches 
in unveiling the difficult but thrilling subject of radioactive particle
decay from nuclei. 
This includes, besides $\alpha$-decay, proton-, two-proton- and heavy-cluster 
emission, as have been outlined above.

\section{General formalism}
In this Section we will briefly review theoretical results acquired before the period covered by 
this report. The resulting  formalism is fundamental to understand the decay
processes to be included here.

There is an important difference between the decays of clusters and one- and two-proton decays. 
In the firt case one has to consider the formation of the clusters, including
alpha and heavier clusters like $^{14}$C, starting from the nucleons
that constitute the decaying mother nucleus. This is an extremely
complicated process which requires the knowledge of the mechanisms that 
induce the cluster formation as well as the description
of the motion of the already formed cluster while departing from the daughter
nucleus. 

The calculation of the decay of the lightest cluster that happens in Nature, that is 
alpha-decay, is already a difficult undertaking which induced
much debate and arguments through the years. A number of theories were
proposed which are by now nearly forgotten. For a review on this one can see Ref. 
\cite{rhoades}. The microscopic treatment of alpha decay required a general framework which
was provided by the introduction of the R-matrix theory as formulated by
Teichman and Wigner \cite{TW}. In this 
formalism the  collision between two nuclei leading to a compound system and
its subsequent decay is
described by dividing the configuration space of the composite system into
an “internal region”, to which the compound state is restricted, and the
complementary “external region”. This division is made
such that in the external region only the Coulomb interaction is important
and the system in the outgoing channel behaves like a two-particle system. 
This is exactly
what occurs in  alpha decay, where the outgoing channel consists of two
fragments, the alpha particle and the daughter nucleus, interacting
through the Coulomb interaction only. The important feature of the formalism
is that the residues of the R-matrix is proportional to the decay width of
the resonance induced by the decay process.
This formalism was applied by Thomas to evaluate the $\alpha$-decay width
in a profound but difficult paper \cite{Thomas}. A more
accessible derivation of the Thomas expression for the width  
can be found in Ref. \cite{MFL}.

Thomas classical expression has the form
\begin{equation}\label{Thomas}
\Gamma_c (R) = \hbar/T = \frac{\hbar^2 k}{\mu}
\frac{R^2 |{\cal F}_c(R)|^2}{F^2_c(R)+G^2_c(R)}
\end{equation}
where $c$ labels the decaying channel, $k$ is the linear momentum carried
by the $\alpha$-particle, $\mu$ is the reduced mass, $R$ is the 
distance between the mass centres of the daughter and cluster nuclei, 
$F_c(R)$ ($G_c(R)$) is the
regular (irregular) Coulomb function corresponding to the two-body system in
the outgoing channel and  ${\cal F}_c(R)$ is the formation amplitude, i. e. 
the wave function of the mother nucleus at the point R.  
It is very important to underline that the distance $R$ in Eq. (\ref{Thomas}) 
corresponds to the matching distance where the internal and external wave
functions coincide.

The way Thomas wrote the width, which, as in Eq. (\ref{Thomas}), may 
indicate that it is $R$-dependent, was the origin of much confusion. This confusion
may have been strengthened by the name used for $R$, namely the ``channel radius". 
There have been authors claiming that since all function in Eq. (\ref{Thomas})
depend exponentially upon $R$ the width itself was strongly $R$-dependent. 
Therefore the formalism was useless, as indeed it would
have been if such dependence existed.  The point $R$ should be chosen outside the range of the nuclear central 
field induced by the daughter nucleus. At this point the $\alpha$-particle is 
already formed. Therefore the width above is {\it independent} of
$R$. This property was often used in microscopic calculations to probe that
the results were reliable. This was earliest done in Ref. 
\cite{PhysRevC.44.545}.

The width (\ref{Thomas}) is valid for the decay of any cluster, not only
alpha-particles (the subscript $c$ in Eq. (\ref{Thomas} refers to "cluster"). 
That equation is often  written in terms of the penetrability 
$P_c (R)$, i. e. the probability that the already formed cluster
penetrates the Coulomb and centrifugal barriers starting at the point $R$. 
It is given by \cite{Thomas}
\begin{equation}\label{penet}
P_c (R) = \frac{kR}{F^2_c(R)+G^2_c(R)}
\end{equation}
and the width becomes
\begin{equation}
\label{gpf}
\Gamma_c (R) = \frac{\hbar^2 R}{\mu} P_c(R)|{\cal F}_c(R)|^2.
\end{equation}

The penetrability is strongly dependent upon the $Q$-value of the decay
channel. It is also strongly dependent upon the distance $R$, a feature
which is shared by the formation probability. When the calculation is 
performed properly, as mentioned above, the two $R$-dependences cancel each other and the evaluated 
half lives are radius independent. 
Effective approaches totally ignore the dependence of the
formation probability upon the distance. Instead the free 
parameters are assumed to somehow take care of it. That is the reason why it
is often stated that the penetrability is overwhelmingly dominant in the
alpha decay process.

The time when Thomas presented his formula, in the 1950s, was the age of 
the shell model. The shell model is, more than a
model, a tool that provides an excellent representation to describe nuclear
dynamics. One would then think that the shell model 
should be capable of taking into account any correlations if the
corresponding basis is large enough.  This was clear to the
pioneers who started to use the shell model for the description of 
$\alpha$-decay.
It aroused optimism at the beginning that very simple shell-model
calculations were able to describe the low energy spectra of nuclei
\cite{Bayman}. Supported by this result, and due to the poor computing 
facilities available at that time, 
 only one configuration 
was included in the first applications of the shell model to the
description of the mother nucleus in $\alpha$-decay. The results were discouraging since the theoretical decay rates 
were significantly smaller than the corresponding experimental values by 4-5 orders of
magnitude \cite{Mang}. This failure arouse doubts about the validity of the
shell model itself \cite{Wilk}. It was later found that the inclusion of
more configurations in the description of the mother nucleus improved the
result dramatically. This, as well as the use of the BCS formalism to
describe superfluid and deformed nuclei, was an important theme during a 
rather long time. A detailed review of these developments can be found in
Ref. \cite{Lovas1998}, where even efforts performed to describe the clustering 
process in heavy nuclei are reviewed. The related subject of clustering in 
light nuclei is reviewed in e.g. Refs. 
\cite{Freer2014,Beck2014,vonOertzen200643,Freer2018}.

\begin{figure}[tb]
	\epsfysize=9.0cm
	\begin{center}
		\begin{minipage}[t]{16.5 cm}
			\centering{  
			\epsfig{file=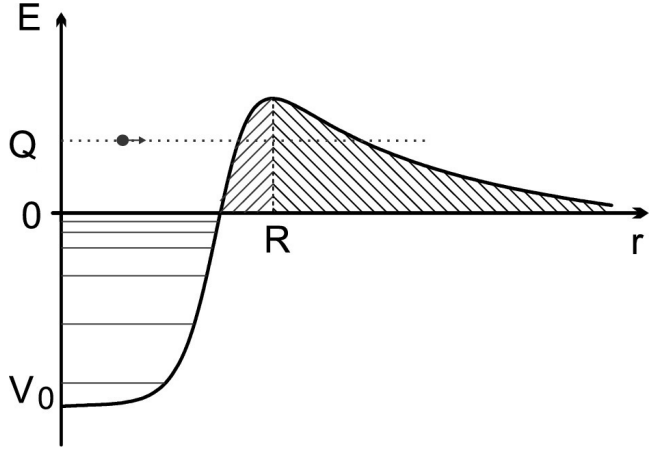,width=8.5cm}}
		\end{minipage}
		\begin{minipage}[t]{16.5 cm}
			\caption{A one-dimensional schematic view of the particle emission process through the Coulomb barrier. Adapted from Ref. \cite{WANG201783}.
				\label{figpq}}
		\end{minipage}
	\end{center}
\end{figure}

 It is
expected that the decays of the proton, the $\alpha$ particle and
heavy clusters can be simultaneously described by a two-step
mechanism, as indicted by the Thomas expression for the width, of first formation and then the penetration of the
particle through a static Coulomb and centrifugal
barriers. This is illustrated in Fig. \ref{figpq}. the charge-particle decay process can be evaluated in two steps in regions divided by the radius $R$: the inner process which describes the dynamic motion of the nucleons composing the emitting-particle inside the nucleus and the possibility for it to be formed and emitted, and the outer process which describes the penetration of the particle through the Coulomb and centrifugal barriers and are independent of nuclear structure effects. The emitting-particle formation amplitude that reflects the overlap between the parent and daughter wave functions and the intrinsic structure of the emitting particle. This scheme avoids the ambiguities of the deduced spectroscopic factor in relation to the surface effects and quantifies in a more precise manner the nuclear many-body structure effects. It is also valid for all charged particle decays. On the other hand, the spectroscopic factor is a model dependent quantity. It makes sense only if the same single-particle wave function are using in both the structure and the decay channel.

\section{Proton radioactivity}
We start this theoretical review of the developments in radioactive
particle decay during the last couple of decades with the microscopic
treatment of proton decay. Among all particle decay processes the simplest one is proton decay since in 
this case one avoids to deal with the most complicated feature of particle decay,
namely the formation and intrinsic structure of the decaying particle. Therefore we
start the discussions of the Thomas formula (\ref{Thomas}) by presenting this 
case, which is just fitted to understand features like 
e. g. the independence of the decay width upon the radius $R$. We will
briefly present the formalism, for details see Ref. \cite{MFL}.

The proton formation amplitude in Eq. (\ref{Thomas}) has the form,
\begin{equation}\label{faout}
{\cal F}_{p}(R)=\int d{\mathbf R} d\xi_d 
[\Psi_d(\xi_d)\chi(\xi_p)Y_l(\mathbf R)]^*_{J_mM_m}
\Psi_m(\xi_d,\xi_p,\mathbf{R}),
\end{equation}
where $d$, $p$ and $m$ label the daughter, proton and mother
nuclei, respectively. The variables $\xi$ label the necessary different degrees of freedom.
The internal wave function of the daughter nucleus is $\Psi_d(\xi_d)$. The
internal proton wave function, $\chi(\xi_p)$,
is the 1/2-spinor corresponding to  the emitted proton, which carries 
an angular momentum $l$. One does not need to consider the intrinsic structure of the emitting-proton
when evaluating the formation probability.

For simplicity we will assume that the decaying (mother) nucleus
is spherical and that it  consists of one proton outside a double closed shell. 
In this simple case the mother nucleus wave function is 
\begin{equation}\label{mother}
\Psi_m(\xi_d,\xi_p,\mathbf{R})=
[\Psi_d(\xi_d)\chi(\xi_p) Y_l(\mathbf R)]_{J_mM_m}\psi_p(R).
\end{equation}
and, therefore, the formation amplitude ${\cal F}_{p}(R)$ is just the 
single-particle wave function $\psi_p(R)$. The index $p$ labels
$(nlj)$, where $n$ is the principal quantum number and $(lj)$ the orbital and
total angular momenta of the proton moving in the central field induced by
the daughter nucleus. 

Since the mother nucleus decays,
the proton has to move in orbits lying in the continuum part of the 
spectrum. To be observable, however, the mother nucleus has to live a time
long enough for the detector to be able to measure it. The shortest mean time
$T$ that one can measure at present is of the order of $10^{-13}$ sec. That
implies that the widest resonance one can measure is of the order (with
$\hbar=6.6 \times 10^{-22}$ Mev sec)
$\Gamma=\hbar/T$=6.6$\times 10^{-9}$ MeV. This is an extremely narrow
resonance and therefore the proton may be considered to be moving in a
bound state. In other words one may consider that the proton wave function 
vanish at long distances. But for our purpose it is  
safer to apply the more realistic outgoing boundary condition. There are a
number of computer codes that allow one to evaluate outgoing wave functions. 
One that has proved to be very stable is the code GAMOW \cite{Gam,BARAN2018185}.
The outgoing boundary condition implies that the calculated energy 
${\cal E}_{nlj}$ is complex, i. e. it is ${\cal E}_{nlj}=
E_{nlj} -i\Gamma_{nlj}/2$, where $E_{nlj}$ is the position and $\Gamma_{nlj}$
the width of the resonance. The imaginary part (i. e.. the width) vanishes
and the real part is negative for bound and antibound states.   
  The evaluation of the wave function ${\cal F}_{nlj}(R)$ is performed by first 
choosing an appropriate central field, for instance a Woods-Saxon potential.
The code provides all possible states, i. e. the resonances, the bound and
the antibound states. For details see Ref. \cite{Gam}. An alternate code has been written by one of us by using the log-derivative method.

Only the Coulomb interaction is 
important at distances $R$ beyond the central field. Therefore the wave function at those distances has the form
\begin{equation}
{\cal F}^{Coul}_{nlj}(R)=N_{nlj}[G_{lj}(R)+iF_{lj}(R)]/R
\end{equation}
where $N$ is a normalization constant and $F$ and $G$ are
the regular and irregular Coulomb functions, respectively. The constant $N$
is determined by matching at the radius $R$ the wave function evaluated inside 
the nucleus, i. e. ${\cal F}_{nlj}(R)$ (Eq. (\ref{faout})) to 
${\cal F}^{Coul}_{nlj}(R)$. The constant $N$ becomes
\begin{equation}
|N_{nlj}|^2=R^2\frac{|{\cal F}_{nlj}(R)]^2}{F^2_{lj}(R)+G^2_{lj}(R)}
\end{equation}
Since $N$ is independent upon $R$, one sees that the Thomas expression 
(\ref{Thomas}) is also $R$-independent.
The constant $|N|^2$ is proportional to the width of the resonance \cite{MFL}. 

It is not difficult to extend the formalism above to deformed nuclei.
Assuming that the daughter nucleus is even-even the angular momentum projected 
wave function is given by 
\begin{equation}
\Psi^{JMK}_m=\sqrt{\frac{\hat J}{16\pi^2}} [{\cal D}^J_{MK}\chi_{_K}
+(-1)^{J+K}{\cal D}^J_{M-K}\chi_{_{\bar K}}]
\end{equation}
where standard notation was used \cite{MFL}.
The intrinsic single-particle wave function $\chi$ can be expanded in
spherical components as,
\begin{equation}\label{limdef}
\chi_{K}(r)=\sum_{j\leq K}\alpha_{lj}(r)[Y_l(\hat r)\chi_{1/2}]_{jK}
\end{equation}
where the orbital angular momentum $l$ is determined by the parity of the
state.

Decays from high lying states are not likely
because electromagnetic transitions from excited states are faster
than proton decay. But it has to be stressed that proton decay transitions 
from excited states has also been measured, for instance  in Ref. \cite{Sobot}.
For simplicity we will consider the most common case of ground state to
ground state transitions.  Therefore it is $J=K$. Since the daughter nucleus is
even-even the corresponding quantum numbers are $J_d=M_d=K_d$=0. As a 
result the angular momentum of the outgoing proton is $j_p=J=K$.

As in the spherical case above, at distance $R$ beyond the deformed mean
field only the Coulomb interaction is active. One thus has,
\begin{equation}
R\chi^{ext}_{_K}(R)=\sum_{lj}
N_{lj}[G_{lj}(R)+iF_{lj}(R)][Y_l(\hat r)\chi_{1/2}]_{jK}
\end{equation}
and as before the set of constants $N_{lj}$ are determined by matching the
external wave function $\chi^{ext}$ to the internal one, Eq. (\ref{limdef}).

Considering the angular momentum constraints discussed above and the
orthogonality of the partial waves, the partial decay width to the channel
$(l_pj_p)$ is
\begin{equation}
\Gamma_{l_pj_p} (R) = \frac{\hbar^2 k}{\mu}
\frac{R^2 \alpha^2_{l_pj_p}(R)}{F^2_{l_pj_p}(R)+G^2_{l_pj_o}(R)}
\end{equation}

This formalism was applied in Ref. \cite{MFL} to study the proton decay from
the ground state of the deformed nucleus $^{113}$Ce. It was found that
proton decay is a powerful tool to probe small components of the deformed
wave function which would be difficult to do with other probes.

\subsection{Systematic studies of Proton radioactivity and nuclear structure}

Proton emission was first observed rather recently in the context of nuclear
radioactivity. This happened in 1970 when the decay of $^{53}$Co from an 
excited high-spin isomer was measured \cite{Jackson,Cerny}. That the
mother nucleus was excited is not surprising  since the proton is bound to the
mother nucleus in its ground state. That is, the proton $Q$-value (which is proportional to the
proton kinetic energy at large distance) corresponding to the decay from 
$^{53}$Co(gs) to the ground state of  the daughter nucleus is negative. 
Therefore the proton can be emitted only from a state which is excited enough.

With the improvement of experimental facilities, which allowed one to
investigate nuclei lying close to the proton drip line, proton emission from
ground states could be measured. But for this to occur more than a decade 
had to pass \cite{Hoffman,Klepper}. Since then many proton radioactivity
cases have been observed.  
Nearly 50 proton decay events 
have been successfully observed in odd-$Z$ elements between $Z=53$ and $Z=83$ 
in the past few decades, 
leading to an almost complete
identification of the proton edge of nuclear stability in this region
\cite{doi:10.1146/annurev.nucl.47.1.541,Blank2008a}. 

Detailed review of the experimental
developments leading to the present status of proton radioactivity can be
found in Refs. \cite{doi:10.1146/annurev.nucl.47.1.541,Blank2008a,Page}. The first developments in the theoretical
treatment of this subject has been reviewed in Refs. \cite{Lovas1998,DELION2006113}.
The proton-emission process can be looked as a quantum
tunnelling through the Coulomb and centrifugal barriers
of a quasi-stationary state. Similar to $\alpha$ and heavy 
cluster decays that will be discussed below, the proton decay process can be divided into an ``internal 
region", where the compound state is restricted,
and the complementary ``external region". This division is such that in the
external region only the Coulomb and centrifugal forces are important and
the decaying system behaves like a two-particle system.

In these proton-decaying nuclei one usually does not have information on the binding
energies and, therefore, on the $Q$-values. But the prevalent quantity in all 
particle decay processes is the $Q$-value. It is therefore not surprising that
there have been  an strong interest in determining the $Q$-values of very 
unstable nuclei. One of the methods to do that use a linear extrapolation 
in cases where at least two consecutive values are known. The
corresponding $Q$-values uncertainties are calculated from the
uncertainties in the individual measured values. The results are
astonishingly good. For instance in the heavy isotopes of $Ta$, $Re$, $Ir$
and $Au$ the root-mean-square deviation is of 34 keV from the extrapolated 
estimates to the corresponding known values \cite{PhysRevC.83.014305}. With
the Q-values thus obtained one evaluates the half lives by using the WKB
approximation.

\begin{figure}
	\centering{
		\includegraphics[width=0.5\textwidth]{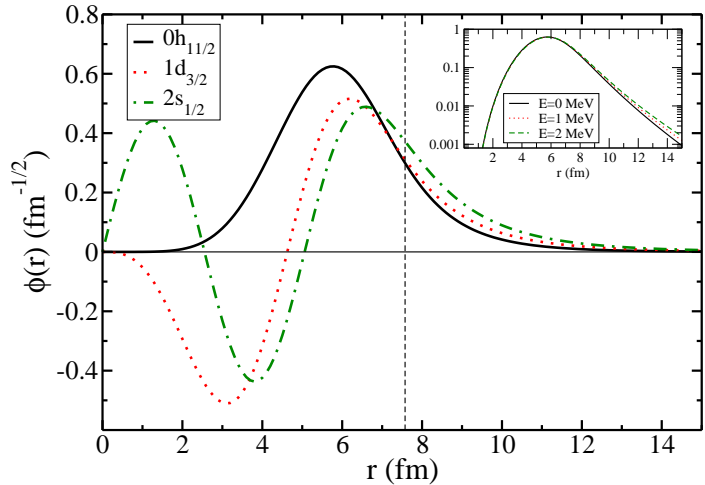}}
	\caption{(Color online) Single-proton wave functions in $^{151}$Lu for 
		different channels. The 
		inserted plot shows the wave functions of the $0h_{11/2}$ orbital with 
		different energies derived by changing the depth of the potential. Taken from Ref. \cite{Qi2012c}.
		\label{single}}
\end{figure}

The WKB approximation was also used to determine the  
partial half-lives for the proton and $\alpha$-decay branches.  One important feature of this work is that 
one could investigate the structure of the decaying nucleus by determining
the orbital from which the decay proceeded. For example, in  $^{160}$Re 
\cite{PhysRevC.83.064320}, it was thus found that the only
possibility for the decay to occur was that the proton was moving on
the $d_{3/2}$ orbital.
As expected, the  decaying single-particle orbitals that have been observed so far from proton-rich neutron between $Z=$ 53 and 83 involve in particular $d_{5/2,3/2}$, $s_{1/2}$, $g_{9/2}$ as well as $h_{11/2}$ and $f_{7/2}$. The decay from the $d_{3/2}$ orbital was expected to be slower than those from the neighbouring $s_{1/2}$ and $h_{11/2}$ orbitals (i.e., with smaller formation probability) \cite{PhysRevC.55.2255,PhysRevC.68.054301,Aberg1997}. However, recent more precise measurement tend to suggest no such strong hindrance
\cite{WANG201783,PhysRevC.96.064307}. In Fig. \ref{single} we plotted the $s$, $d$ and $h$ components of the single-particle wave functions in $^{151}$Lu as an example, which indeed show similar amplitude at the nuclear surface. That study also indicates that the 
nucleus may not be of moderate oblate deformation as suggested earlier \cite{PhysRevC.91.044322}. The ambiguity here is that, since the proton decay involves the major component of the Nilsson orbital, it will not be sensitive to the change in deformation even if the nucleus is modestly deformed.

Another feature that has to be considered is the influence (if any) of the  $Q$
value upon the formation probabilities. The
$Q$ value determines the penetrability and, therefore, the radioactive decay
process. The question is whether even the spectroscopic quantities are
affected by the $Q$ value. 
To analyse this we show  in the inserted plot of Fig. \ref{single}
the wave  functions of the $0h_{11/2}$ orbital under different energies. These were
obtained in Ref. \cite{Qi2012c} by changing the depth of the potential. As perhaps expected, the tail of the wave function changes dramatically as a function of the energy of the state or the depth of the potential, leading to dramatically different decay rates. On the other hand,
the formation amplitude (or more exactly, the single-particle wave function of the emitting particle) at the nuclear surface is not sensitive to changes
in the energy, i.e., in the $Q$ values. This is an important conclusion one can also draw from more complex calculations. It is important in particular in the theoretical description of proton-decay half-lives. If the calculated half-life varies as a function of certain inputs like nuclear deformation, one should carefully analyse the reason behind the variance is due to the change in nuclear structure (proton formation amplitude) or simply the decay $Q$ value which is relatively less sensitive to nuclear structure.

There have been extensive theoretical and experimental efforts 
studying the rotational bands of proton emitters as well as the influence
of triaxiality upon proton decay, particularly  $^{141}$Ho and the triaxially deformed nucleus 
$^{145}$Tm, as can be found in Refs.  
\cite{PhysRevC.58.R3042,PhysRevLett.86.1458,PhysRevLett.99.082502}.
$\gamma$ rays from excited states feeding proton-emitting ground- 
or isomeric-states  have been observed in $^{112}$Cs \cite{PhysRevC.85.034329}, 
$^{117}$La \cite{Liu201124}, $^{171}$Au \cite{Back2003}, 
and $^{151}$Lu \cite{Procter201379,PhysRevC.91.044322}. A multiparticle spin-trap $19^-$ isomer was discovered in $^{158}$Ta in 
Ref. \cite{PhysRevLett.112.092501}. The state is unbound to proton decay but 
shows remarkable stability. 
Structure calculations have been carried out for 
those nuclei. In Ref. \cite{PhysRevC.89.014317} the rotational band in 
$^{141}$Ho is described using the
projected shell model by taking deformed Nilsson quasi-particle orbitals as 
bases. The $^{145}$Tm is well described as the coupling of  deformed 
rotational core and the odd proton within the particle-rotor framework in 
Ref. \cite{PhysRevC.78.041305}.  In  Ref. \cite{danielK} the Schr\"odinger
equation corresponding to a triaxial potential was solved 
by using a coupled-channel approach. It was thus shown that the angular 
distribution corresponding to transitions to the
ground state is not sensitive to nuclear structure details, a feature which
is at variance with the $\alpha$-decay case. Instead, the decay width is very 
sensitive to triaxial deformations. It was thus concluded that proton decay 
is a powerful tool to
determine spin, as well as to uncover triaxial shapes in nuclei.
These studies reveal the importance of
proton decay as a deeper probe for nuclear structure properties. 

The proton
formation probability can indeed depend upon the
deformation of the decaying nucleus. In a well-deformed nucleus the decay
can proceed through one of the spherical components of the deformed orbit,
which can be very small in the case of large deformations. Therefore the
formation probability will be small. As a result, the decay will be  very much hindered. On the contrary, in spherical or weakly
deformed nuclei the decay proceeds through the only component that is
available or the large component and, as a result, the formation probability is large. Therefore 
proton decay is an important tool to investigate nuclei which cannot be
reached otherwise, especially nuclei which are beyond the drip lines.
For example, the decay of the
$h_{11/2}$ component of the Nilsson orbital $11/2^- [505]$ will not be as hindered as the decay from other smaller components.

An interesting case is the proton decay from the nucleus $^{109}_{53}$I 
\cite{Procter2011118} for which the lowest collective band starting from $7/2^+_1$ and the inner-band E2 
transition properties are observed to be very similar to those of ground state band in the even-even nucleus
$^{108}$Te with one less proton \cite{PhysRevC.84.041306} as well as those of the $7/2^+_1$ band in $^{109}_{52}$Te 
\cite{PhysRevC.86.034308}. Such a similarity indicates that the odd proton in $^{109}$I, which  
occupies the $g_{7/2}$ orbital, is weakly coupled to the $^{108}$Te daughter 
nucleus like a spectator. This simple scheme is also supported by complex large-scale shell-model and pair 
truncated shell model calculations \cite{PhysRevC.88.044332}.
The ground state of $^{109}$Te is 
predicted to be dominated by the coupling of a $d_{5/2}$ neutron to $^{108}$Te. 
Based on systematics of proton decay half-lives \cite{Qi2012c} and the level 
structure of I isotopes from Ref. \cite{nudat}, a similar $5/2_1^+$ state is 
also expected to be the ground state of $^{109}$I. However, it was not seen in 
the life-time measurement of Ref. \cite{Procter2011118}.

Another interesting topic is the competition between $\alpha$ and proton decays from the same nucleus. This has been observed in several nuclei including $^{185}$Ti, $^{177}$Tl and $^{171}$Au \cite{PhysRevC.63.044304,PhysRevC.59.R2979,Back2003}. There is no microscopic model description of that competition available so far. In addition, the significantly hindered proton decay from the intruder $1/2^+$ state in $^{185}$Ti has been observed in Ref. \cite{PhysRevLett.76.592}.

\subsection{Semi-empirical description of proton radioactivity}

A  simple formula to evaluate the half live of a mother nucleus against
proton decay was presented in Refs. \cite{Delion2006,DELION2006113}. This formula enables
the precise assignment of spin and parity for proton decaying states. The
only quantities that are needed are the half-life of the mother nucleus and
the proton $Q$ value. As a function of these quantities, corrected by the
centrifugal barrier, the experimental data of proton emitters with
Z$>$ 50 lie along two straight lines which are correlated with two regions 
of quadrupole deformation. This can be used in experimental searches in a
manner similar as the Geiger-Nuttall (GN) law for alpha decay is used.

Within the R-matrix framework, the logarithm of the decay half-life can be 
approximated by the so-called universal decay law (UDL) as \cite{Qi2012c}
 \begin{eqnarray}
\label{gn-3} 
\log T_{1/2}  &=&a\chi' + b\rho' +d l(l+1)/\rho' + c,
\end{eqnarray}
where $a$, $b$, $c$ and $d$ are constants determined by fitting 
available experimental data. $\chi'=A_{pd}^{1/2}Z_pZ_dQ_p^{-1/2}$, where
$A_{pd}=A_dA_p/(A_d+A_p)=A_d/(A_d+1)$ and $Q_p$ is the proton Q-value. 
$\rho'=\sqrt{A_{pd}Z_pZ_d(A_d^{1/3}+A_P^{1/3})}=\sqrt{A_{pd}Z_d(A_d^{1/3}+1)}. $
The coefficients $a$ to $d$ can be determined by fitting available 
experimental data. The UDL was firstly proposed to describe $\alpha$ and cluster decays (see below). It turns out to work well also for proton decays and fission.

\begin{figure}
	\centering{
\includegraphics[width=0.45\textwidth]{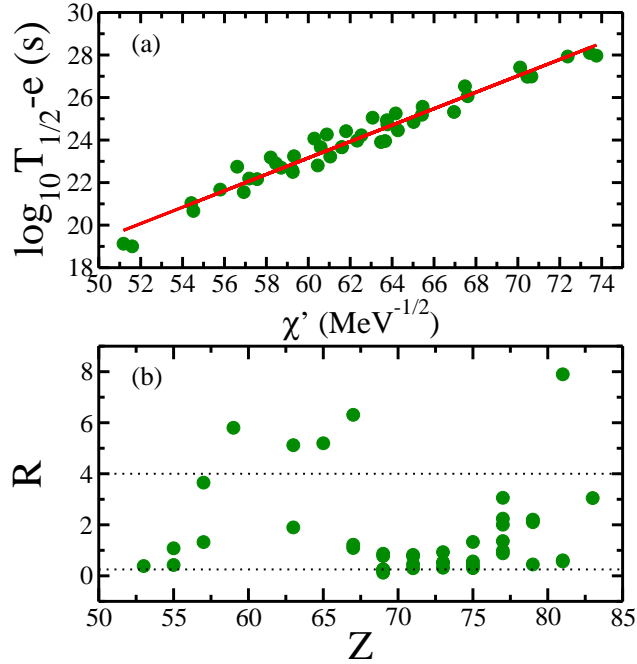}}
\caption{(Color online) Upper: UDL description of proton decay half-lives. 
Dots are experimental values. $e$ is given by 
$e=b\rho' +d l(l+1)/\rho' + c$. 
Lower: The ratio ${\cal R}=T^{{\rm Expt.}}_{1/2}/T^{{\rm Cal.}}_{1/2} $ as a function of the
charge number $Z$. Taken from Ref. \cite{Qi2012c}.  \label{fig1p}}
\end{figure}

The corresponding calculations 
reproduce 
well  available experimental data, as illustrated in Fig. \ref{fig1p} where
the upper panel depicts the quantity 
$ \log T^{{\rm Expt.}}_{1/2}- [b\rho' +d l(l+1)/\rho' + c]$ as a function of $\chi'$.
In the lower part of that Figure, the discrepancy between
experimental and calculated half lives, i.e., the ratio
${\cal R}=T^{{\rm Expt.}}_{1/2}/T^{{\rm Cal.}}_{1/2} $, is plotted as a function of the 
emitter charge numbers $Z$.
It is seen that most of the data can be reproduced by the calculation within a 
factor of 4, i.e., with $0.25\leq {\cal R}\leq4$. Larger discrepancies are seen for
emitters between $63\leq Z\leq 67$ and the isomeric $h_{11/2}$
hole state in  the $Z=81$ nucleus $^{177}$Tl, where the experimental decay 
half life is underestimated by the calculation by a factor of about 8.

\begin{figure}
\begin{center}
\includegraphics[width=0.45\textwidth]{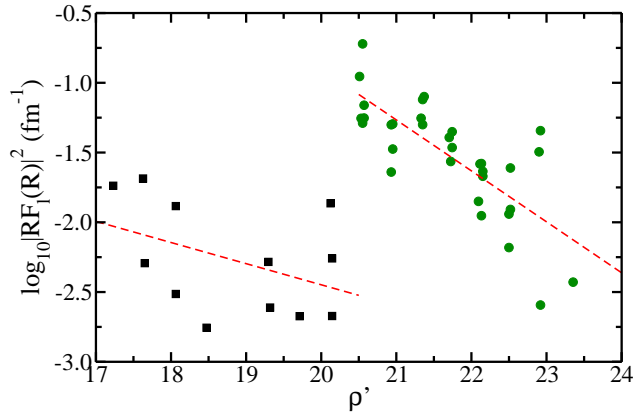}
\end{center}
\vspace{-0.8cm}
\caption{(Color online) The logarithm of the proton-decay formation 
probabilities $\log_{10} |RF_l(R)|^{2}$ extracted from experimental data as a 
function of $\rho'$ from \cite{Qi2012c}.  Squares correspond to nuclei with 
$N<75(Z\leq67)$ while circles are for $N\geq 75(Z>67)$.
\label{fvsrho}}
\end{figure}

The proton formation amplitudes $F_l(R)$ were 
extracted from the experimental half lives in Ref. \cite{Qi2012c} by taking $R=1.2 (A_d^{1/3}+1)$. They are plotted 
in Fig.~\ref{fvsrho} as a function of $\rho'$, where
one can notice two clearly defined regions. The region to the left of the figure corresponds to the decays of well 
deformed nuclei where the decay mostly involve small and low $l$ components of the deformed single-particle orbital. The proton formation probabilities decreases for these
nuclei as $\rho'$ increases. Then, suddenly, a strong transition occurs at the 
nucleus $^{144}_{~69}$Tm at $\rho'$=20.5. Here the formation probability 
acquires its maximum value, where the experimental uncertainty regarding the 
half-life (from where the formation probability is extracted) is still quite 
large, and then decreases again as $\rho'$ increases. 
The reason of the tendency of the formation probability in the figure is
related to the influence of the deformation as discussed above: In the left region of Fig. 
\ref{fvsrho},  the decays of the deformed nuclei proceed through  
small spherical components of the corresponding deformed orbitals and, 
therefore, the formation probabilities are small.
The right region of Fig. \ref{fvsrho} involves the decays of spherical orbits
as well as major spherical components of deformed orbitals  which give large 
proton formation amplitudes. 

\begin{figure}
	\centering{
	\includegraphics[width=0.45\textwidth]{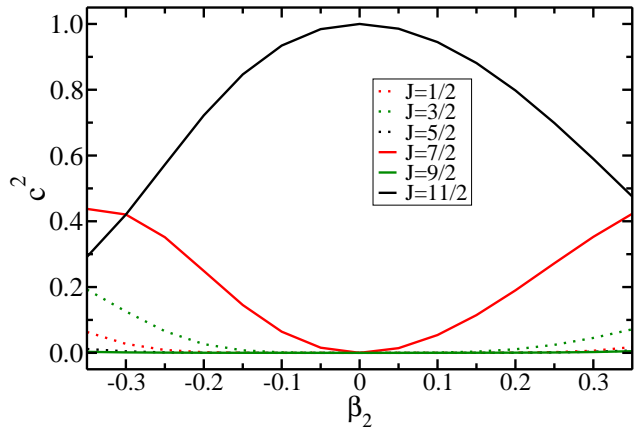}}
	\caption{(Color online) The squares of the expansion coefficients $c$ in terms of spherical components for the Nilsson orbit $1/2[550]$ in $^{145}$Tm as a function of deformation $\beta_2$  calculated with the deformed Woods-Saxon potential. \label{deformation}}
\end{figure}

The deformed single-particle orbital can be expanded in terms spherical harmonic oscillator single-particle wave functions. As an example, in Fig. \ref{deformation} we show the expansion of the $\Omega=1/2$ state, i.e., $1/2[550]$ out of $h_{11/2}$, where many orbits with $j\leq\Omega$ contribute. It is seen that the largest components correspond to the $h_{11/2}$ and $f_{7/2}$. The decay of the $f_{7/2}$ component can be favoured due to the smaller centrifugal barrier although the corresponding coefficient is relatively smaller.

\begin{figure}
	\centering{
\includegraphics[width=0.45\textwidth]{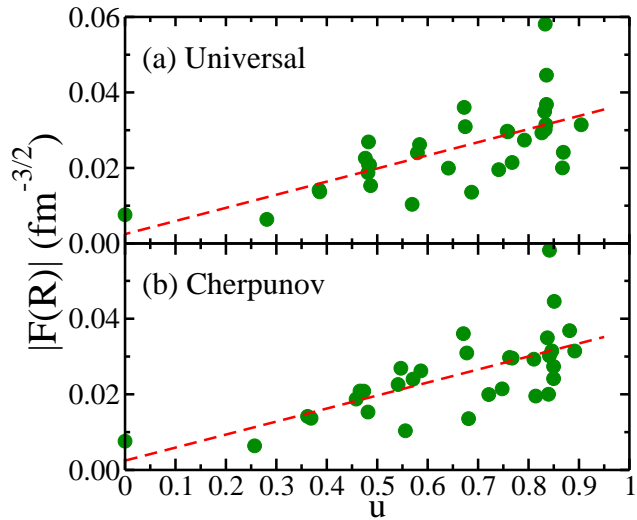}}
\caption{(Color online)
The formation amplitudes $|F_l(R)|$ extracted from experimental data for proton
decays of nuclei  $N\geq 75(Z>67)$ as a function of $u$ calculated from 
BCS calculations using different Woods-Saxon mean fields. From Ref. \cite{Qi2012c}. 
\label{fvsu}}
\end{figure}

In 
the BCS approach the formation amplitude at a given radius $R$ is
proportional to the product of the occupancy $u$ times the single-proton wave 
function $\psi_p(R)$. Therefore the tendencies seen in Fig. \ref{fvsrho} may
be due to the BCS amplitudes or the radial wave functions. 
In Fig. \ref{fvsu} the formation probabilities $|F_l(R)|$ extracted 
from experiment for the case of proton
decays corresponding to nuclei with $N\geq 75(Z>67)$ are plotted as a function of $u$. 
The $u$ values were calculated by using a Woods-Saxon potential with
the  universal and Cherpunov parameters which give similar results. A striking feature is that the values of
of the formation probabilities increases with $u$. One can thus conclude that the fluctuations in the experimental 
formation amplitudes found above
for nuclei with $N\geq75$ are mainly due to fluctuations of the $u$ values.
It is also true for cases which departs from the UDL and
correspond to the decays of hole states. This occurs in the isomeric state of 
$^{177}$Tl (as already pointed out above) and also in the ground state 
of $^{185}$Bi.

\section{Two-proton decay}

One would expect that the most likely form of two-nucleon decay is the
emission of deuterons, which are bound and would be as easy to detect as 
protons. However that is not the case since so far there is no known 
nucleus that emit deuterons. In other words, the deuteron 
Q-value is always negative for all known nuclei. Although proton decay exists the
decaying nucleus becomes stable against deuteron decay due to the large 
binding energy of the added neutron.
One may argue that this feature would be even more conspicuous in the decay
of $\alpha$-particles, with two neutrons on top of two-protons. The reason
why this is not the case is that the binding energy of the
$\alpha$-particle is very large, with a value of $B_\alpha$ = 28.3 Mev.
The deuteron, instead, has a binding energy of $B_d$=2.2 MeV. This
hughe difference in binding energies explains why $\alpha$-decay is the most
likely form of particle radioactivity. As an example one may consider the
classical example $^{212}$Po(gs) $\rightarrow\alpha$+$^{208}$Pb(gs). The 
Q-value in this case is
$Q_\alpha$=8.95 MeV. Instead for a deuteron decaying from $^{212}$Po(gs), i.
e.  $^{212}$Po(gs) $\rightarrow d$+$^{210}$Bi(gs; $1^-$) the
Q-value is $Q_d$=-8.6 MeV. One sees that if the deuteron would have had the
same binding energy as the alpha particle, then it would decay from 
$^{212}$Po(gs) with a Q-value of 17.5 MeV, which is nearly double as much as
the one carried by the $\alpha$-particle. It is worthwhile to point out that
in this hypothetical case
the deuteron decay would be somehow hindered by the small $l=1$ centrifugal 
barrier.

The same mechanisms that forbid the decaying of deuterons also prevent 
$^{3}$H (triton) and $^{3}$He decays. This remarkable feature was already 
noticed in Ref. \cite{gol60}. In this pioneering paper, in which one- as well 
as two-proton decay processes were predicted, Goldansky pointed out that a  
very curious effect emerges, namely that in isotopes  which are stable against 
proton and alpha decay, two-proton decay may be observed.
Therefore two-proton radioactivity  is a very exotic  
mode of decay which is energetically possible in some nuclei lying beyond
the proton drip line only. The 
$^6$Be
nucleus is the lightest two-proton ground-state emitter with a $Q$ value $Q_{2p}=1.372$ MeV. Moreover, since the Coulomb barrier would hinder 
the decay of protons one expects that two-proton decay will be possible only 
in light and medium nuclei \cite{RevModPhys.84.567,Blank2008a,0034-4885-71-4-046301}, including $^6$Be, $^7$B (1.42 MeV), $^8$C (2.11 MeV) and $^{12}$O (1.638 MeV) where the numbers in the brackets are the corresponding $Q_{2p}$ values extracted from Ref. \cite{1674-1137-41-3-030003}. On the experimental side, the 2p decay was firstly observed in $^{45}$Fe \cite{gio2002,pft2002} and then in $^{19}$Mg\cite{PhysRevLett.99.182501}, $^{48}$Ni \cite{PhysRevC.72.054315,PhysRevC.83.061303}
, $^{54}$Zn \cite{blank2005,PhysRevLett.107.102502}, and  $^{67}$Kr\cite{PhysRevLett.117.162501}.   The possible observation of two-proton decay from intermediate-mass nuclei was discussed recently in Ref. \cite{JOSS2017703}.

As in all forms of radioactive particle decay the two-proton decay rates are 
expected to be extremely sensitive to the corresponding
separation energy (that is to the two-particle Q-value) and therefore a 
reliable estimate of the Q-values  would give a measure of the possibility 
that the two-proton decay channel is dominant. This was the main theme
in one of the first microscopic treatments of two-proton decay \cite{bro91}.
Within the framework of the shell model the process of direct two-proton decay 
of nuclei with $Z = 22-28$ on the proton drip line was considered. On the basis 
of $0d_{3/2} - 0f_{7/2}$ shell-model mass extrapolations the nuclei
$^{39}$Ti, $^{42}$Cr, $^{45}$Fe, $^{48}$Ni and $^{49}$Ni were found to be 
bound to single-proton decay but unbound to two-proton decay. 
The spectroscopic factors and lifetimes were evaluated assuming that the
decaying two protons are clustered with an internal
motion of the two protons (diproton) in a $0s$ state. Using the so-called 
cluster overlap approximation \cite{chu78}, which essentially consists 
of the overlap of the two-proton shell-model wave function 
with the diproton cluster wave function, all spectroscopic factors are near
unity. This is not in agreement with experimental data. One failure maybe
related to the evaluated binding energies. The diproton approximation may
also be incorrect. All these will be important in the developments that
follows, as seen below.  

A similar calculation was performed in Ref. \cite{orm96}, but with the
improvement that the shell model treatment of all nuclei
in the study were done coherently. In particular, the Coulomb energy shifts 
were computed using
the same shell  model space and interaction for both the ‘‘purely’’ fp-shell
and the cross-shell nuclei. This maybe the reason why the results here are
more in agreement with experimental data than in \cite{bro91}. 
Thus, it was correctly predicted that $^{45}$Fe was a good candidate to be 
observed experimentally.

Another similar calculation but performed within the framework of the 
Hartree-Fock-Bogoliubov and relativistic mean field, including various 
effective interactions, was performed in Ref. \cite{naz96}. Here the main
conclusion regarding two-proton emission was that diproton emission half-lives 
depend mainly on the two-proton separation energy and very weakly on the
intrinsic structure of diproton emitters. It was also found a very weak 
dependence of the decay width upon the details of the proton
potential. The paper concludes by stating that the results of the 
calculations justifies the simple estimates of the Refs. \cite{bro91,orm96}
mentioned above. 

An approach that relies on the properties induced by the pairing interaction
upon the ground state of decaying even-even nuclei was introduced in Ref. 
\cite{del13}, where the decays of $^{45}$Fe and $^{58}$Ni were analysed.
The determining variables in this treatment are the distances $r_1$ and $r_2$
of the protons from the centre of the mother nucleus, and the angle $\theta$
between the corresponding vectors $\vec r_1$ and $\vec r_2$. The formalism 
takes into account that due to the pairing interaction the two 
protons are clustered on the surface of the mother nucleus. At the same time
one uses the property that the protons occupy paired states. Therefore
initially they are at the same distance from the centre of the mother
nucleus and carry the same energy. The evolution of these three degrees of
freedom, i. e. $r_1$, $r_2$ and $\theta$, are
followed according to the dynamics determined by the Schr\"odinger equation.
It was thus found that the half lives are strongly dependent upon the 
strength of the pairing interaction (see Figs. 7 and 8 of Ref. \cite{del13}). 
That is, two-proton decay is an excellent tool to probe the pairing force. 
An important prediction is that the decay width is strongly 
peaked around the symmetric configuration $r_1 = r_2$, in the angle interval 
$\theta = 45^o \pm 2^o$.

Although these calculations treat correctly the spectroscopic quantities
related to the mother nuclei, the decaying process itself, including  the
relative motion of the decaying two protons, was inadequate. A proper
treatment of this starts by noticing that the two-proton decay process may 
occur through three possible mechanisms: 
(i) sequential  emission of protons via an intermediate state (this was
called "democratic decay" in Ref. \cite{boc89}, a name used by some authors
afterwards), (ii) simultaneous emission of protons
and (iii) diproton emission, as the one already discussed above, 
i.e. emission of a $^{2}$He cluster with very
strong pp correlations. 

It is difficult to distinguish experimentally among these decay
possibilities. Thus, one of the first observation of two-proton emission was the 
decay from a
narrow resonance at 7.77 MeV in $^{14}$O \cite{bai96}. In Fig. 4 of that
paper the two-proton decay is seen to proceed as a sequential emission. 
Instead, in the two-proton decay of $^{31}$Ar \cite{axe98} the simultaneous 
emission of the two protons fits the data well. But sequential decays 
might also describe the experimental data. The $\beta$-delayed two-proton emission from $^{22}$Al was observed recently in Ref. \cite{WANG201812}. Based on Monte Carlo simulations, they claim that the involved two-proton decay (from the excited states of $^{22}$Mg to $^{20}$Ne) can be explained as a mixture of di-proton and sequential decays.
 
Another example is the decay of $^{18}$Ne($1^-$), which is a resonance 
lying at 6.15 MeV \cite{gom01}. As clearly seen in Fig. 4 of this paper the 
experimental data does not differentiate between diproton $^{2}$He emission and 
direct three-body decay (which here is called "democratic decay"). The possible diproton decay from $^{17}$Ne was studied in Ref. \cite{PhysRevC.96.025807} where an upper limit for the decay width is provided. The two-proton decay from the first excited state of 
$^{16}$Ne($2^+$) was reported in Ref. \cite{PhysRevLett.113.232501,PhysRevC.92.034329}, which again showed patterns of both sequential and prompt two-proton decays. The sequential decay from that nucleus was calculated in a simple potential model in Ref. \cite{PhysRevC.96.064313} and in Ref. \cite{GOLUBKOVA2016263}. The unbound $^{15}$Ne was observed in Ref. \cite{PhysRevLett.112.132502}.

The prominent feature in these examples is that the experiments guide the
theoretical undertakings carried out to describe this complicated decay mode.
An example of this is the decay from $^{45}$Fe, which was measured at Ganil 
\cite{gio2002} and GSI \cite{pft2002}. The decay from $^{54}$Zn was also 
measured at Ganil \cite{blank2005} were even the previous results corresponding 
to $^{45}$Fe could be confirmed \cite{dos2005}. A detailed report of the 
experimental efforts can be found in Ref. \cite{gio2013}.
In order to describe these decays in Ref. \cite{bro03} a diproton emission
was assumed, which from a theoretical viewpoint is the simplest form of decay
among the three cases mentioned above. The results of the calculation agree
well with the experimental data if some reasonable assumptions are
fulfilled. In Ref. \cite{PhysRevLett.117.162501}, the 2p decay from  $^{67}$Kr was observed.
The decay energy was determined to be 1690(17) keV, the
 2p
decay branching ratio is 37(14)\%, and the half-life 
is 7.4(30) ms which is considerably below earlier theoretical expectations. The observation of prompt 2p decay in that experiment was questioned in Ref. \cite{PhysRevC.95.021601}. Meanwhile, in Ref. \cite{PhysRevLett.120.212502} it was argued that 2p decay can be significantly increased by the deformation effect which leads to a deformed $Z = 36$ subshell and large $l=1$ proton components at large oblate deformation.
Such expectations should be compared with nuclear structure model calculations.

A more rigorous treatment of the decay should consider the two-proton emission 
within a realistic three-body model. This was proposed in Ref. \cite{gri00}.
In this model the hyper-spherical harmonic method formulated in Ref. 
\cite{zhu93} is used. 
Within this formalism, which includes all two-proton decay modes, 
the “diproton” model is compared with the three-body calculation
corresponding to the decay from the ground states of $^{19}$Mg and $^{48}$Ni.
It was found that the decay width provided by the diproton model is much
larger than the one predicted by the three-body calculation, as seen in Fig.
2 of Ref. \cite{gri00}. As we will see below, this limitation of the diproton
model is not seen in all cases. This three-body model was used to analyse a large number of two-proton decay
cases in the past \cite{gri02,gri03a,gri03b,gri03c} and even at present (see
Ref. \cite{gri17} and references therein). This is because the formalism 
contains all ingredients determining the decay process. However its
application is not easy, which explains why mainly only one group uses it, as
can be gathered from the references regarding this method. 

A number of other approaches have been proposed. In Ref. \cite{tom14} a
time-dependent method was employed to analyse the role played by  diproton
correlation in two-proton decay from the proton reach isotope $^{6}$Be.
The two-proton emission is
described as a time evolution of a three-body metastable state. Introducing
a realistic model Hamiltonian which reproduces well experimental two-proton
decay widths it was shown that a strongly correlated diproton emission is a
dominant process in the early stage of the two-proton emission. When the
diproton correlation is absent, the sequential two-proton emission competes
with the diproton emission, and the decay width is underestimated. This
feature shows  that the diproton model may not be as limited as indicated
by the three-body model mentioned above. 
In Ref. \cite{tom17} the same time-dependent treatment of $^{6}$Be is
applied to analyse the influence of a schematic density dependent pairing 
force. It is found that with such simple pairing force it is impossible to
describe simultaneously the two-proton decay width, the two-proton Q-value
and the two-nucleon scattering length. It is concluded that to achieve a
complete agreement with all those experimental quantities the pairing force
has to be elaborated farther. But this would  harm considerably the simplicity 
of the model and no additional development took place.

Since the two protons in the decay process may be trapped within the mother
nucleus during a long time, one may expect that long-lived mother nuclei
should show a halo corresponding to the two protons orbiting the daughter
nuclei before decaying. This was analysed in Ref. \cite{sax17}.

Another related subject is the possible existence of two-neutron decay and even tetra-neutron resonance. Those states can be analysed in principle in the same framework as mentioned above but influence of the continuum may become more important there.

Concluding this Section one sees that two-proton decay has shown to be a
powerful tool to study properties of nuclei lying on and beyond the proton
drip line.

\section{$\alpha$ decay}
There have been long-standing experimental interest in conducting nuclear alpha decay studies which 
not only carry important information on nuclear structure but are also very useful even today for isotope 
identification via $\alpha$-decay tagging
\cite{Duppen18,DEVARAJA2015199,PhysRevC.96.014324,SUN2017303,YANG2018212,PhysRevC.94.054301,PhysRevC.96.064314,PhysRevC.97.064602,PhysRevLett.115.242502}.
Moreover,
the importance of $\alpha$ particle capture reactions (or the inversed $\alpha$-decay process) for 
nucleosynthesis has been investigated during a long time. One of the most prominent cases
in this line is the famous 
${\alpha}$-capture to  the so-called Hoyle state in $^{12}$C, which is essential to the 
nucleosynthesis of carbon. The direct 3$\alpha$ decay or breakup from  $^{12}$C has been measured 
recently in Refs. \cite{PhysRevLett.119.132501,PhysRevLett.119.132502}, which
provides constraints for the many theoretical models, both microscopic and empirical, that have 
been applied to study the clustering property of the nucleus.

Most $\alpha$ emitters concern proton-rich or neutron-deficient nuclei. In principle, it can also be relevant for the astrophysical rapid neutron capture process (r-process) in  actinides like Pb and Bi \cite{2018arXiv180504637H} and in superheavy nuclei \cite{PhysRevC.84.044617}. The competition between fission
and $\alpha$  decay under typical r-process conditions was studied recently in Refs. \cite{PhysRevC.85.025802,PhysRevC.97.034323}.
The relevance of $\alpha$ decay of $^{210}$Po in the termination of the slow neutron capture process (s process) was re-measured recently in Ref. \cite{PhysRevC.97.034303}. 

As already pointed out in the Introduction $\alpha$-decay is not only the
most common of all particle decay processes but it is also, through the
understanding of the penetration of the $\alpha$-particle through the
coulomb and centrifugal barriers, the most
important event that confirmed the probabilistic interpretation of
quantum mechanics. The amount of work related to this subject is huge but
there have also been many reviews describing that work. Among the latest of
these is Ref. \cite{Lovas1998}. Here we will only
report the advances that took place in this field since that time.
There have been significant developments during this period, in the
microscopic as well as in the effective approaches to this problem. We will
first report the microscopic developments.  

\subsection{The microscopic description of $\alpha$ decay}
Before entering into the developments of this difficult and fascinating 
subject during the last decades we will very briefly describe the background 
upon which those developments are founded. 

The microscopic treatment of $\alpha$-decay started with the study of
scattering processes where a compound nucleus may decay into channels 
$c$ and $c'$. For our purposes  the main feature of these studies was the
parametrization of the corresponding S-matrix as formulated in Ref. 
\cite{brwi}. The resulting Breit-Wigner formula reads,
\be
\label{BW}
S_{cc'}=\delta_{cc'}-i\frac{\Gamma_c^{1/2}\Gamma_{c'}^{1/2}}
{{\cal E}-E+i\Gamma /2} 
\ee
where $E$ and $\Gamma$ are the position and width of the resonance induced
by the trapping of the compound nucleus before decaying into the various
open channels. The decay into channel $c$ is characterized by the partial
decay width $\Gamma_c$. Therefore the total width is $\Gamma=\sum_c
\Gamma_c$.   

The poles of the S-matrix are the complex energies ${\cal E}=E-i\Gamma /2$. 
At these energies the compound nucleus lives a mean time 
$T=\hbar/\Gamma$. Since the nucleus decays the corresponding incoming wave
is negligible compared with the outgoing wave. At the limit of an infinite
value of $T$ the width vanishes and the compound nucleus is bound. 

Eq. (\ref{BW}) was based on an intuitive argument. A strict  quantum mechanics
derivation of this formula was performed within the framework of the
R-matrix theory in Ref. \cite{TW}. As pointed out in the Introduction, this
formalism was used by Thomas to derive the expression of the 
$\alpha$-decay width given by Eq. (\ref{Thomas}).   

For the case of $\alpha$-decay the formation amplitude of the alpha particle
is given by 
\begin{equation}\label{foram}
{\cal F}_c(R)=\int d{\mathbf R} d\xi_d d\xi_{\alpha}
[\Psi_d(\xi_d)\phi(\xi_{\alpha})Y_l(\mathbf R)]^*_{J_mM_m}
\Psi_m(\xi_d,\xi_{\alpha},\mathbf{R}),
\end{equation}
where $\xi_d$ and $\xi_{\alpha}$ are the internal degrees of freedom
determining the dynamics of the daughter nucleus and the $\alpha$-particle
respectively. The wave functions $\Psi_d(\xi_d)$ and
$\Psi_m(\xi_d,\xi_{\alpha},\mathbf{R})$ correspond to the daughter and
mother nuclei. The $\alpha$-particle wave function has the form of a
$n=l=0$ harmonic oscillator wave function in the neutron-neutron relative 
distances $r_{nn}$, as well as in the proton-proton distance $r_{pp}$ and 
in the distance between the mass centres of the $nn$ and $pp$ pairs
$r_{np}$, i. e. \cite{TonAri},
\be
\label{ta}
\phi(\xi_{\alpha})=\sqrt{\frac{1}{8}}(\frac{\nu_\alpha}{\pi})^{9/4}
exp[-\nu_\alpha(r_{nn}^2+r_{pp}^2+2r_{pn}^2)/4]\xi_\alpha
\ee
where $\xi_\alpha$ is the $\alpha$-spinor corresponding to the lowest
harmonic oscillator wave functions. That is the orbital angular momenta 
are $l_{nn}$=$l_{pp}$= $l_{np}$=0 and the same for the spin part. Therefore 
the total angular momenta are $L=S=0$.
The quantity  $\nu_\alpha$ is the $\alpha$-particle harmonic 
oscillator parameter. 

The calculation of the formation amplitude (\ref{foram}) is the most difficult 
task in the evaluation of the width. From a microscopic point
of view one has to be able to describe the mother wave function well outside
the nuclear surface, where at the matching point $R$ only the Coulomb 
interaction is present. During a time in the early 1970's there was
a long argument about the importance of the Pauli principle acting upon 
nucleons in the $\alpha$-cluster and those in the daughter nucleus 
\cite{flies}. But eventually it was shown that the influence of Pauli 
exchanges is negligible at these long distances. For details see e. g. 
Ref. \cite{Lovas1998}.

The most difficult challenge in the calculation of the formation amplitude 
is to describe properly the small but crucial daughter times $\alpha$-cluster 
component in the mother wave function $\Psi_m(\xi_d,\xi_{\alpha},\mathbf{R})$.
As seen from Eq. (\ref{foram}) it is this component that contributes
to the formation amplitude. In other words one should be able
to describe at the point $R$ the clustering of the four nucleons that
eventually becomes the $\alpha$-particle. 
This task was pursued by many researches \cite{Lovas1998}.
While struggling with this problem a number of other important nuclear
properties where found. Below we will present these developments which occur
through studies performed within the two most current microscopic
frameworks, namely the shell model and the BCS theory.

\subsection{Shell model treatment of the formation amplitude\label{shellmo}}

The shell model provides an excellent representation to describe nuclear
properties. In the case of the $\alpha$-formation amplitude the main region
that such representation should describe is at the matching point $R$, as
discussed above. 
Therefore the representation should  include states which are not 
negligible at such distances. One thus sees from the beginning that high
lying single-particle states, which extend far out in space, should be an
important part of the shell-model representation.
But this representation should also 
be able to describe the clustering of the four nucleons that constitute the
$\alpha$-particle. Here is the core of the problem facing microscopic
calculations of cluster decay. 

The problem of describing processes on the nuclear surface is old. At the
beginning of two-particle transfer studies one faced the same problem. The
calculation of the two-particle transfer cross section required a precise
knowledge of the corresponding form factor on the target nuclear surface.
Since the asymptotic behaviour of the form factor is determined
by the two-particle separation energy one chose as binding energy for all
single particle states half that energy \cite{BayKal,IbaBay}. This gave rise
to the introduction  of the Sturm-Liouville representation to describe 
processes occurring on the nuclear surface. In this representation one uses
eigenstates of a realistic central potential, e. g. Woods-Saxon, which has
eigenvalues with proper asymptotic behaviour. But all these eigenstates are
evaluated at the same bound energy, for instance at half the energy of the
pair of nuclei to be transferred. Details about the Sturm-Liouville
representation can be found in Ref. \cite{JBang}. Although this
representation is well suited to describe the asymptotic behaviours of
nuclear functions, its application is not very convenient as compared with
harmonic oscillator (ho) representations. Besides the easiness of dealing
with ho eigenfunctions, they  allow
one to evaluate the $\alpha$ formation amplitude analytically. This is
because the $\alpha$-cluster function itself is a product of ho functions.
Therefore most microscopic calculations of $\alpha$-decay have been
performed by using ho representations. We will see this below. We will also
see the advantage of other representations introduced in order to describe
processes in the continuum part of nuclear spectra.     

The most simple case in the calculation of the formation amplitude 
(\ref{foram}) is when the numbers of neutrons and protons in the daughter
nucleus are both magic numbers. This implies within the
shell model that the daughter nucleus has no effect in the formation
amplitude if, as it is usually the case, no core excitations are considered.
The shell model description of the mother
nucleus can be performed in terms of uncorrelated two neutron and two
proton states or in term of correlated (physical) ones. This is possible
because the uncorrelated and correlated states are related by a unitary
(rotational) transformation in the Hilbert space. 
The preferred decaying nucleus fulfilling these condition is $^{212}$Po(gs).  
The corresponding wave function can then be written within the shell model as,
\begin{equation}
\label{msmwf}
|^{212}{\rm Po}(gs)\rangle=\sum_{\alpha_2 \beta_2} X(\alpha_2\beta_2;gs) 
|^{210}{\rm Pb}(\alpha_2)\otimes^{210}{\rm Po}(\beta_2)\rangle
\end{equation} 
where $\alpha_2$ ($\beta_2$) labels two-neutron (two-proton) states.
Due to the pairing correlation one expects that in the above expression the
components with $\alpha_2=gs$ and $\beta_2=gs$  are overwhelmingly dominant.
Therefore one may write,
\begin{equation}
\label{msmpai}
|^{212}{\rm Po}(gs)\rangle=|^{210}{\rm Pb}(gs)\otimes^{210}{\rm Po}(gs)\rangle
\end{equation} 

We will give a brief introduction to calculations based in the expansion 
(\ref{msmpai}). A review, including
references, can be found in \cite{Lovas1998}.

In the first alpha-decay evaluation of $^{212}$Po(gs) using this expression
was found that the inclusion of 
many configurations increased the width by many orders of magnitude.
This can be understood since on the surface of the nucleus, where the 
$\alpha$-particle is formed, the continuum part of the single-particle
representation (or very high lying shells in a bound representation) is
important. But even including up to 16 major harmonic oscillator shells the
absolute decay width was smaller than the experimental one in spherical
nuclei \cite{TonAri}. 

It was also found that the huge increase of the decay width with increasing
number of configurations was mainly due to the clustering of 
the two neutrons and two protons induced by the valence shells. 
The pioneers in $\alpha$-decay \cite{Ras65} already suspected that the 
increase of the decay width with the number of configurations was
due to the clusterization thus described.
But it has to be noticed that higher lying
shells contribute to the clustering to a lesser extent \cite{CLUS}.

Yet with all
shells included, that is the valence plus the high lying shells, the decay 
width was still too small by one order of magnitude. Taken into
consideration only one configuration in Eq. (\ref{msmpai}) implies that the 
neutron-proton interaction is neglected. To correct this deficiencies other 
configurations (such as 
$|^{210}{\rm Pb}(2^+_1)\otimes^{210}{\rm Po}(2^+_1)\rangle$) were also
taken into account, but the corresponding  calculations did not improve 
significantly. This is expected since neutrons and protons occupy different
major shells with different parity and the neutron-proton interaction matrix
elements are therefore hampered. 
In view of these drawbacks it was assumed that the only possibility left to 
increase the calculated value of the width was to include the nuclear 
continuum, and the neutron-proton clustering, in a more
realistic fashion than the one provided by the bound ho states. This we will 
review below.

The evaluation of $\alpha$ formation amplitude involves the evaluation of the overlap between the corresponding proton and neutron radial functions in the laboratory framework with
the $\alpha$-particle intrinsic wave function as defined in the centre of mass framework (see, e.g., Ref. \cite{PhysRevC.69.044318}). The transformation can be relatively easily handled if the radial wave functions are defined within the harmonic oscillator basis due to its intrinsic simplicity. This is also the season that the harmonic oscillator representation is used in most \textit{ab initio} and shell-model configuration interaction calculations.
More realistic calculations have been based on phenomenological Woods-Saxon and Nilsson single-particle single-particle states. A single particle basis consisting of two different harmonic oscillator representations was introduced in Ref. \cite{PhysRevC.54.292}.
An additional attractive pocket potential of a Gaussian form was introduced on top of the Woods-Saxon potential in  Ref. \cite{PhysRevC.87.041302} in order to correct the asymptotic behaviour of the  $\alpha$ formation amplitude. The mixture of shell model and cluster wave functions was considered in Ref. \cite{VARGA1992421} and was applied to describe the decay of the ground state of $^{212}$Po.  the amount of core + clustering in the parent state and the corresponding spectroscopic factor that can reproduce experimental decay half-life are found to be 0.30 and 0.025, respectively.

\begin{figure}
\begin{center}
\includegraphics[scale=0.4]{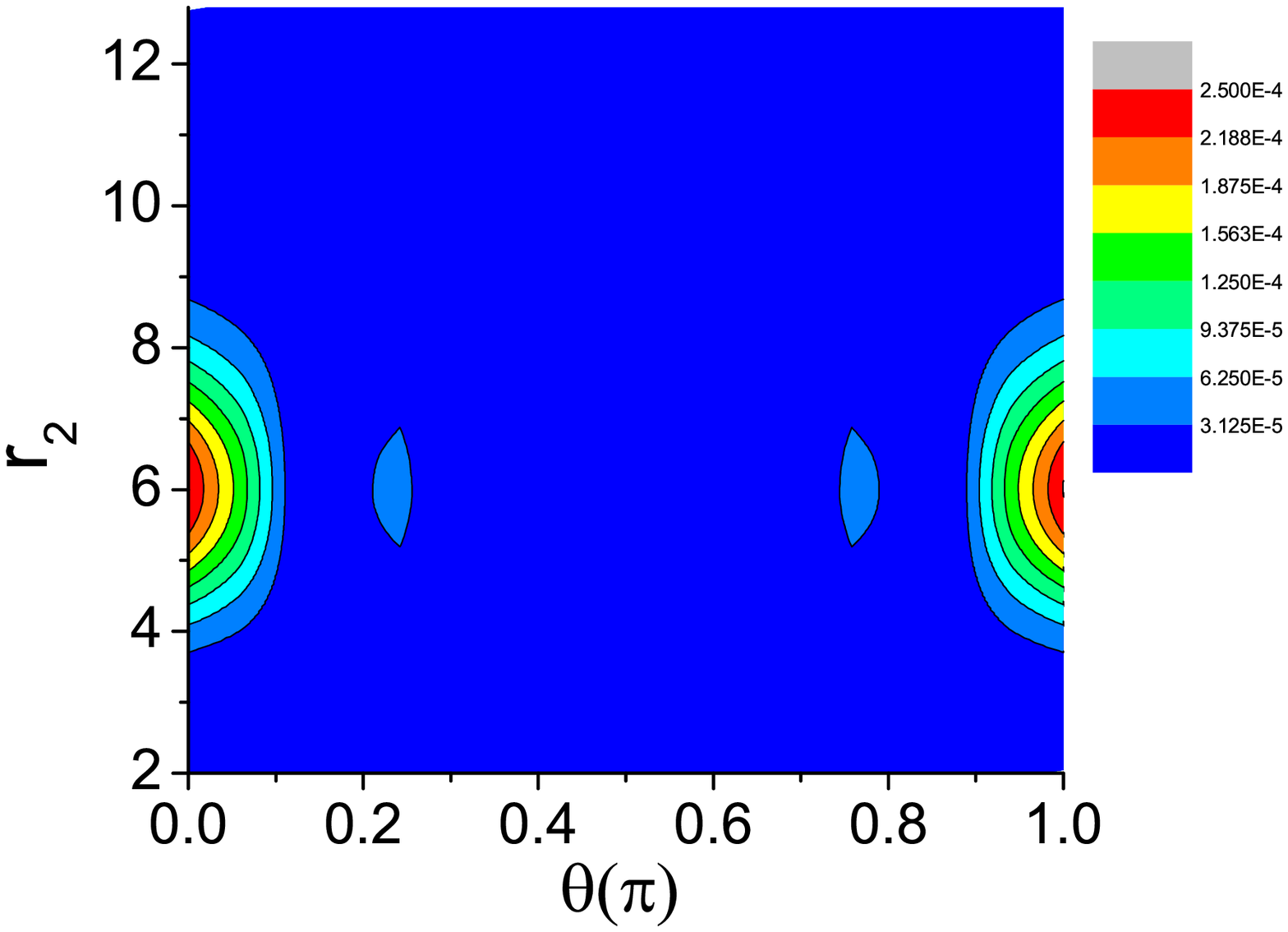}
\includegraphics[scale=0.4]{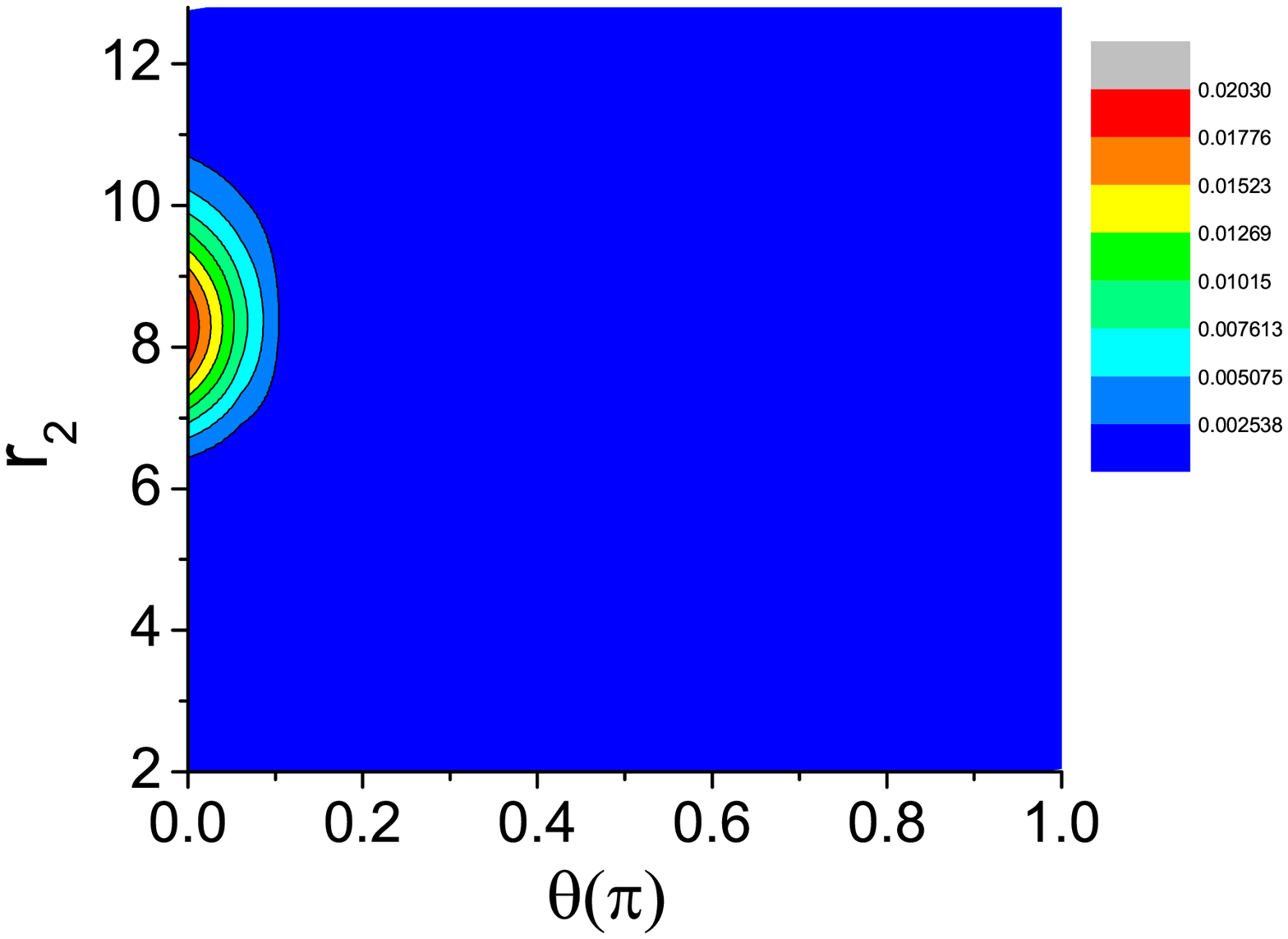}\\
\end{center}
\caption{The square of the two-boday wave
function $|\Psi_{2}(r_1,r_2,\theta)|^2$ with $r_1=9$~fm for the two protons in $^{210}$Po calculated within one orbital (left) and with all shells (right).  Taken from Ref. \cite{Qi2010c}.}\label{wf210}
\end{figure}

The fundamental role of configuration mixing was only confirmed by actual 
large-scale calculations \cite{TonAri,jan79,Qi2010c}.
The physics behind the enhancement 
induced by configuration mixing is that, with the participation of high-lying 
configurations, the pairing interaction clusters the two neutrons and the two 
protons on the nuclear surface, as can be seen from Fig. \ref{wf210} where the calculated  two-body wave
function $|\Psi_{2}(r_1,r_2,\theta)|^2$ for the two protons in $^{210}$Po within model spaces containing one orbital and all the orbitals above $Z=82$ are compared. This wave function has the form,
\begin{eqnarray}
\label{sinwf}
\Psi_2(r_1,r_2;\theta_{12})=\frac{1}{4\pi}
\sum_{p\leq q}\sqrt{\frac{2j_p+1}{2}} X(pq;{\rm gs})
\varphi_p(r_1)\varphi_q(r_2)P_{l_p}(\cos\theta_{12}),
\end{eqnarray}
where $\varphi$ is the single-particle wave function and $P_l$ is the Legendre polynomial
of order $l$ satisfying $P_l(\cos0)=1$ (notice that for
the ground states 
studied here it is $l_p=l_q$). The 
two-neutron and two-proton wave-function terms add up constructively 
in the surface region when many shells are involved. It was found that the mechanism that induces
clustering is the same that produces the pairing collectivity, which is
manifested in an strong increase in the form factor corresponding to the
corresponding transfer cross section. This property gives rise
to a giant pairing resonances (see below), which corresponds to the most collective of
the pairing states lying, as the standard (particle-hole) giant resonances,
high in the spectrum.

\begin{figure}
\begin{center}
\includegraphics[scale=0.4]{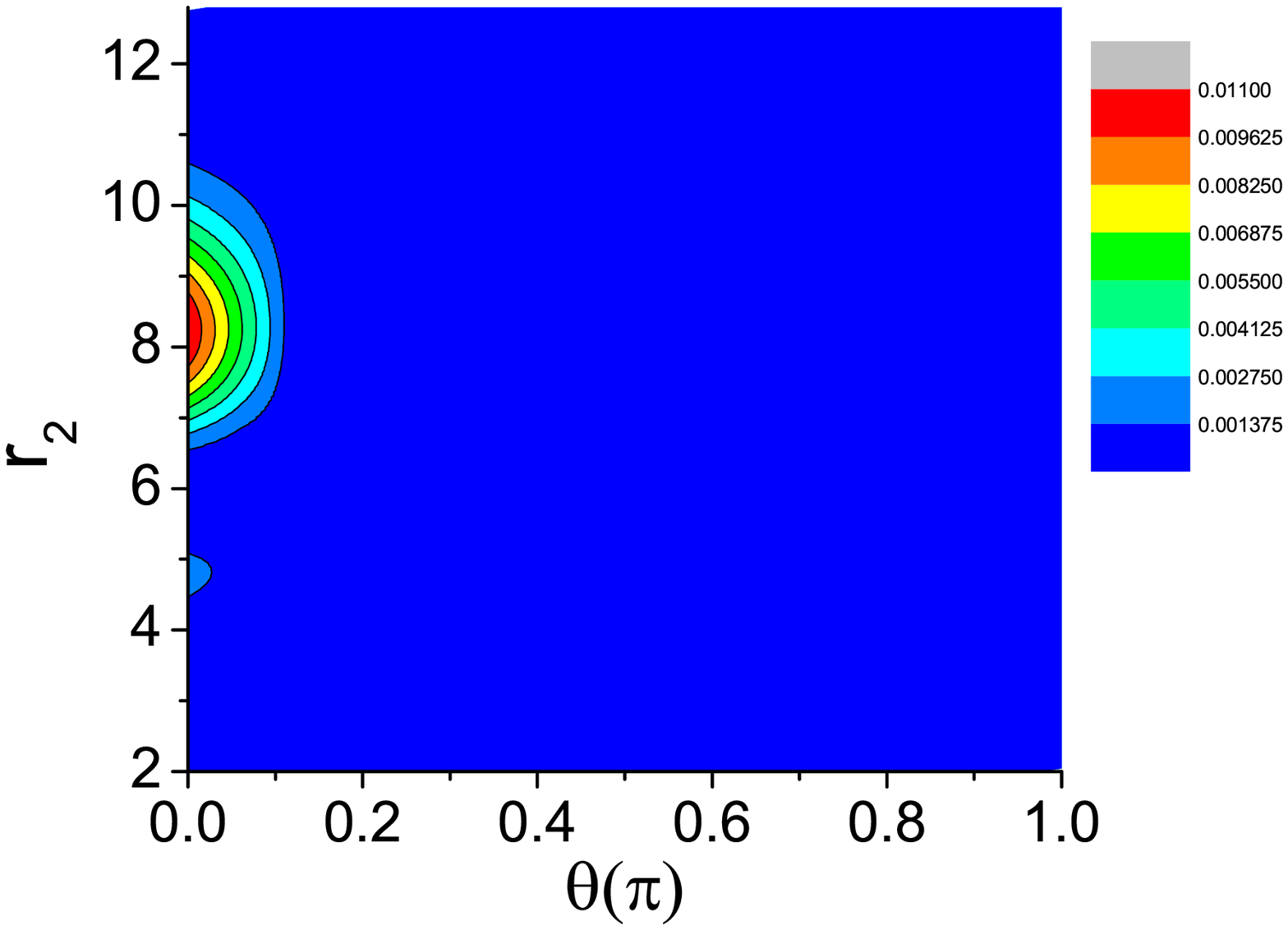}
\includegraphics[scale=0.4]{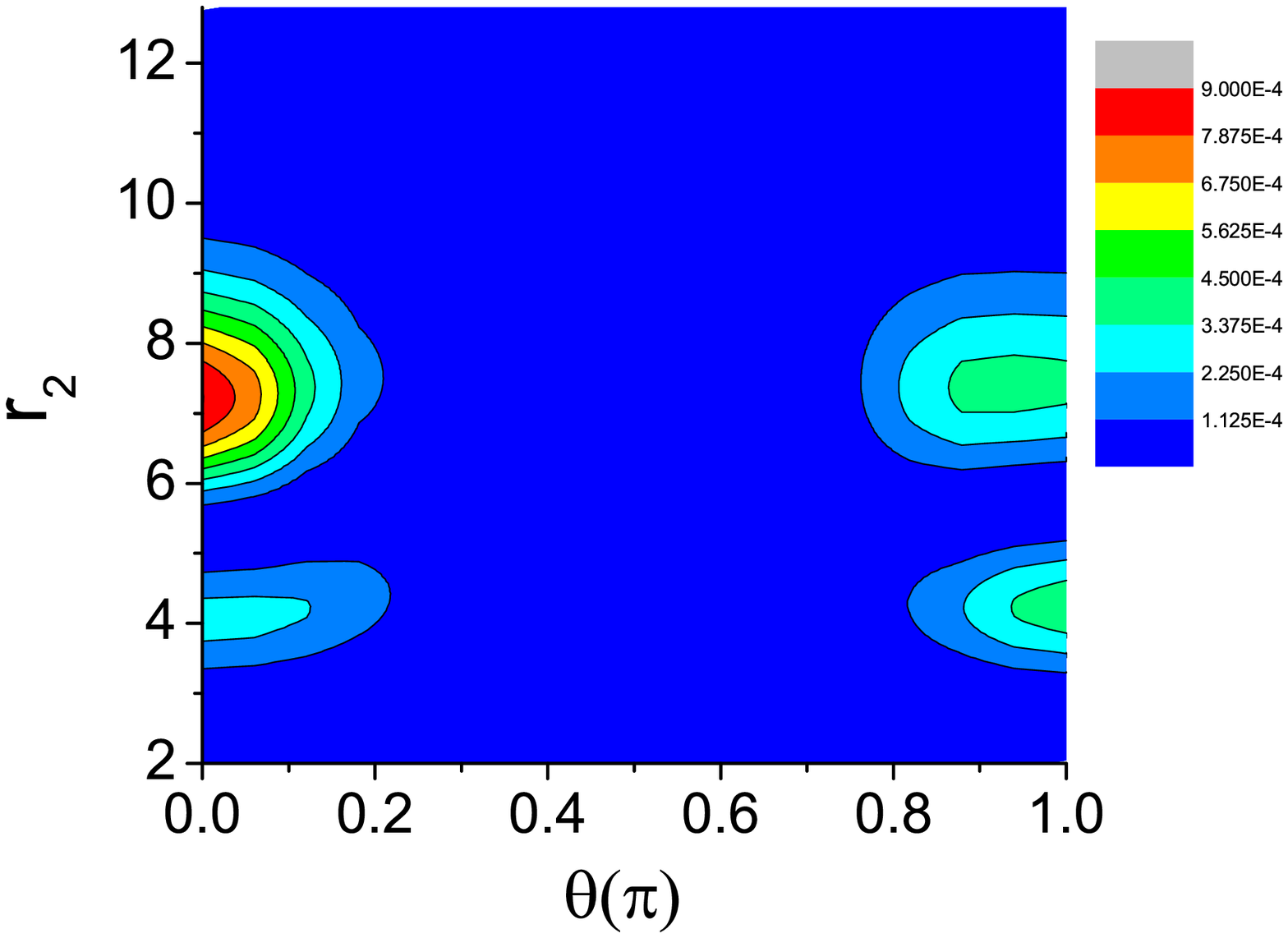}\\
\caption{The square of the two-boday wave
function $|\Psi_{2}(r_1,r_2,\theta)|^2$ with $r_1=9$~fm for the two neutrons in $^{210}$Pb (left) and two neutron holes in
$^{206}$Pb (right). Taken from Ref. \cite{Qi2010c}.}\label{wfvsr}
\end{center}
\end{figure}

For non-degenerate systems the pairing collectivity
manifests itself through the correlated contribution from many configurations, which is induced by the non-diagonal matrix elements of the pairing interaction in a shell-model context.
For two particles in a non-degenerate system with a constant pairing, the energy can be evaluated through 
the well known relation \cite{Changizi2015210},
\begin{equation}
\label{dren}
    G\sum_{i}\frac{2j_i+1}{2\varepsilon_i-E_2}=2.
\end{equation}
The corresponding wave function amplitudes are given by 
\begin{equation}
X_i=N_n\frac{2j+1}{2\varepsilon_i-E_2}
\end{equation}
where
$N_n$ is the normalization constant. All amplitudes $X_i$ contribute to the 
two-particle clustering with the same phase due to the strongly attractive nature 
of the pairing interaction. The correlation energy induced by 
the monopole pairing corresponds to the difference
\begin{equation}
\Delta^ {(3)}_{C}=\varepsilon_{\delta}-\frac{1}{2} E_2,
\end{equation}
where $\delta$ denotes the lowest orbital. As the gap $\Delta$ increases the 
amplitude $X_i$ becomes more dispersed, resulting in stronger two-particle spatial
correlation. This difference, or more exactly 
$\Delta-1/2G$ with the self energy removed, is an important measure of the 
two-particle spatial correlation at the surface, reflected in a corresponding  clustering of the
two nucleons forming the pair (see, e.g., Ref. \cite{Qi2010c}).
This clustering induces an increase in the strength of the corresponding 
pair-transfer reaction. 

 In Fig. \ref{wfvsr} we compared the two-neutron wave functions of $^{210,206}$Pb. A striking feature thus found is that the two-neutron correlation is significantly larger in $^{206}$Pb than in $^{206}$Pb in relation to the larger pairing gap in the former case.

\begin{figure}
	\begin{center}
		\includegraphics[scale=0.45]{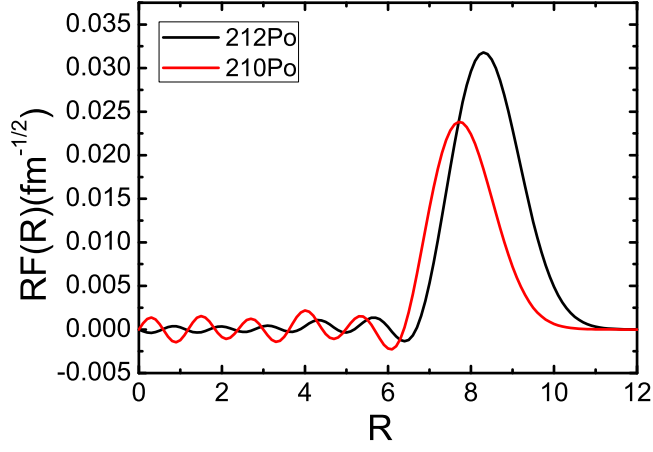}\\
		\caption{(Color online) The $\alpha$-formation amplitudes
			$R\mathcal{F}_{\alpha}(R)$ corresponding to the nuclei 
			$^{212}$Po(gs) and $^{210}$Po(gs). Taken from Ref. \cite{Qi2010c}.}
		\label{form}
	\end{center}
\end{figure}

Within the shell model The amplitudes $X$ of the four-particle state $\alpha_4$ in $^{212}$Po vanish if this interaction is neglected. Then
only one of the configurations in Eq.~(\ref{msmwf}) would appear. This
is done, for instance, in cases where the correlated four-particle
state is assumed to be provided by collective vibrational states.
The corresponding formation amplitude acquires the form,
\begin{eqnarray}\label{po212}
{\cal F}_{\alpha}(R;^{212}{\rm Po(gs)}) =
\int d\mathbf{R}
d\xi_{\alpha}\phi_{\alpha}(\xi_{\alpha})
\Psi(\mathbf{r_1}\mathbf{r_2};^{210}{\rm Pb(gs)})
\Psi(\mathbf{r_3}\mathbf{r_4};^{210}{\rm Po(gs)}),
\end{eqnarray}
where $\mathbf{r_1},\mathbf{r_2}$ ($\mathbf{r_3},\mathbf{r_4}$)
are the neutron (proton) coordinates and $\mathbf{R}$ is the centre
of mass of the $\alpha$ particle.

The decay of the
nucleus $^{210}$Po(gs) leads to the daughter nucleus $^{206}$Pb(gs), which
is a two-hole state. The formation amplitude becomes,
\begin{eqnarray}\label{po210}
{\cal F}_{\alpha}(R;^{210}{\rm Po(gs)}) = \int d\mathbf{R}
d\xi_{\alpha}\phi_{\alpha}(\xi_{\alpha})
\Psi(\mathbf{r_1}\mathbf{r_2};^{206}{\rm Pb(gs)})\Psi
(\mathbf{r_3}\mathbf{r_4};^{210}{\rm Po(gs)}).
\end{eqnarray}
With this expression for the formation amplitude, the experimental 
half-life can be reproduced if a large number of
high-lying configurations is included. The calculated formation amplitude for $^{210,212}$Po are plotted in Fig. \ref{form}.

\begin{figure}
	\begin{center}
		\includegraphics[scale=0.45]{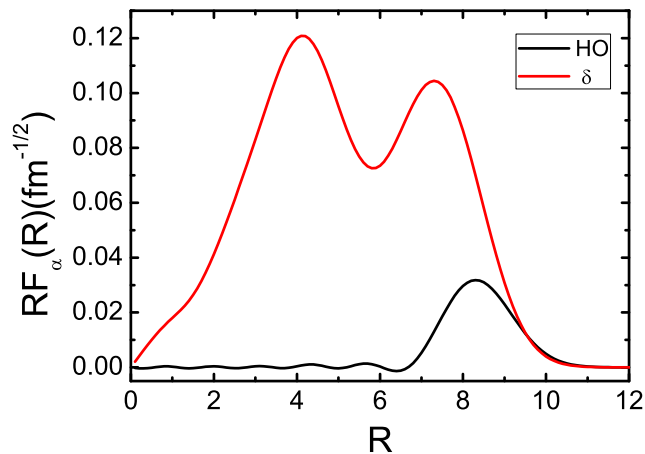}\\
		\caption{(Color online) Comparison of the $\alpha$ formation amplitude
			$RF_{\alpha}(R)$ of $^{212}$Po calculated from the $\delta$ function
			approximation and the Gaussian form of the $\alpha$ intrinsic wave
			function.}\label{comp}
	\end{center}
\end{figure}

If we assume that the intrinsic wave function of the
$\alpha$ particle can be approximated by a $\delta$ function, a even
simpler expression exist for the $\alpha$ formation amplitude which reads,
\begin{equation}\label{delta}
F_{\alpha}(R) = \frac{16\pi^2}{R^4}\left(\frac{
	s_{\alpha}^3}{3}\right)^{3/2}\Phi_{2\pi}(R,R,0)\Phi_{2\nu}(R,R,0),
\end{equation}
where $S_{\alpha}=\sqrt{20}/3R_{\alpha}$ and we take
$\mathbf{\hat{r}_1} =\mathbf{\hat{r}_3} =\mathbf{\hat{z}}$.
In Fig.~\ref{comp} we plotted the formation amplitude evaluated with this approximation, in comparison
with that calculated with the realistic $\alpha$ intrinsic wave function. The $\delta$-function approximation
ignores the fact that the four nucleons forming the $\alpha$ particle only strongly clustered at the nuclear surface and overestimates strongly the $\alpha$ particle formation probability inside the nucleus. It may be interesting to point out here that the result from the simple calculation in Fig.~\ref{comp} indicates clearly the  four body spatial/clustering correlation inside does not necessary mean that $\alpha$ or $\alpha$-like clustering has occurred inside the nucleus. This is one confusion we often see in theoretical studies of nuclear clustering.

\vskip 3mm
\subsubsection{Limitations of the shell-model} 
\vskip 3mm
The shell model has been extremely successful to explain and predict nuclear
properties. Features like the structure of nuclear spectra,
particularly high-spin states, 
nuclear reactions, processes in the continuum part of the spectrum, nuclear
deformations and so on,  could be well explained by the shell model. For a 
review see Ref. \cite{Andres}.
Even $\alpha$-decay processes, including half lives of
excited states and relative decay widths, could be well understood
through the shell model as seen in Subsection \ref{finest}. 
For an early
importance of the shell model in $\alpha$-decay see Ref. \cite{Lovas1998}.
Yet, the absolute decay with of the heavy nucleus $^{212}$Po(gs) could be
calculated only within an order of magnitude. More striking is that the
clustering properties of neutrons and protons that constitute the
alpha-particle can be explained by the shell-model.

One may argue that it is the difference of neutron and proton numbers in 
$^{212}$Po which is responsible for the shortcomings of the shell-model in
explaining the absolute decay width. This is not the case, as shown in the
next subsection.

The first experimental indication that the shell model alone could not 
describe properties in $^{212}$Po did not come from $\alpha$-decay probes
but rather from electromagnetic transitions. This was done in Ref. 
\cite{astier}, where excited states in that nucleus were populated by 
$\alpha$-transfer using the $^{208}$Pb($^{18}$O,$^{14}$C) reaction. Their 
de-excitation $\gamma$-rays were studied and several levels were found to 
decay by an unique low energy $E1$ transition populating the yrast state
with the same spin value. Their lifetimes were measured and it was
discovered that the
transitions were very enhanced. These results, which could no be explained
within the standard shell-model,  were found to be consistent with an alpha
cluster structure. This gives rise to states with non-natural parity.  

As we have described above it had been known for a long time that a necessary 
requirement to properly describe particle emission and transfer processes 
is that the basis wave functions follow correct asymptotic values. In 
$\alpha$-decay this feature seems to be even more remarkable and it is in
fact at the origin of the deficient description of the decay process by 
using the standard shell model. A successful solution of this problem was
presented in Ref. \cite{varga}, where the decaying state was described as a
combination of a shell-model wave function plus a cluster component.
The important feature of this approach is that the cluster
component is expected to take care of the high-lying shell-model
configurations and, therefore, the shell-model component is evaluated
within a major shell only. The cluster component is expanded in terms
of shifted Gaussian, and the coefficients are found by diagonalizing the
residual two-body interaction.

This method was applied in Ref. \cite{ddelion} to describe the experimental
features of Ref. \cite{astier}. 
By using a shifted Gaussian component in the single-particle wave
functions it was possible to describe the $\alpha$-decay process, while the
shell-model part of the wave function explained well the corresponding 
$B(E2)$ transitions.

This deviation of the pure shell model is a disadvantage which would make
the shell model less appealing if the mixing of shell model and cluster
components should be a general trait. Fortunately it is not. However, attempts
were done trying to include the effects induced by the cluster component 
within a pure shell-model representation. This implies that the standard
(e. g. Woods-Saxon) central potential has to be modified. The modification
consists in adding an attractive pocket 
potential of a Gaussian form localized
on the nuclear surface. The eigenvectors of this new mean field 
provides a
representation which retains all the benefits of the standard shell model
while at the same time reproducing well the experimental absolute $\alpha$-decay
widths from heavy nuclei \cite{dellio}. Although one can in this way
obtain results similar to the ones provided by the shell-model plus
cluster representation, the application of the method is cumbersome 
and no farther application was reported.
But this confirms the limitations of the shell model in explaining 
absolute decay widths in $\alpha$-decay.

\subsection{Significance and outcome of the Continuum treatment}
The study of the influence of the continuum upon alpha decay gave rise to
the appearance of new features which are apparently unrelated to the alpha
decay process. We will analyze these features case by case.
\vskip 3mm
\subsubsection{ Giant pairing resonances}

The first attempt to consider the continuum in $\alpha$-decay was
related to the inclusion of the neutron-proton interaction. As
discussed above, the most important states in the formation of the 
$\alpha$-particle are the isovector $0^+$ pairing states, which in our case are
$^{210}$Pb(gs) and $^{210}$Po(gs), due to their neutron-neutron and 
proton-proton clustering features. It was therefore assumed that the 
neutron-proton clustering should also proceed through an isovector pairing 
neutron-proton state. Such a state cannot be built upon the valence shells in 
this case, since 
they correspond to the principal quantum number N=5 for protons and N=6 for 
neutrons carrying opposite parities. Therefore the lowest isovector pairing 
neutron-proton state should be formed by protons and neutrons  moving in the 
N=6 shell. This state had not been observed but was assumed to lie at 5 MeV
above the ground state, i. e. above the state $^{210}$Bi($1^{-}$;gs). The
corresponding wave function was obtained by using a pairing force, adjusting
the pairing strength to fit the energy of the lowest $0^+$ state thus
calculated to lie at 5 MeV. This wave function showed to
have strong clustering features, as expected. Including this state in the
basis of Eq. (\ref{msmpai}) one obtained the alpha clustering as well as the 
experimental value of the decay width. However, this was accomplished by 
adjusting the components of the wave function in an
unrealistic fashion \cite{gordana}. But the idea that there should exist a 
neutron-proton isovector pairing state at high energy in nuclei with proton
number differing from the neutron number prompted the possibility of
considering this neutron-proton state as the isobaric analog to the
neutron-neutron ground state. In our case the state $^{210}$Bi($0^+_1$)
should be analog to the state $^{210}$Pb($0^+_1$;gs). As a result there
should be another isobaric analog state corresponding to two-proton 
excitations. In our case this should be an state lying at about 10 MeV (the 
gap corresponding to two major shells) above the ground state of 
$^{210}$Po(gs). The same should be valid in the nucleus $^{210}$Pb. Here
there should be a collective isovector pairing state at about 10 MeV. This
is analog to the particle-hole collective excitations, were e. g. the
isovector dipole giant resonance in $^{208}$Pb lies at about 10 MeV (the real 
figure is 13.5 MeV).

To verify the existence of the high lying collective pairing state a
calculation was performed by using a pairing interaction and a large
single-particle representation \cite{marek}. The three isobaric states
discussed above were evaluated. The ones in $^{210}$Bi and in $^{210}$Po
show strong clustering features, even more than in $^{210}$Pb(gs). They are
built mainly upon high lying single-particle states and are strongly excited
in two-particle transfer reactions. Therefore they can be considered
pairing giant resonances (GPR). In fact this state had been predicted before
just as an analogue to the particle-hole giant resonance \cite{besbro}.

The state $^{210}$Pb(GPR) was also found to be strongly pairing collective 
lying at an energy 0f 11.4 MeV. This prompted an intense experimental
activity looking for this GPR, but without any success \cite{Zybe}. However
the calculation  of this GPR was later confirmed in an independent work 
\cite{Fortu}. To probe the importance of the continuum in this very high
lying two-neutron state a calculation using the Berggren representation was
performed. As will be seen in the next item below, the Berggren
representation was also introduced in relation to alpha decay. It is very well 
fitted to take into account the escape process of particles lying high in the 
continuum. It was thus found that in the state $^{210}$Pb(GPR) the neutrons
tend to scape the nucleus since there is no Coulomb barrier to trap them. As 
a result that state  is a very wide
resonance. Therefore it can be considered a part of the continuum
background rather than an observable state \cite{Willy}.

Yet the GPR was still been sought but in lighter nuclei. It was thus
proposed that it should exist in the nucleus $^{132}$Sn \cite{Augus}.
Finally, signatures of the GPR were found in $^{14}$C and in $^{15}$C
\cite{Cappu}.     
\vskip 3mm
\subsubsection{ The Berggren representation} 
\vskip 3mm
The failure of bound representations to explain the width of $\alpha$-decay
resonances brought up the question whether the continuum should be included
explicitly. An important step in the study of the continuum in many-body
problems was given by the introduction of the Gamow resonances
\cite{Zel,Tamas}. These are solutions of the time-independent Schroedinger
equation with purely outgoing waves at large distances. These resonances,
together with the proper continuum and bound states, were used by Berggren
as a representation to write the  single-particle Green function
\cite{berggr}.
However the first time that the Berggren representation was applied did not
concern alpha-decay but rather particle-hole giant resonances \cite{curut}.
This was because at that time there were a large amount of experimental data 
and open questions related to giant resonances and the continuum. But the
introduction of the Berggren representation was followed by many
applications. It was the origin of what eventually would be called "Shell
model in the complex energy plane" or "Gamow shell model". A review of this
development can be found in Ref. \cite{Mic09}. 

This representation was used to evaluate the alpha formation amplitude
corresponding ´to the $\alpha$-decay of $^{212}$Po(gs) \cite{silvia}. One thus
succeed in describing the clustering up to large distances. For this a
large number of configurations had to be included in the representation.
However, the maximum value of the formation amplitude was about the same as
the one obtained by using a bound representation. As a result, the
disagreement between the calculated and experimental alpha decay width
persisted. 

Yet, in Ref. \cite{Rolo1} a similar calculation provided a good agreement
with experiment. But this paper was strongly criticized, as incorrect, in Ref.
\cite{lovas1}. The authors of \cite{Rolo1} answered in
\cite{Rolo2}. This in fact left the question open on whether the shell 
model is indeed able to describe absolute decay widths. It seemed that a 
more radical solution to this problem was needed, as described in the next 
item below.

\subsection{The Geiger-Nuttall law and its generalizations\label{geinut}}

The huge  range of $\alpha$ decay half-lives can be modelled through the 
Geiger-Nuttall law~\cite{gn1,gn2}, which shows a striking correlation 
between the half-lives of radioactive
decay processes and the decay $Q_{\alpha}$ values. The decay half life is
predicted by this law to be,
\begin{equation}\label{gn-o}
\log T_{1/2}=\mathcal{A}Q_{\alpha}^{-1/2}+\mathcal{B},
\end{equation}
where $\mathcal{A}$ and $\mathcal{B}$ are constants that can be determined 
by fitting to experimental data.
The Gamow theory reproduced the Geiger-Nuttall law very well by describing the 
$\alpha$ decay as the tunnelling through the Coulomb barrier.

\begin{figure}
\begin{center}
\includegraphics[scale=0.65]{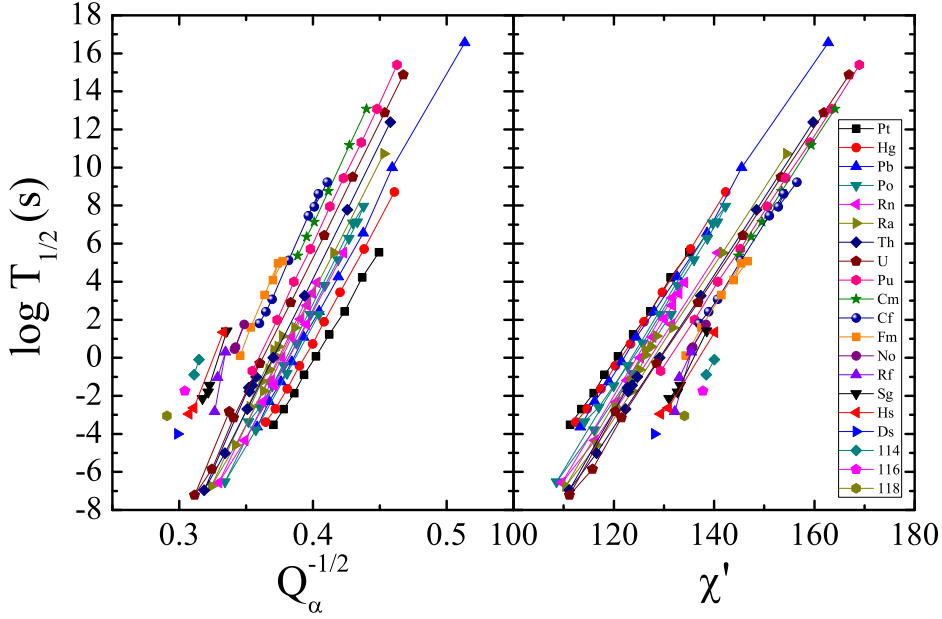}\\
\vspace{-0.5cm}
\end{center}
\caption{Systematics of $\alpha$ decays from even-even nuclei with
$Z=78-118$. The $Q_{\alpha}$ values are in unit MeV. The quantity $\chi'$ (in MeV$^{-1/2}$) is defined as $\chi' = Z_cZ_d\sqrt{A/Q_c}$ (see text). Taken from Ref. \cite{Qi2009}. \label{fig1alpha}}
\end{figure}

The Geiger-Nuttall law in
the form of Eq.~(\ref{gn-o}) has limited prediction power since its
coefficients change for the decays of each isotopic
series, see Fig. \ref{fig1alpha}. Intensive work have
been done trying to generalize the Geiger-Nuttall law for a
universal description of all detected $\alpha$ decay
events \cite{PhysRevC.80.024310,0954-3899-39-1-015105}. One of the most 
known generalization is the  Viola-Seaborg law \cite{VIOLA1966741} which for 
even-even nuclei reads
\begin{equation}\label{gn-v}
\log T_{1/2}=(aZ_d+b)Q_{\alpha}^{-1/2}+bZ_d+d
\end{equation}
where $a$, $b$ and $d$ are constants and $Z_d$ the charge number of the daughter 
nucleus.

The importance of a proper treatment of $\alpha$ decay was attested in 
Refs. \cite{Qi2009,QI2009a} which shows that the different lines can be 
merged into a single line. 
In this generalization the penetrability is
still a dominant quantity where $H^+_0(\chi,\rho)$ can be well
approximated by an analytic formula
\begin{equation}
H^+_0(\chi,\rho) \approx (\cot
\beta)^{1/2}\exp\left[\chi(\beta-\sin\beta\cos\beta)\right].
\end{equation}
 By defining
the quantities $\chi' = Z_{\alpha}Z_d\sqrt{A_{\alpha d}/Q_{\alpha}}$ 
and $\rho' =
\sqrt{A_{\alpha d}Z_{\alpha} Z_d(A_d^{1/3}+A_{\alpha}^{1/3})}$ where 
$A_{\alpha d}=A_d A_{\alpha}/(A_d+A_{\alpha})$,
one gets, after some simple algebra,
\begin{eqnarray}\label{gn-2}
\log T_{1/2}=a\chi' + b\rho' + c,
\end{eqnarray}
where $a$, $b$, $c$ are constants to be determined.

One thus obtained a generalization of the Geiger-Nuttall law (called UDL)
which holds for all isotopic chains and all cluster radioactivities.
Eq.  (\ref{gn-2}) reproduces well most available
experimental $\alpha$ decay data on ground-state to ground-state 
radioactive decays. 

The UDL works not only for alpha decay but also for proton decay (see above) and heavier cluster decays (see below).
\subsubsection{Extraction of the formation probability from experimental half-lives}
The success of the Geiger-Nuttall law and UDL is mainly due to the small variations
of the $\alpha$-particle formation probability when going from a
nucleus to its neighbours, as compared to the penetrability. 
In the logarithm scale of the Geiger-Nuttall law, the differences in the
formation probabilities are usually small fluctuations along the straight
lines predicted by that law.

\begin{figure}
\begin{center}
\includegraphics[width=0.45\textwidth]{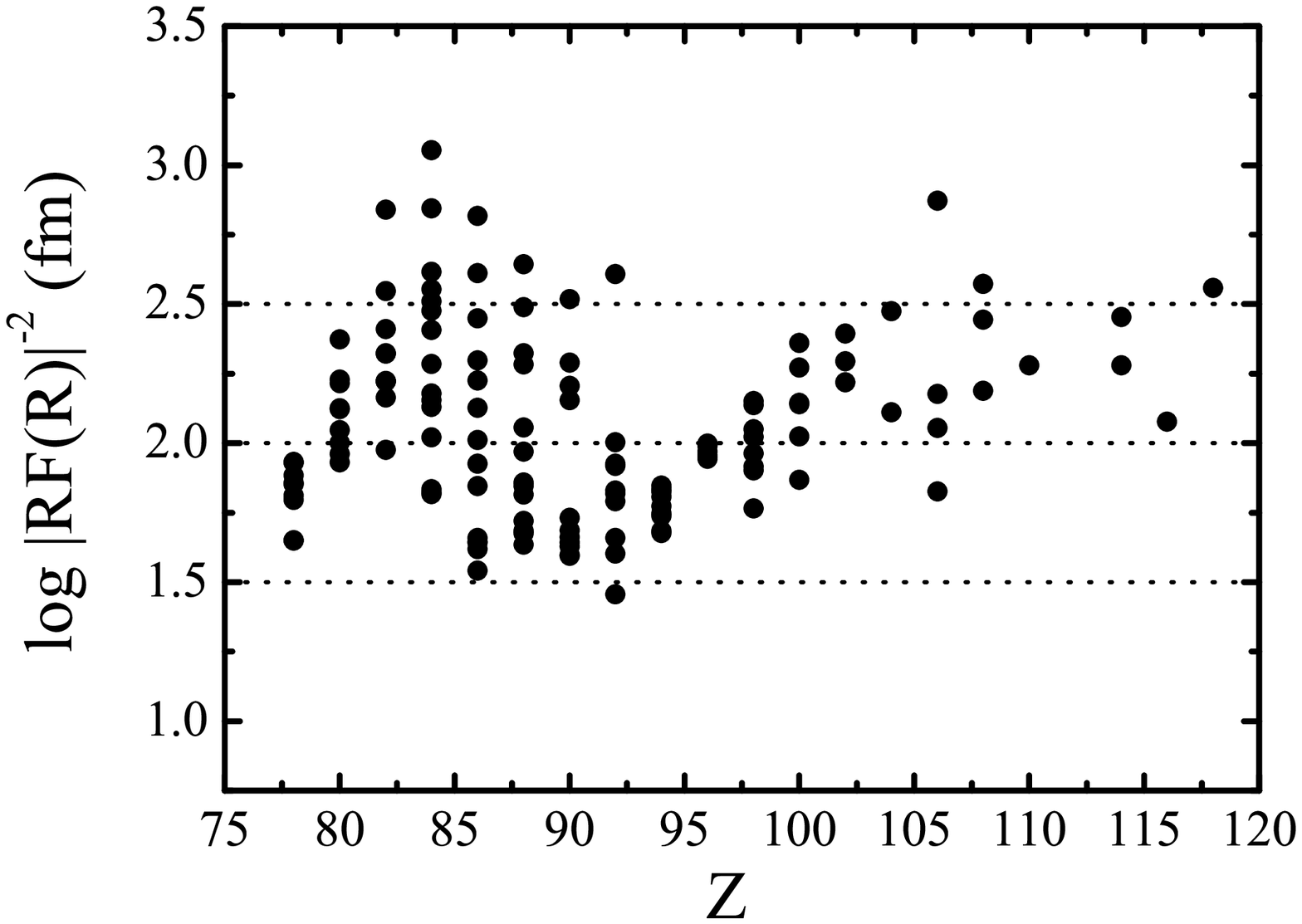}
\includegraphics[width=0.45\textwidth]{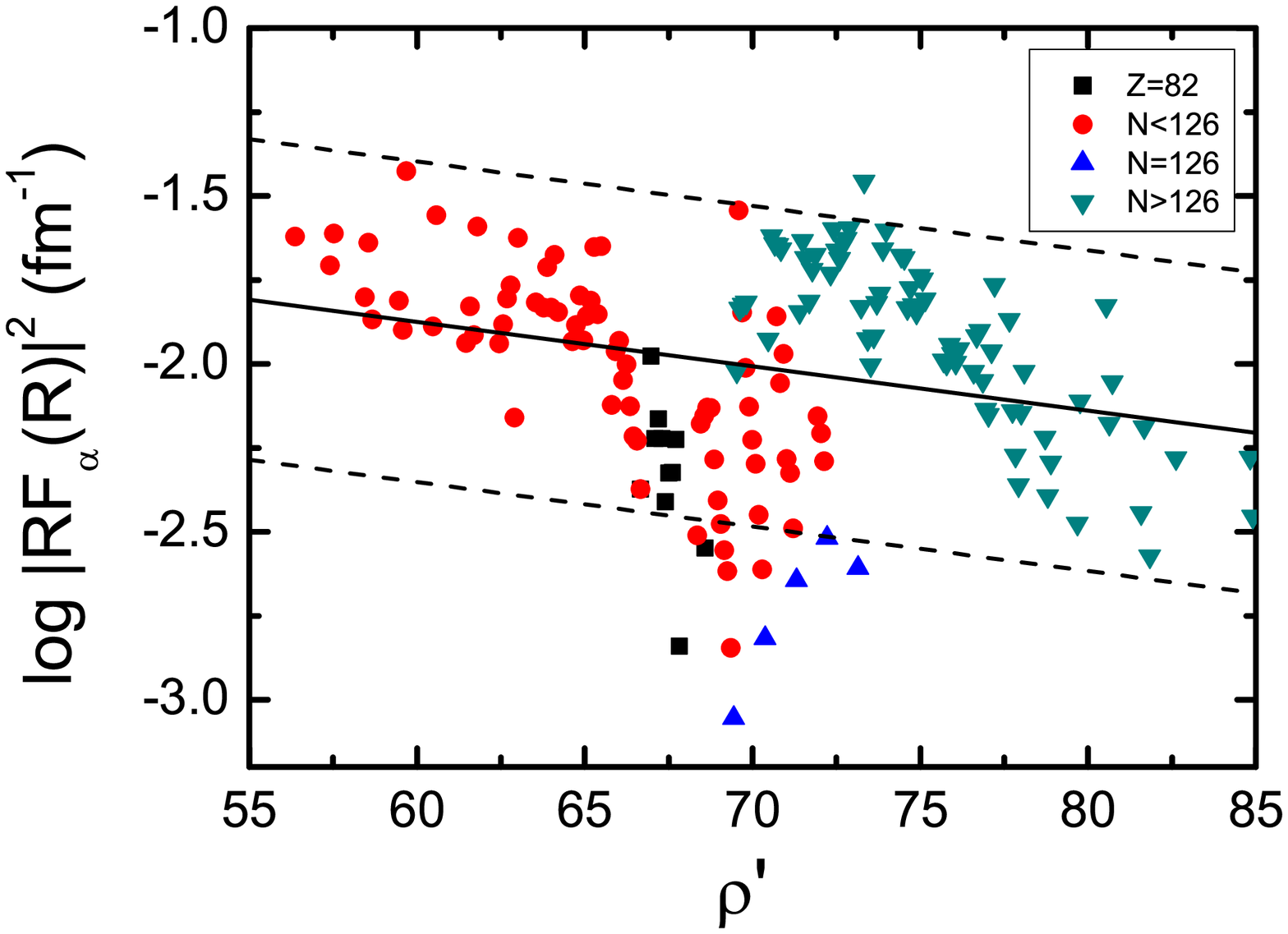}
\end{center}
\caption{The $\alpha$ decay formation amplitudes $\log |RF(R)|^{-2}$
as a function of the charge number of the mother nucleus
$Z$ (left) and 
$\rho'$. Right plot from Ref. \cite{Qi2010c}.}\label{formation}
\end{figure}

The formation amplitude ${\cal F}_{\alpha}(R)$ can be extracted from the 
experimental half-lives $T^{{\rm Expt.}}$ corresponding to ground state 
to ground state transitions 
\begin{equation}
\log |R{\cal F}_{\alpha}(R)|=\frac{1}{2}\log \left[ \frac{\ln
	2}{\nu}|H^+_0(\chi,\rho)|^2\right] - \frac{1}{2}\log T^{{\rm
		Expt}}_{1/2}.
\end{equation}
This was done in Refs. \cite{Qi2010c,Qi2009,QI2009a,Andreyev2013,Qi2014203}. In Fig. \ref{formation} we plotted the formation probability for known alpha decays. They follow roughly a linear behaviour as a function of $\rho'$ which is the key for the success of UDL.

\begin{figure}
		\begin{center}
	\includegraphics[scale=0.45]{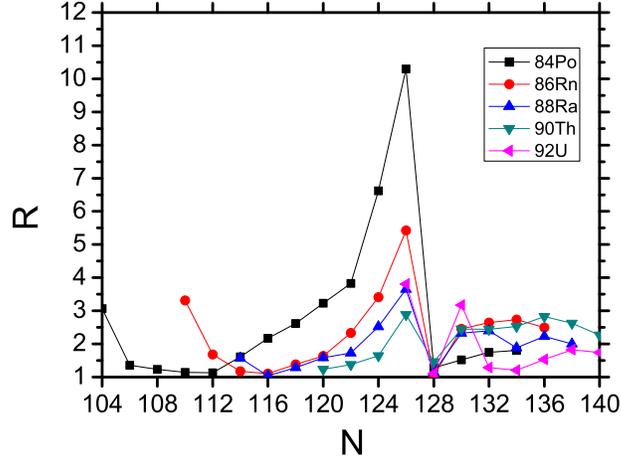}\\
		\end{center}
	\caption{Left:  Discrepancy between experimental decay half-lives and UDL calculations as a function of the neutron
		number of the mother nucleus N.}\label{fvsrp}
\end{figure}

It was found that although the UDL reproduces nicely most available
experimental $\alpha$ decay data, as expected, there is a case where it fails by a
large factor. This corresponds to the $\alpha$ decays of nuclei with 
neutron numbers equal to or just below $N=126$~\cite{Qi2010c,qi10a}, as can be seen from the left panel of Fig. \ref{fvsrp} where we plotted the discrepancy between experimental and calculated $\alpha$ half-lives. 
The reason for this large discrepancy is that in $N\leq126$ nuclei 
the $\alpha$ formation amplitudes
are much smaller than the average quantity predicted by the UDL. The case that shows the most significant hindrance corresponds to the $\alpha$ 
decay of the nucleus $^{210}$Po. It was found that the formation amplitude in $^{210}$Po is hindered 
with respect to the one in $^{212}$Po due to the hole character of the neutron states in
the  $^{210}$Po case. This is a manifestation of the pairing mechanism that induces
clusterization, which is favoured by the presence of high-lying
configurations (see Sec. \ref{bcss} below). Such configurations are more accessible in the
neutron-particle case of $^{212}$Po than in the neutron-hole case of
$^{210}$Po. As a result, the neutron pairing correlation and eventually the two-neutron and $\alpha$ clustering are significantly enhanced in $^{212}$Po in comparison with those in $^{210}$Po. 

\begin{figure}
	\begin{center}
		\includegraphics[scale=0.5]{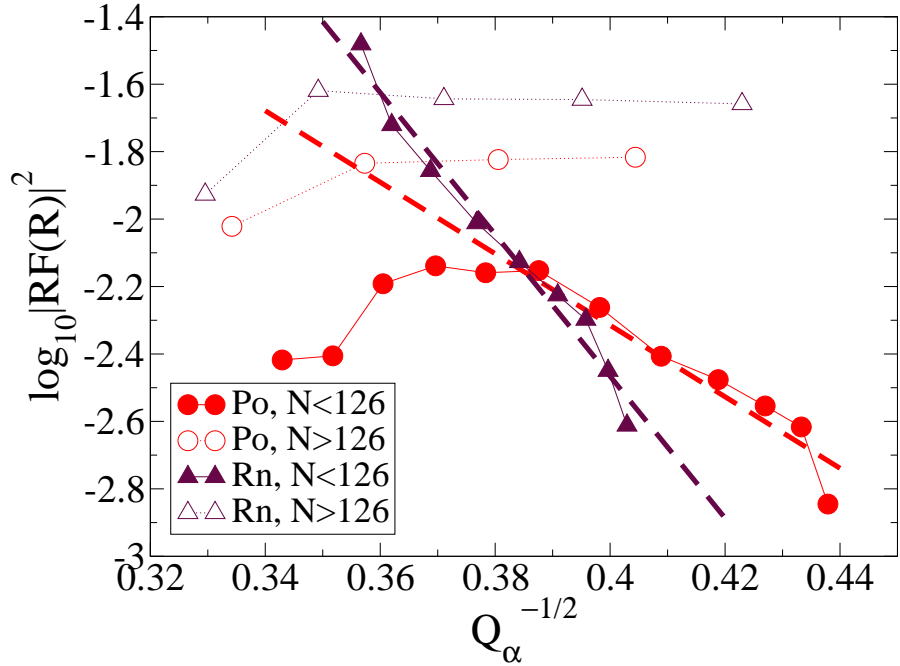}
	\end{center}
	\caption{\label{fvsq} $log_{10}|RF(R)|^2$ for Po (circle) and Rn (triangle) isotopes with $N<126$ (closed symbols) and with $N>126$ (open symbols) as a function of $Q_{\alpha}^{-1/2}$. The dashed lines are to guide the eye. From Ref. \cite{Qi2014203}.}
\end{figure}

\subsubsection{Limitations of the Geiger-Nuttall law}

The origin and physical meaning
of the coefficients $\mathcal{A}$ and $\mathcal{B}$ in the Geiger-Nuttall 
law can be deduced by comparing Eq. (\ref{gn-o}) and (\ref{gn-2}).
These coefficients are determined from experimental data and show a linear 
dependence upon $Z$. 
The need for a different linear $Z$ dependence of the coefficients 
$\mathcal{A}$ and $\mathcal{B}$ in different regions of the nuclear chart 
was discussed in Ref. \cite{Qi2014203}, which is related to the generic form 
of the $\alpha$ formation probability as will be discussed just below.
When the dependence of $log_{10}|R{\cal F}_{\alpha}(R)|^2$ on the neutron 
number is not linear or constant, the Geiger-Nuttall law is broken.  This 
also explains why the Geiger-Nuttall law works so well for nearly all 
$\alpha$ emitters known 
today, as the data within each isotopic chain are limited to a region 
where $log_{10}|R{\cal F}_{\alpha}(R)|^2$ is roughly a constant or behaves 
linearly with $N$.

\begin{figure}
	\centering{
	\includegraphics[width=0.6\textwidth]{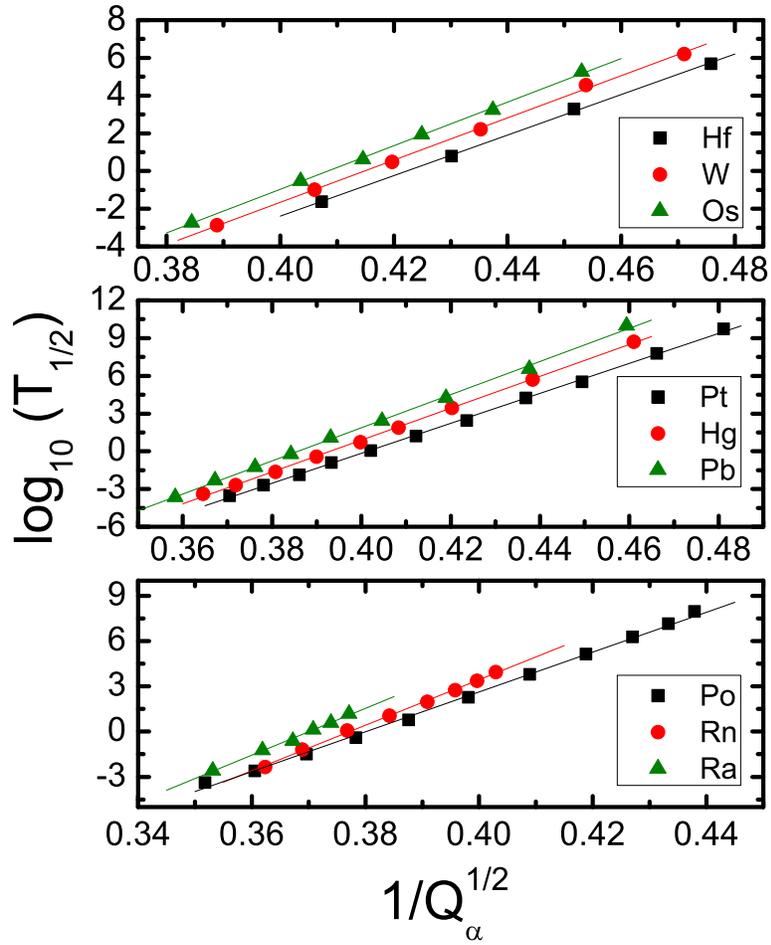}}
	\caption{\label{GNexp}(Color online) The logarithms of experimental
		partial half-lives (in
		sec) corresponding to the $\alpha$ decay of heavy nuclei as a function
		of $Q_\alpha^{-1/2}$, where $Q_\alpha$ is the alpha-decay $Q$-value
		(in MeV). The solid lines are determined by fitting the GN law to data. Only cases with $N<126$ are presented.}
\end{figure}

We notice that, for the $\alpha$ decays of nuclei with 
neutron numbers equal to or just below $N=126$,  the UDL fails by a
large factor but,on the other hand, the normal GN law seems to work, as illustrated in Fig.~\ref{GNexp}. The fact that the GN law is not ``broken" at the first glance, is simply related to the $Q_{\alpha}$ (as well as $Q_{\alpha}^{-1/2}$) values exhibit a quasi linear pattern as a function of rising neutron number when  approaching the $N=126$ shell closure. Therefore $log_{10}|RF(R)|^2$ and  $log_{10}(T_{1/2})$ will still depend linearly on $Q_{\alpha}^{-1/2}$.  This is shown in Fig. \ref{fvsq} where  the logs of the $\alpha$ formation probabilities $|RF(R)|^2$ for polonium and radon isotopes are plotted as a function of $Q_{\alpha}^{-1/2}$. Thus the GN law can still hold. However, it should be pointed out that the linear patten seen in the figure does not necessary mean that there is any correlation between the two quantity.

In Fig.~\ref{GNexp}, the values of
the coefficients ${\cal A}$ and ${\cal B}$ change for the decays of each
isotopic series, given rise to the different lines seen in the Figure.
The coefficients may also change within a single
isotopic chain when magic numbers are crossed. 

\begin{figure}
	\centering{
	\includegraphics[width=0.65\textwidth]{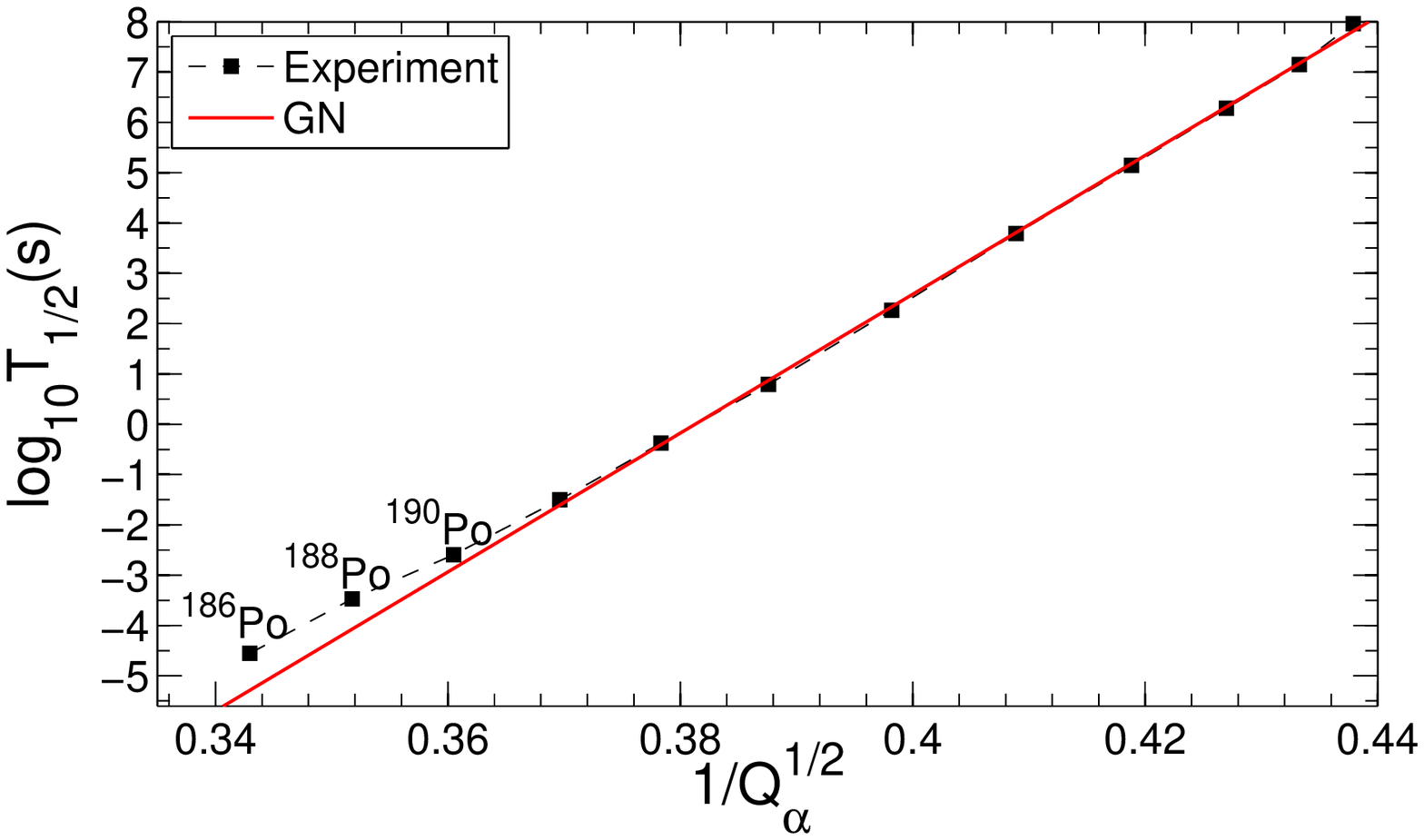}}
	\caption{(Color online) The logarithms of the $\alpha$ decay half-lives of neutron-deficient Po isotopes as a function
		of $Q_\alpha^{-1/2}$. \label{Po}}
\end{figure}

For the polonium isotopic chain with $N<126$, as can be seen from Figs. \ref{fvsq} \& \ref{Po},  the linear behaviour of 
$log_{10}|R{\cal F}_{\alpha}(R)|^2$ breaks down below $^{196}$Po.
As a result, the Geiger-Nuttall law is broken in the light polonium 
isotopes. In $^{186}$Po, the width difference between the experimental result and that predicted by the GN Law is as large as one order of magnitude \cite{Andreyev2013}. This violation is induced by the strong suppression of the 
$\alpha$ formation 
probability due to the fact that the deformations (or shell-model 
configurations) of the ground states of the lightest $\alpha$-decaying
neutron-deficient polonium isotopes ($A < 196$)  are very different from 
those of the daughter lead isotopes (see Sec. \ref{finest}).
One may say that the decay from  $^{186}$Po is strongly hindered that the ``average" behaviour predicted by the GN law.
On the other hand, we emphasize here that it is important to have an overall understanding of all alpha decay properties using UDL before one can draw conclusions on the enhancement or hindrance of the alpha formation probability of specific alpha decay processes. If one just look at the GN description of light Po isotopes, one may draw to the wrong conclusion that the alpha decay from light Po isotopes should be much hindered than heavier ones which, however, is not the case as can be seen from the figure. This is related to the fact that the generic behaviour of the alpha formation is washed out in the GN law description which is essentially a local theory with parameters fitted to each certain isotopic chain.

\subsubsection{Generic form of the $\alpha$ formation probability}

\begin{figure}
\begin{center}
\includegraphics[scale=0.5]{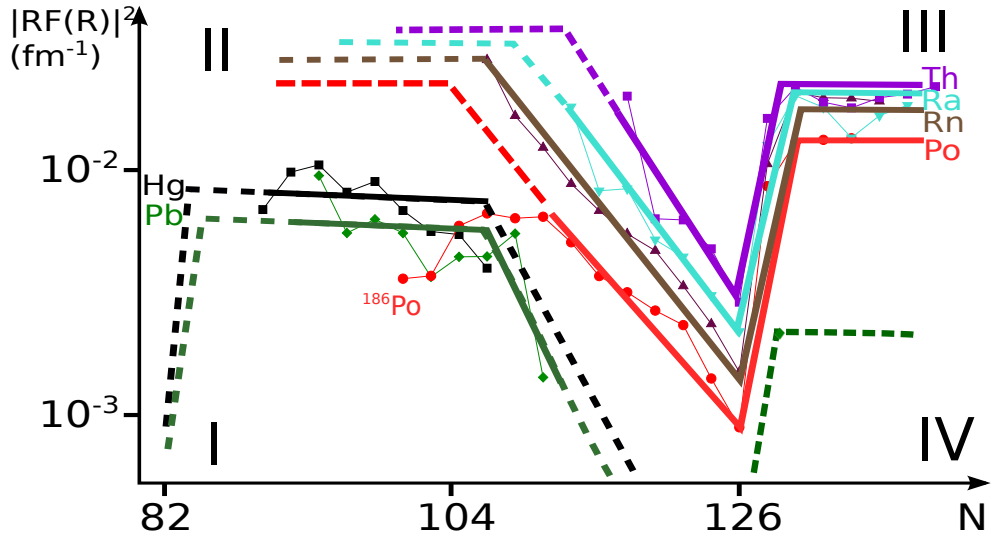}
\end{center}
\vspace{-0.2cm}
\caption{\label{gene} A  generic representation of the generic form of the evolution of the $\alpha$ formation probabilities $|RF(R)|^2$. Thick solid  lines are for isotopes, where experimental data are available and dashed lines are extrapolations to the regions with the yet unavailable data.  Taken from Ref. \cite{Qi2014203}.
}
\end{figure}

A generic form for the $\alpha$-particle formation amplitude as a function of nucleon (proton or neutron) number was proposed in Refs. \cite{Andreyev2013,Qi2014203} based on  experimental values \cite{nudat} and 
calculations performed within the framework of the seniority scheme. This is shown in Fig. \ref{gene}: When the nucleons are filling a new major closed  shell (e.g. $N$ between 82 and 126) the $\alpha$-particle formation amplitude is nearly constant as high-$j$ orbitals are filled first. As soon as the low-$j$ orbitals are filled, the formation probability smoothly reduces until one reaches again a closed proton or neutron configuration, i.e. the upper boundary of the major shell. Here a minimum is reached. Crossing the closed shell induces a steep increase and the approximately constant trend mentioned above continues. However, when strong particle-hole excitations across closed shells are encountered, this 'generic' form of the $\alpha$-particle formation probability is altered as one clearly sees in the light polonium isotopes.  Such effects, however, do not invoke a disappearance of the influence of the $Z=82$ shell gap on the $\alpha$-decay probability.

Guided by recent experimental findings, the $\alpha$-decaying nuclei was
divided into four regions: \\
I) $N\leq126$, $Z\leq82$; \\
II) $N\leq126$, $Z>82$;\\
III) $N>126$, $Z>82$;\\
IV) $N>126$, $Z\leq82$.\\
Except for $^{210}$Pb, $\alpha$ decay has not yet been observed for nuclei in region IV. The need for a different linear $Z$ dependence of the coefficients $A$ and $B$
of Eq. (\ref{gn-o}) in the four regions of the nuclear chart is obvious.

\begin{figure}
	\begin{center}
		\includegraphics[scale=0.3]{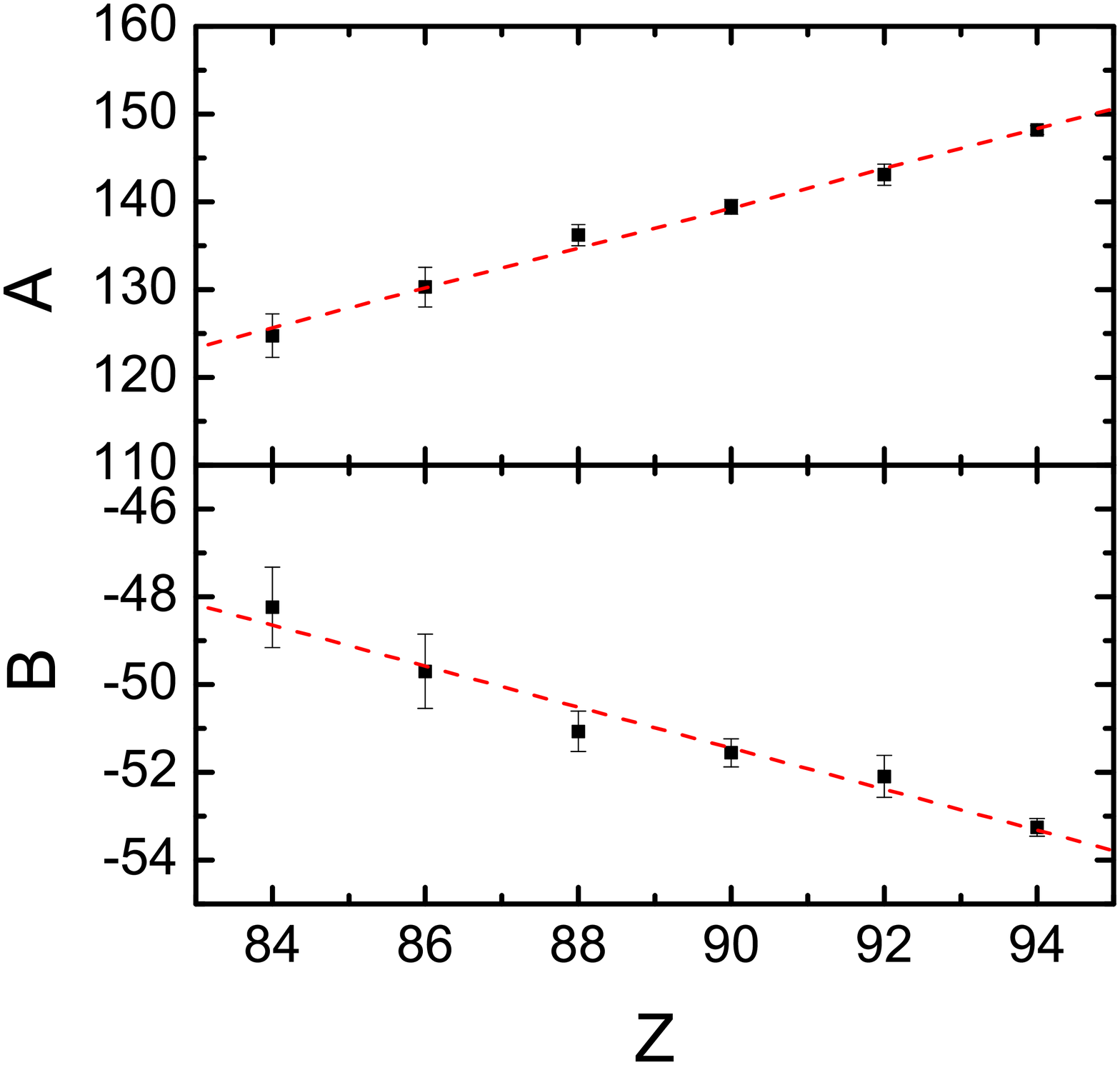}\includegraphics[scale=0.3]{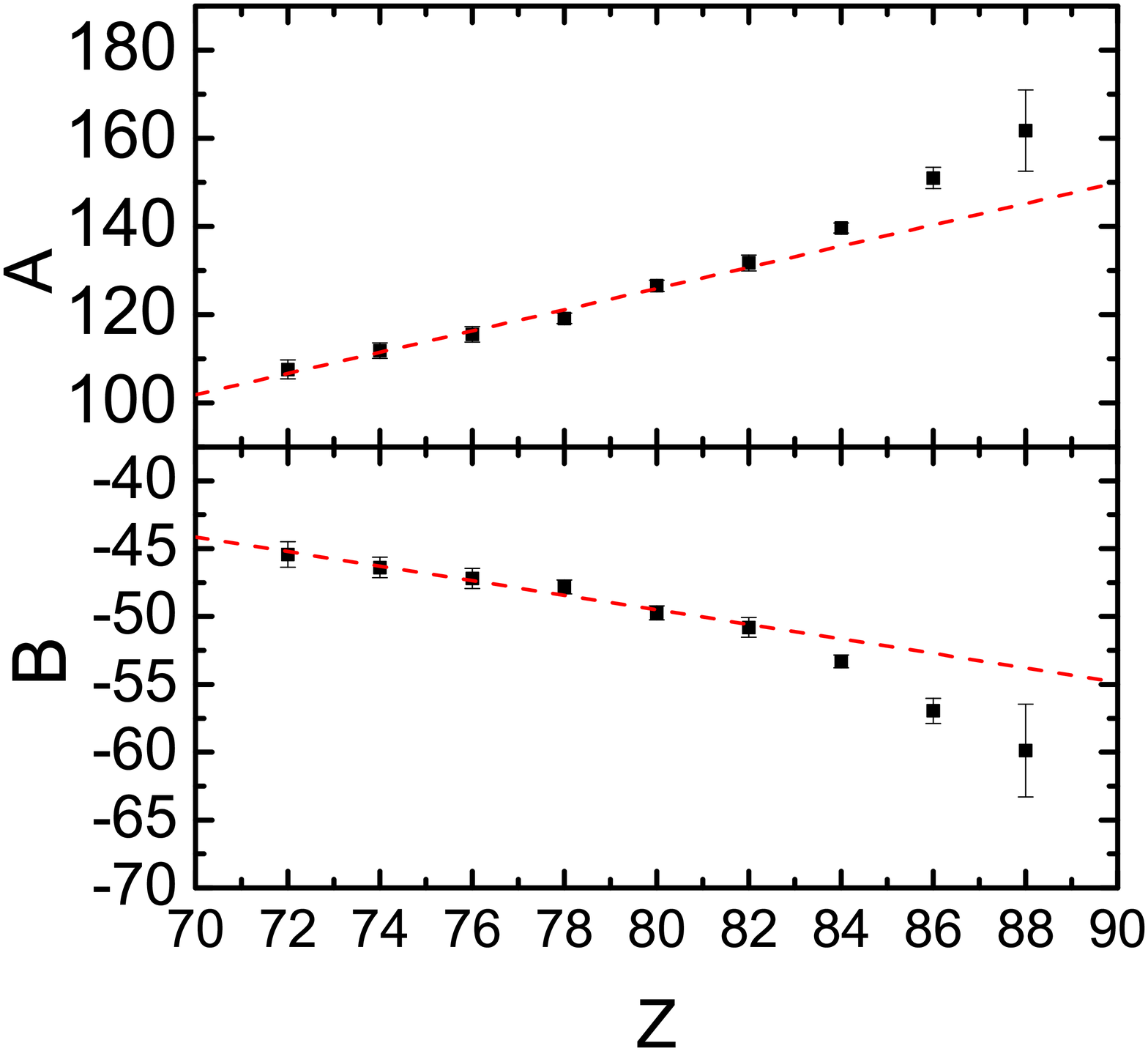}
	\end{center}
	\caption{\label{abngt126} (Color online)  The coefficients ${\cal A}$ and $|{\cal B}|$ of the GN law as a function of $Z$ for $\alpha$ decays in the Pb
		region corresponding to neutron numbers $N>126$(left, Region II) and $N<126$ (right, Region I\& II).}
\end{figure}

In comparison with those in region III for which the formation probabilities are nearly constant or only weakly depend on $Q_{\alpha}^{-1/2}$, the data in region II show an exponential dependence. The other isotopic chains in region II show a similar linearly decreasing behaviour  of $log_{10}|RF(R)|^2$ as a function of $Q_{\alpha}^{-1/2}$, however, with different slopes. As a result, the GN law remains valid for isotopic chains in region II. However, the corresponding values of ${\cal A}$ and $|{\cal B}|$ will increase with $Z$ beyond the trend observed in regions I and III, as can be seen in Fig. \ref{abngt126}.
In regions I and III, Both ${\cal A}$ and $|{\cal B}|$ show a linear behaviour as a function of $Z$.

Approaching the $N = 126$ shell closure with increasing neutron number, a strong and exponential decrease of the  formation
probability is observed. In spite of a variation of $|RF_{\alpha}(R)|^2$ over one order of magnitude, the GN law and the ${\cal A}$ and $|{\cal B}|$ linear dependence upon $Z$ are still valid. This has no real physical meaning, but is a consequence of the specific dependence of the $|RF_{\alpha}(R)|^2$ on $Q_{\alpha}$. 

\subsection{$\alpha$ decays of $N\sim Z$ nuclei}

One would expect that the influence of the neutron-proton (np) interaction 
upon alpha decay would be better taken into account in nuclei with the number of
neutrons close to the number of protons. In such a case the shell model
predicts that neutrons and protons move in the same shells and, therefore,
the isovector monopole np interaction would not be hindered by constraints
like parity restrictions. As we have seen, in heavy nuclei the
neutron-neutron and proton-proton interaction, when treated properly, induce
the nn and pp clustering \cite{Qi2010c}. Yet the decay width evaluated by
including many configurations and even the proper continuum was too small. 
One may think that the reason of this failure is that the np interaction is not 
treated properly in the cases where the number of neutrons differs much from
the number of protons.  Therefore, the $\alpha$ decays from $N\sim Z$ 
nuclei can provide an ideal test ground for our understanding of the np
correlations.  This includes the isovector as well as the
isoscalar pairing mode. The intense efforts to elucidate this problem 
can be attested by the long list of references, e. g. Refs. 
\cite{Ced11,Frauendorf201424,Qi11,Xu2012,Qi2015}.

The many undertakings on the pn interaction in $N\sim Z$ nuclei and its effect
upon $\alpha$-decay were hindered by the lack of experimental data.
Finally an experiment was performed in which the isotopes $^{109}$Xe and 
$^{105}$Te were identified \cite{PhysRevLett.97.082501} through the
detection of the $\alpha$-decay chain $^{109}Xe \rightarrow ^{105}Te \rightarrow 
^{101}Sn$.  
The half-lives of the two $\alpha$-decays were determined to be $13 \pm 2$
ms and $620 \pm 70$ ns for $^{109}$Xe and $^{105}$Te, respectively. 
It was also found that the reduced $\alpha$-decay widths relative to
$^{212}$Po was enhanced in the two cases, which prompted the name of
``superallowed $\alpha$-decay transitions". This was aptly applied in  this region 
of $N\sim Z$ where such a feature was expected. The authors of that
experiment stated that 
and attempt to measure the $\alpha$-decay chain 
$^{108}Xe \rightarrow ^{104}Te \rightarrow ^{100}Sn$ were to be expected.
This was indeed observed very recently in Ref. \cite{Auranen2018}. In addition, the $\alpha$ decays of $^{114}$Ba 
\cite{Mazzocchi200229,PhysRevC.94.024314} and light Xe and Te isotopes have also been observed
\cite{Seweryniak2006,Janas2005,Liddick2006}. An experimental search for $^{113}$Ba was presented recently in Ref. \cite{Xiao2017}.

\begin{figure}
\begin{center}
\includegraphics[width=0.6\textwidth]{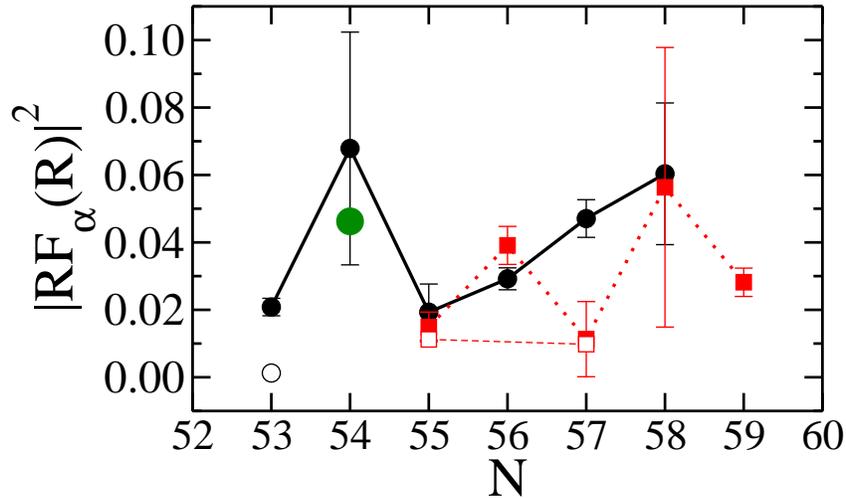}
\end{center}
\vspace{-0.6cm}
\caption{(Color online) $\alpha$-decay formation amplitudes 
$|RF_{\alpha}(R)|^{2}$ extracted from experimental 
data \cite{nudat,PhysRevLett.105.162502} as a function $N$ for neutron-deficient 
Te (circle) and Xe (square) above $^{100}$Sn. Open symbols correspond to the decays 
of $\alpha$ particles carrying orbital angular momentum $l=2$. The figure was taken from Ref. \cite{QI201677}. The new data on $^{108}$Xe from Ref. \cite{Auranen2018} is added and highlighted in green. 
\label{fvsrpnz}}
\end{figure}

In Fig. \ref{fvsrpnz} is compared the $\alpha$ formation probabilities of nuclei just 
above $^{100}$Sn. The $\alpha$ formation probabilities of those nuclei follows the 
general average mass-dependence trend of $\alpha$ formation probability systematics 
but shows rather large fluctuations and uncertainties. Contrary to what is stated in Ref. Ref. \cite{Auranen2018}, it is still difficult to 
determine whether there is indeed an extra enhancement in those transitions. Further 
experimental investigation
is essential in clarifying the issue. It may be useful to mention here that the 
systematics of formation probabilities for available $\alpha$ decays shows an 
increasing trend as the mass number decreases. This is related to the fact that the 
size of the nucleus also gets smaller, which favours the formation of $\alpha$ 
particles on the surface.

The robustness of the $N=Z=50$ shell closures has fundamental influence  on 
our understanding of the structure of nuclei around the presumed doubly magic 
nucleus $^{100}$Sn. 
It was argued that $^{100}$Sn may be a soft core in analogy to the soft 
$N=Z=28$ core $^{56}$Ni. It seems that such a possibility can be safely ruled 
out based on indirect information from recent measurements in this region 
\cite{PhysRevC.84.041306,Back2013,10.1038/nature11116,PhysRevLett.110.172501,
PhysRevC.91.061304}. It is still difficult to measure the single-particle 
states outside the $^{100}$Sn core. The neutron single-particle states 
$d_{5/2}$ and $g_{7/2}$ orbitals  in $^{101}$Sn, which have been expected to 
be close to each other, were observed by studying the $\alpha$-decay chain 
$^{109}$Xe$\rightarrow$ $^{105}$Te $\rightarrow$ $^{101}$Sn 
\cite{PhysRevLett.97.082501}. 
In Ref. \cite{Seweryniak2006}, the nucleus $^{105}$Te was also populated and 
one $\alpha$ transition was observed.
A prompt 171.7 keV $\gamma$-ray transition was observed in Ref. 
\cite{PhysRevLett.99.022504} and was interpreted as the transition from the 
$g_{7/2}$ to the $d_{5/2}$ orbital, which was assumed to be the ground state. 
On the other hand, two $\alpha$ decay events from $^{105}$Te were observed in 
Ref. \cite{PhysRevLett.105.162502} with the branching ratios (energies) of 
89\% (4711 keV) and 11\% (4880 keV). Based on those observations and on the 
assumption that the ground state of $^{105}$Te has spin-parity $5/2^+$, a flip 
between the $g_{7/2}$  and $d_{5/2}$ orbitals was suggested.
This information was used in the optimization of the effective shell-model 
Hamiltonian for this region \cite{PhysRevC.86.044323}. 

The influence of np correlation upon the formation of $\alpha$ particles in
$^{220}$Ra and $^{116,108}$Xe was studied in Ref. \cite{DELION99} 
within the framework of a generalized BCS approach in an axially deformed 
Woods-Saxon potential. Only diagonal terms between proton and neutron orbitals
with the same angular-momentum projections were considered and a modest enhancement 
of the clustering was found in $^{116,108}$Xe.
In Ref. \cite{Qi2015} the nn and pp two-body clustering in $^{102}$Sn 
and $^{102}$Te was analysed within the framework of the shell-model.
The correlation angle between the clustered nucleons was investigated 
by switching on and off the np correlation . It was found that when a large
number of configurations is included there is a significant enhancement of 
the four-body clustering at zero angle when the np interaction is switch on. 
This is an indication of the importance of the np interaction in the
clustering process.

One would have expected an intense theoretical activity dealing with the 
superallowed $\alpha$-decay region. But this was not the case. In Ref. \cite{MahBhaGam15}
an empirical Gamow inspired calculation was performed analysing
$\alpha$-emitters close to the $N=Z$ line.
As usual in this type of calculations the novel ingredient introduced as
compared with previous like calculations was the $\alpha$-daughter interaction.
Here a double folded potential was used. 
The calculations were found to be close to the experimental data. 
Also the half-lives of some of the neutron-deficient I, Cs, and
Ba isotopes were predicted to be longer than 1 ms. 

Another theoretical paper concerned with this region is Ref. \cite{monika}.
Here the structure and the $\alpha$-formation probability
corresponding to the nuclei $^{212}$Po and $^{104}$Te were studied
using a full microscopic formalism within the shell-model. It was found 
that the proton-neutron 
correlations are much more important in $Te$ than in $Po$. This is 
expected, since the active neutrons and protons  move in the same orbits 
in $Te$, but in different ones in $Po$. As a result the state
$^{212}$Po(gs) becomes a near pure monopole isovector state, while the state
$^{104}$Te(gs) presents strong mixing with other (multipole) states. Another 
consequence of the proton-neutron correlations is that the $\alpha$-particle 
formation probability in $^{104}$Te is 4.85 times larger than that in $^{212}$Po, 
thus attesting that in the Te region there is a superallowed $\alpha$-decay 
transition.
Yet the $\alpha$-decay width is too small as compared with the
expected 
experimental value. This is another strong indication that the standard
shell model representation needs some additional ingredient.  On the other hand, the proton-neutron BCS calculations presented recently in Ref. \cite{PhysRevC.94.034319} tend to argue that
the proton-neutron pairing correlations have a small influence on the alpha particle formation in $N=Z$ nuclei above $^{100}$Sn.

\subsection{BCS treatment of alpha decay and pairing correlation}
   \label{bcss}
The BCS approach was introduced just in the beginning of the microscopic
studies of $\alpha$-decay \cite{Ras65,mrkgl,Mang1964,Soloviev1962202}.
The great appealing of this approach resides in its ability to transform the
wave function of a correlated system into a pure quasiparticle 
configuration. This is a consequence of the dominant isovector pairing force
in nuclei. A many-nucleon superconducting nucleus is controlled by 
very few quasiparticle degrees of freedom. This made it possible to evaluate 
nuclei with nucleon number lying between magic numbers even with limited
computing facilities, as those available five decades ago. 
An early and  detailed calculation of decay widths using this formalism 
was presented in Ref. \cite{PhysRev.181.1697}. Hundreds of transitions were
calculated and the empirical trends were well reproduced. 
     
The isovector pairing correlation is behind the success of the BCS theory in 
nuclei. As already mentioned above, this
correlation manifests itself through the 
coherent contribution of a large number of shell-model configurations. The isovector pairing correlation highly enhances the calculated 
$\alpha$-decay width and is indeed  the mechanism governing the 
formation of $\alpha$ particles at the nuclear surface. 
This feature is  also responsible for the two-particle clustering, which 
is manifested in a strong increase in the form factor of the
two-particle transfer cross section in transfer reactions between 
collective pairing states. As we have seen, this also gives rise to giant 
pairing resonances, which correspond to the most collective of the isovector
pairing states.

\begin{figure}
\begin{center}
\includegraphics[scale=0.75]{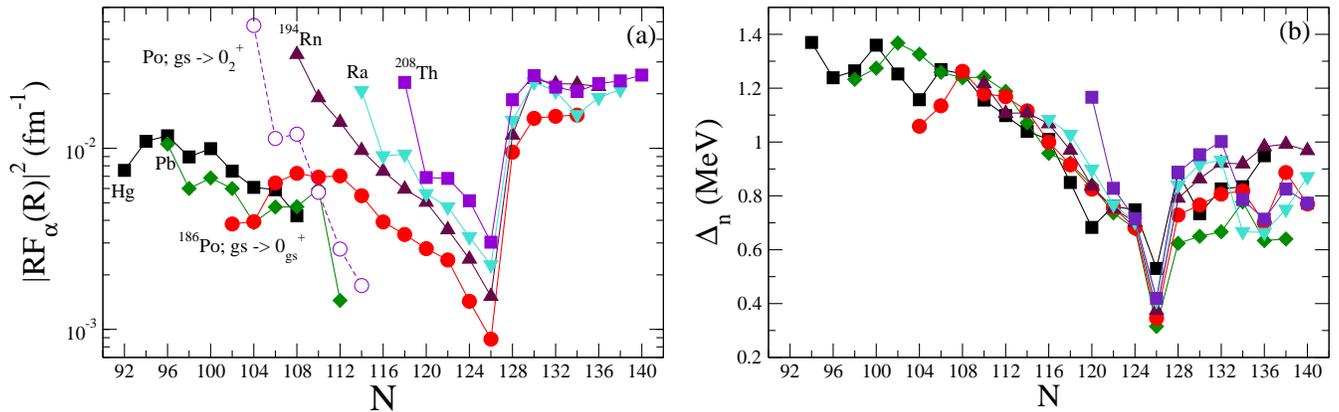}
\end{center}
\vspace{-0.8cm}\caption{(color online).  Upper panel: $\alpha$-particle
formation probabilities for the decays of the even-even isotopes 
as a function of the neutron numbers $N$ of the mother nuclei. 
 Lower panel: Neutron pairing gaps in even-even lead to Thorium nuclei
extracted from experimental binding energies. From Ref. \cite{Andreyev2013}.
}
\label{fap}
\end{figure}

Fig. \ref{fap}a shows the formation probabilities $|RF_{\alpha}(R)|^2$ 
extracted from the experimental half-lives from known
ground state to ground state $\alpha$-decay transitions in even-even 
isotopes from $N=92$ to 140. From the trend of 
$|R{\cal F}_{\alpha}(R)|^2$ around 
the neutron shell closure at $N=126$, one can deduce a global trend. Below
the shell closure, $|R{\cal F}_{\alpha}(R)|^2$ decreases as a function of
rising neutron number, reaching its lowest values at the shell closure. When
the shell closure is crossed, a sudden increase in 
$|R{\cal F}_{\alpha}(R)|^2$  
is observed. It is followed by an additional but smaller increase and 
finally saturation occurs. The  $\alpha$-particle formation amplitudes for 
nuclei $^{162}$W, $^{162}$Hf \cite{PhysRevC.92.014326} and $^{193}$At 
\cite{Ket2003} are systematically larger than those of neighbouring nuclei, 
which is not understood and needs further investigation.

Within the BCS approach the two-particle formation amplitude is proportional
to $\sum_k u_kv_k$ where $u_k$ and $v_k$ are the standard occupation numbers. To this
one has to add the overlaps of the corresponding proton and neutron radial functions with
the $\alpha$-particle  wave function on the nuclear surface,
which do not differ strongly from each other for neighbouring nuclei. The BCS pairing gap is given by
$\Delta=G\sum_k u_kv_k$,
where $G$ is the pairing strength. It indicates that
the $\alpha$ formation amplitude is proportional to the product of the proton and the neutron pairing gaps which can serve as a signature of the
change in clusterization
as a function of the nucleon numbers. To probe this conjecture one may compare the
formation probabilities extracted from the experimental half-lives to the
corresponding pairing gaps. The latter can readily be obtained from the 
experimental binding energies as 
\cite{Andreyev2013,Satula1998,Changizi2015210,PhysRevC.91.024305,Qi2012g,PhysRevC.91.024305,CHANGIZI201697}
\begin{equation}\label{dexp}
\Delta_n(Z,N)=\frac{1}{2}\left[B(Z,N)+B(Z,N-2)-2B(Z,N-1)\right].
\end{equation}
The empirical pairing gaps are shown as a function of the neutron number in 
Fig. \ref{fap}b.
One indeed sees a striking similarity between the tendency of the pairing 
gaps in this figure with the $\alpha$-particle formation probabilities.
This similarity makes it possible to draw conclusions on the
tendencies of the formation probabilities. The near constant value of 
$|R{\cal F}_{\alpha}(R)|^2$ for neutron
numbers $N\leq114$ is due to the influence of the $i_{13/2}$ and other 
high-$j$ orbitals. As these highly degenerate shells
are being filled the pairing gap and
the formation probability
should remain constant, as indeed they do in Fig. \ref{fap}.
A quite sharp decrease of 
formation probability and pairing gap happens as soon as the low-$j$ 
orbitals like $2p_{3/2}$, $1f_{7/2}$ and $2p_{1/2}$ start
to be filled. Finally, when we reach $N=126$, the pairing gap reaches its 
lowest value.
The possible influence of the $Z=82$ shell closure on the $\alpha$ 
formation probability and the robustness of the shell was also discussed in 
Ref. \cite{Andreyev2013}, which was questioned based on earlier measurements 
on the $\alpha$ decays of neutron-deficient Pb isotopes 
\cite{PhysRevC.60.011302}.

The role of the pairing interaction in multi-quasiparticle isomeric states 
and  the reduction of pairing in those states on $\alpha$-decay half-lives 
was examined in Ref. \cite{PhysRevC.90.044324}.

Primary time-dependent Hartree-Fock calculations for $\alpha$ decay and $\alpha$ capture were already carried out in Ref. \cite{San83} with a simplified Skyrme plus Yukawa potential. No spin-orbital field was considered in that paper.  Significant progress has  been made in the development of nuclear energy density functional approaches which are now able to provide a reasonable description
of ground state binding energies
and densities throughout the nuclear chart, even though the description of the single-particle spectroscopy is still less satisfactory. The Skyrme-Hartree-Fock single-particle wave functions were applied to calculate the $\alpha$ formation amplitudes in even-even nuclei in Refs.
\cite{PhysRevC.88.064316,1402-4896-89-5-054027} and even-odd nuclei \cite{PhysRevC.92.014314}. However, the calculated formation amplitude is still several of orders of magnitude too small in comparison to experimental data. The application of the recently refined functional seems to make the discrepancy even worse \cite{1402-4896-89-5-054027}. Further investigation along this line would be interesting to understand the origin of the discrepancy, which may shed additional light on the constraint of the density functional.

\subsection{Nuclear deformation and $\alpha$-decay }

The treatment of $\alpha$-decay processes from deformed nuclei began soon
after  Bohr and Mottelson proposed the rotational model 
\cite{PhysRev.103.1298}. In contrast to the case of spherical nuclei the WKB
approximation as used by Gamow \cite{Gamow1928} and most other calculations
of alpha decay since then, cannot be applied. This is because only when
there is no interference between two degrees of freedom
one can perform the integration on one variable independently of the other.
Instead, in the case of deformed potentials the Hamiltonian includes terms 
where the radial and angular variables interfere with each other. This
difficulty was overcome by generalizating the WKB approximation to deformed
systems, as done by Bohr and Mottelson \cite{bohr1998nuclear2} and described
in detail in Ref. \cite{froman}. In this theory the formation amplitude
enters in a more complicated fashion than in the Thomas formulation described
above. For a comprehensive presentation of the theory and its application see 
Ref. \cite{PhysRevC.44.545}. This method was used extensively 
\cite{PhysRev.181.1697,PhysRev.103.1298,PhysRevC.69.044318,PhysRev.112.512}, 
in particular to describe anisotropy in $\alpha$-decay 
\cite{deinli,PhysRev.124.1512,PhysRevC.2.2379,PhysRevC.71.044324,
PhysRevLett.77.36}.

The exact treatment of the $\alpha$-decay process in deformed
nuclei requires a coupled channel (CC) approach. A CC as well as a WKB
calculation of the decay
width in the transition $^{152}$Dy $\rightarrow\alpha$+$^{148}$ Gd was
performed in Ref. \cite{delwkb}. As seen in Table I of this reference the
WKB width coincides with the exact one for small deformations, but already
for $\beta$=0.3 the WKB width, $\Gamma_{WKB}$, is nearly double the exact
one, $\Gamma_{cc}$. For $\beta$=0.8 it is
$\Gamma_{WKB}/\Gamma_{cc}$=65.3. This is a drawback which, together
with the difficulties of the method, prompted the proposal of other 
approaches, like the one in Refs. \cite{berg1,berg2,berg3} where, in 
addition, a critical assessment of the WKB approximation is presented. 

The coupled-channel approach provides a microscopic foundation to
study the multi-dimensional decay and reaction processes. Usually,
the wave function is propagated numerically starting with proper
initial conditions.
The coupled-channel equations can be written in matrix notation as
\begin{equation}
\Psi''(r)=\left[\frac{2\mu}{\hbar^2}V(r)-k^2I\right]\Psi(r).
\end{equation}
Here $V(r)$ denotes the potential matrix with the dimension $N$. The
radial wave function $\Psi(r)$ is a $N\times N$ square matrix with
each column being a linearly independent solution of the equations.
Usually the wave functions are propagated starting with $N$ initial
conditions (through, e.g., the Numerov procedure). However, this
propagation is numerically quite unstable since at the classically
forbidden region the propagation of the corresponding exponentially
growing components destroys the independence of the solutions.

Instead, one can propagate the set of coupled equations for the
logarithmic derivative
$Y(r)=\Psi'(r)\Psi(r)^{-1}$,
\begin{equation}\label{log-deri}
Y'(r)=\left[\frac{2\mu}{\hbar^2}V(r)-k^2I\right]-Y^2(r).
\end{equation}
The log derivative matrix in above Ricatti-type equation can be
propagated directly without the stability problem met in the
integration of the coupled-channel equation. We have developed an invariant embedding procedure to solve above log-derivative equation.
In the CC formulation one expands the deformed wave function in terms of its
spherical components $g_c$, where the index $c$ labels the quantum numbers
of the spherical wave, including the angular momentum $l$. Therefore the 
asymptotic  form of these radial components is as in the spherical case, i. e.,
\be
\lim_{r\rightarrow \infty} g_c(r) = N_c H^{(+)}_l (\chi, K_cr) 
\ee
where, as in the spherical case, $\chi$ and $K_c$ are the Coulomb parameter 
and the wave number, respectively. Also as in the spherical case,
the constant $N_c$ is obtained by matching
the external and internal wave functions. A detailed presentation of the 
CC method can be found in Ref. \cite{Delion2010book}.

In spite of the shortcomings of the WKB approach it was found that a
simplification of its formulation  provides decay widths
which are very close to the coupled-channel values in the laboratory frame
\cite{PhysRevC.92.051301}. This has an influence in recent developments
in modern laser facilities. Such facilities  will allow one to probe nuclear
properties by using strong electromagnetic fields \cite{lasernuclei,Zamfir}.
This prompted an intense research activity trying to find means to exploit
these facilities in the pursue of yet unknown nuclear features.
The most relevant of these inquiries so far is the study of the
nuclear behaviour under strong laser fields. Those studies use the modified
WKB approximation of Ref. \cite{PhysRevC.92.051301}.
Thus, in Ref. \cite{dorulase} it is found the remarkable property that as a 
consequence of the laser field
the penetrability of the $\alpha$-particle through the Coulomb barrier may be 
reduced by many orders of magnitude. Due to this
property one may treat nuclear wastes in dedicated facilities by exposing them
to strong laser fields. The resulting decay of alpha particles through the
weak Coulomb barrier would transform the dangerous isotopes
contained in the waste into lighter short living nuclei, thus eliminating one
of the major hinders in the use of nuclear power. 
This conclusion was confirmed by an independent study \cite{laser2}.
Moreover, in another recent publication it is shown that besides alpha-decay 
proton- and cluster-emission may be important modes of decay under strong
magnetic fields and, therefore, as a means of reducing 
nuclear wastes \cite{renlase}. In this paper it is even found that  
proton-decay  may become the main form of decay even when normally alpha-decay 
would be dominant. This is an important point which may trigger new inquiries 
in the old subject of radioactive particle decay.   

\subsection{Alpha decay as a probe to nuclear shape changes and shape
coexistence\label{finest}}

The $\alpha$ decays from ground states to  excited states (fine structure) as 
well as the decays from excited states  are usually less favoured than 
ground-state to ground-state decays. Already in the beginning of microscopic 
studies of $\alpha$-decay the concept of reduced width, and
the ratio between reduced widths for transitions to
ground and excited states, was introduced to estimate the strength of a 
transition \cite{PhysRev.113.1593}. The usefulness of these early notions 
can be pondered  by their importance and use even today, as e. g. in Ref. 
\cite{PhysRevC.83.064320},

Similar early studies were  performed in relation to transitions to vibrational 
states \cite{Sandulescu1965404} and, more recently, in 
\cite{peltonen,PhysRevC.75.054301,PhysRevC.64.064302}. There have also been
a large amount of work dealing with
transitions to and from rotational states \cite{PhysRev.181.1697,
PhysRevC.78.034608,PhysRevC.73.014315,PhysRevC.81.064318}. A recent review on the application of the coupled channel approach to the $\alpha$ decay fine structure can be found in Ref. \cite{0954-3899-45-5-053001}.

Usually the ground state to ground state (gs to gs) alpha-decay transitions in
even-even nuclei are much more probable than the corresponding transitions to
excited states. This feature depends strongly upon the Q-value,
i. e. upon the penetrability of the alpha-particle. The Q-value contains no
information on the structure of the nuclei involved in the transition, like
pairing collectivity.
This information pertains to the formation probability, as we have seen in
Subsection \ref{shellmo}. To extract information about the characteristics  of
the nuclei connected by the $\alpha$-decay process a quantity called "hindrance
factor" (HF) was defined in Ref.  \cite{Delion1995a}. The HF is the ratio of 
the formation probabilities corresponding to the gs to gs divided by the gs to 
excited state transitions. In that reference the HF were evaluated microscopically without
including continuum configurations. Yet the results agree well with the
corresponding experimental data. One may argue that this good agreement is 
because the continuum affects equally all states. Therefore in the ratio of 
the HF the contribution of the continuum is cancelled. However, that is not
the case in the calculations of Ref. \cite{peltonen} where absolute
transitions were studied without including the continuum  either. Yet the 
results also agree well with experiment.
One therefore concludes that the shortcomings of the shell model mentioned
in Subsection \ref{shellmo} affects only $0^+_1$ collective pairing states.

Systematic evaluations of the $\alpha$-decay fine structure were also done 
recently in Refs. \cite{PhysRevC.92.021303,Delion20151}.  It was found that 
the $\alpha$ decays to excited states also follow the Viola-Seaborg law
discussed above (see Eq. (\ref{gn-v})).

The $\alpha$ decay transitions of neutron-deficient nuclei around $Z=82$ are 
of particular interest because through such transitions one can uncover
features like the co-existence of states with different 
shapes \cite{RevModPhys.83.1467}. Thus, three low-lying $0^+$ states in 
$^{186}$Pb  were observed following the $\alpha$ decay of $^{190}$Po(gs) 
\cite{10.1038/35013012}. These three states have been found to be of quite
different shapes, namely spherical, oblate and  prolate.
The transition $^{187}$Po(gs)$\rightarrow\alpha$+$^{183}$Pb(gs), where 
$^{183}$Pb(gs) is spherical, 
was observed to be strongly hindered \cite{PhysRevC.73.044324} whereas the 
decay to a low-lying excited state at 286 keV was favoured. Based upon a 
potential energy surface calculations, the $^{187}$Po ground state and the 
286 keV excited state in $^{183}$Pb were interpreted as of prolate shape. The 
decay to the $^{183}$Pb ground state is hindered since this state has a 
spherical shape which is different from that of the ground state. The 
difference in the shapes indicates that the configurations of the mother and 
daughter nuclear wave functions are very different. As a result, the 
$\alpha$ formation amplitude is significantly reduced. 

In another experiment the HF corresponding to the $\alpha$ decay of the 
isomeric state  in $^{191}$Po was shown to have the same origin, i. e. the
mother and daughter nuclei have different shapes \cite{PhysRevLett.82.1819}. 
The HF corresponding to the $\alpha$ 
decays of neutron-deficient even-even nuclei around $Z=82$ was measured in 
Ref. \cite{PhysRevLett.72.1329}.
The $\alpha$ decays to and from the excited $0^+_2$ states in Po, Hg and Rn 
isotopes were studied in Refs. 
\cite{Delion1995a,PhysRevC.54.1169,PhysRevC.90.061303}. These states are 
described as the minima in the potential energy surface  provided by the 
standard deformed Woods-Saxon potential. A simple approach was also presented 
in Ref. \cite{Karlgren2006} to evaluate the HF by taking the ratio 
between the wave function amplitudes for the transitions to the ground and 
excited $0^+$ states of the daughter nucleus obtained from potential energy 
surface calculations.

In a recent experiment, the fine structure has been observed in the $\alpha$ decay of the high spin  isomeric states in nuclei
$^{155}$
Lu($25/2^-$)
and 
$^{156}$Hf($8^+$)
\cite{PhysRevC.98.024321}. A significantly strong hindrance is seen for the decay to the seniority $v=3$ and 2 states in the corresponding daughter nuclei. These lighter nuclei can be evaluated within the framework of large-scale shell model \cite{PhysRevC.94.064311}. The main configuration of the $25/2^-$ isomer in $^{155}$
Lu has been assigned as $\pi (h_{11/2})^3\otimes \nu f_{7/2}h_{ 9/2}$. The decay of such a configuration to the ground state ($11/2^-$) of $^{151}$Tm is hindered by a factor of nearly 20 in comparison with the g.s. to g.s. transition in relation to the broken neutron pair in the decay state. The decay to other $v=3$ states is hindered by factors upto 300.

In conclusion of this Subsection, the study of fine structure in nuclei has
shed new light on nuclear structure studies. It has  provided a decisive 
evidence of shape coexistence in nuclei and, also very important, has shown 
that the shell model representation is able to properly describe 
$\alpha$-decay transitions in nuclei.

\section{Heavy cluster decay}

In 1977 W. Greiner and A. Sandulescu analysed in detail the mass and charge 
distributions of fission fragments as a function of the positions of the 
fragmentation valleys \cite{aurel}. They found that asymmetric shapes
are generated mainly by the minima in the potential and that there might 
occur a new and highly asymmetric fission
in which one of the fragments is close to the double magic nucleus
$^{208}$Pb. This would correspond to a new form of spontaneous particle
emission where the decaying particle would be much heaver than $^{4}$He.

And indeed this spontaneous emission of clusters heavier than the $\alpha$ particle 
was observed seven years later from a systematic study 
of the properties of nuclei heavier than lead \cite{Nature1984}. The
decaying cluster was identify as the nucleus $^{14}$C. It was observed in 
competition  with $\alpha$-decay from the
mother nucleus $^{223}$Ra leaving behind the
daughter nucleus $^{209}$Pb, as predicted in Ref. \cite{aurel}. 
The branching ratio for emission of $^{14}$C relative to 
$\alpha$-particles was about $10^{-9}$. The corresponding relative decay 
widths  was measured to be $\Gamma (^{14}C)/\Gamma
(\alpha) = (6.1\pm 1.0)\times 10^{-10}$ \cite{price1985,hourani1985}.
This was an outstanding experimental feat and was claimed as the greatest 
achievement in radioactive decay since Becquerel \cite{maddox}. Since then the emission of heavier clusters including $^{20}$O, $^{23}$F,$^{22-26}$Ne, $^{28,30}$Mg, and $^{32,34}$Si have been observed.

Many calculations were done soon afterwards trying to understand the
mechanisms that induce the decay of heavy clusters as well as to explore the
possibility of finding by-products of the decay process. Thus, 
a calculation performed by using Eq. (\ref{gpf}) for the decay width from 
$^{222}$Ra
showed that the relative formation probability of $^{12}$C with respect to
the formation of $\alpha$-clusters
is $|{\cal F}(^{12}C/{\cal F}(\alpha)|^2=3.3\times 10^{-3}$ \cite{Irma}. It was
also found that this ratio diminishes strongly with the size of the cluster.
For example, in the decay $^{216}Rn \rightarrow ^{8}Be+^{208}Pb$ it is
$\Gamma (^{8}Be)/\Gamma(\alpha) = 3.8\times 10^{-15}$, 
$|{\cal F}(^{8}Be/{\cal F}(\alpha)|^2=5.7\times 10^{-3}$, while for the decay
$^{247}Bk \rightarrow ^{46}Ar+^{201}Au$ it is
$\Gamma (^{46}Ar)/\Gamma(\alpha) = 6.6\times 10^{-19}$  and 
$|{\cal F}(^{46}Ar/{\cal F}(\alpha)|^2=3.0\times 10^{-9}$.

Microscopic calculations as those performed in $\alpha$-decay have
not been possible in the case of heavy cluster-decay. There was an attempt to
evaluate the emission of $^{14}$C from $Ra$ isotopes \cite{deinli1}. In order 
to make the calculation feasible crude approximations had to be applied. As
a result the evaluated quantities disagreed greatly from experiment and no attempt to
perform such calculations was attempted afterwards. Thereafter the
calculations were done within effective frameworks, as those discussed in
the previous Section.

The fission-like approach of cluster decay which, as discussed above, is at
the origin of cluster decay processes, has been very successful in describing
cluster decay. There are
many works performed within this framework. The most recent ones are in
\cite{jaj2918,psg2018,PoGh2018}. In the references therein one can find the
developments in this subject during the last years. In this context it is
important to underline that a recent and important example which uses the 
Gamow approach is 
in \cite{manjsow}. The importance of this paper is that it shows that in
superheavy nuclei the dominant decay mode is $\alpha$-decay, in
contradiction of the fission-like study of cluster decay performed in Ref.
\cite{PhysRevLett.107.062503}. 

Cluster states in nuclei were investigated since a long time \cite{brink66}.
One of these investigations had an impact in future studies of cluster decay
\cite{buDoVa75}. The model proposed in this reference was first applied in 
$\alpha$-decay \cite{buJoMePe95} and then in clusterization in general \cite{ipwbm12}
and references therein.  

\subsection{Description of the cluster decay with generalized GN law}
 
\begin{figure}
	\begin{center}
			\includegraphics[scale=0.45]{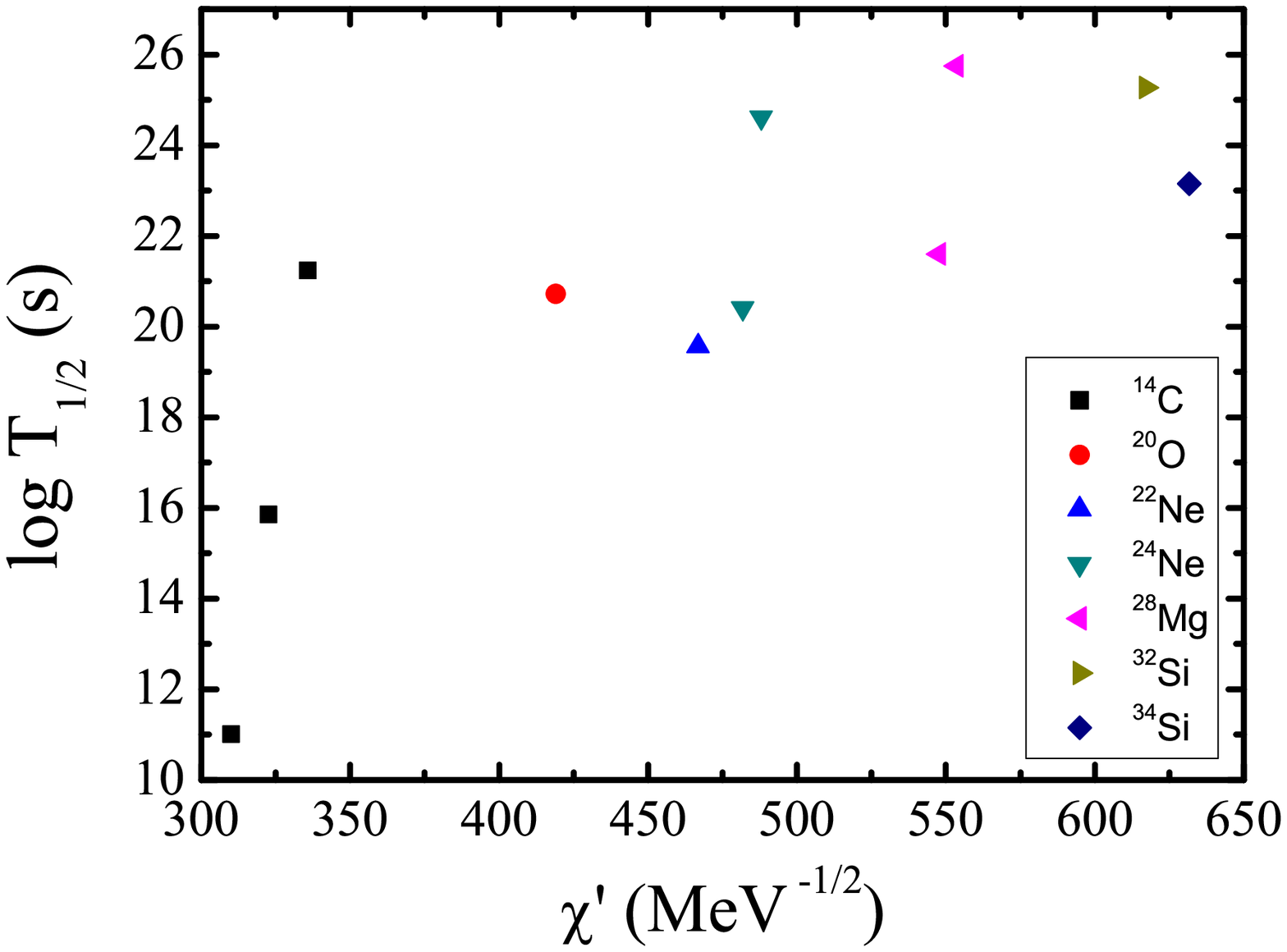}\\
		\includegraphics[scale=0.4]{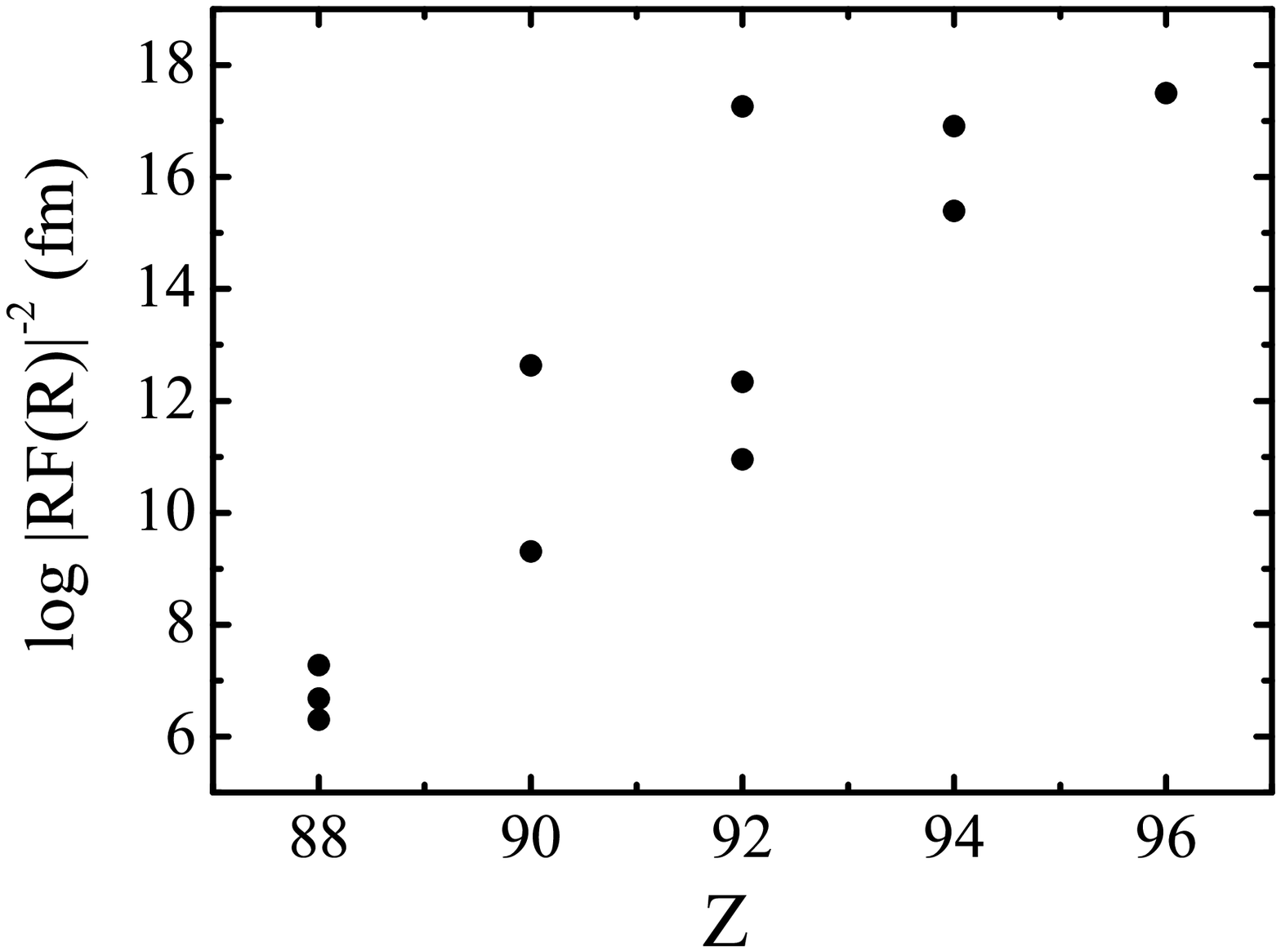}
		\includegraphics[scale=0.4]{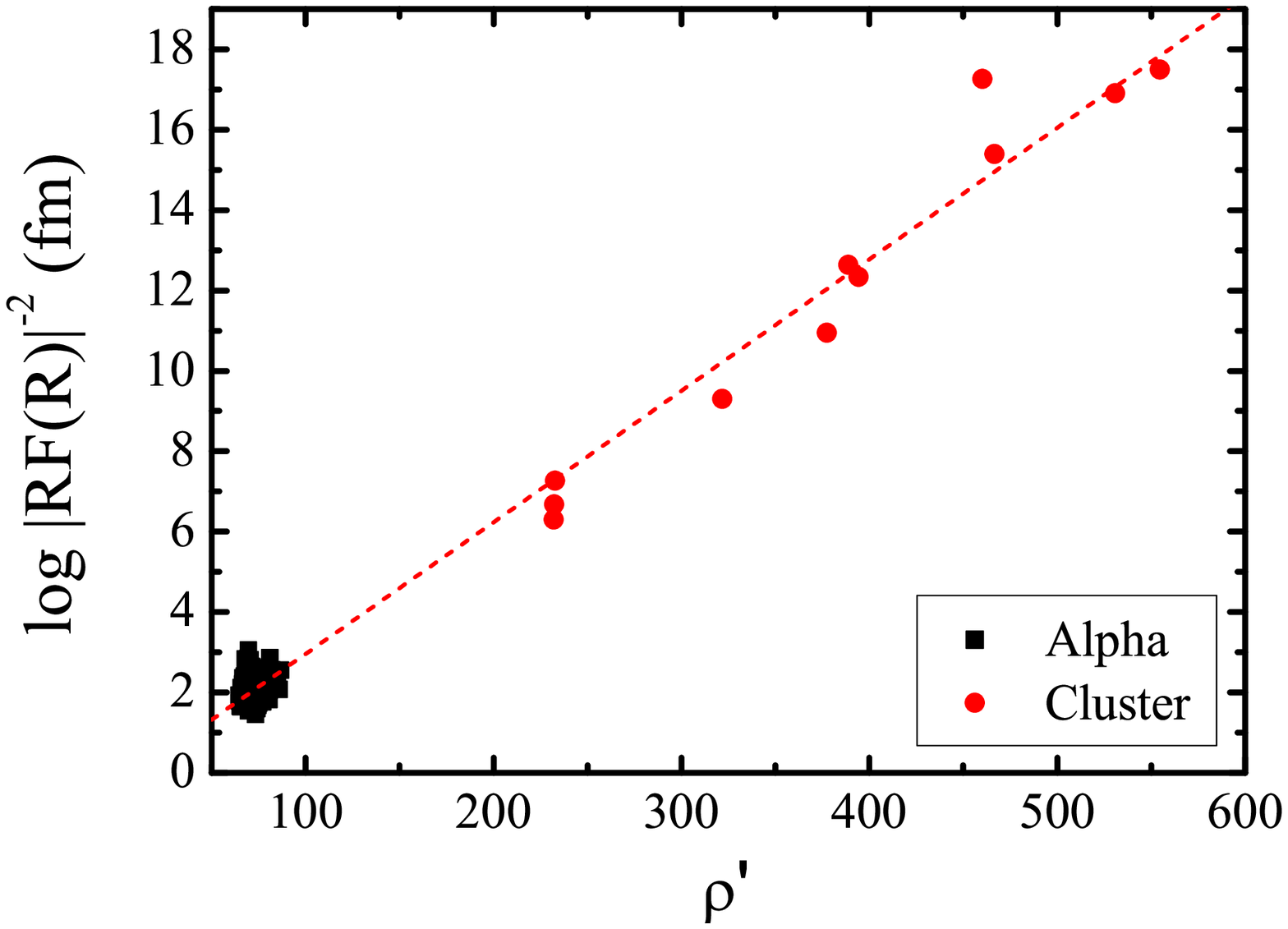}
	\end{center}
	\caption{Upper: Systematics of cluster decay half-lives for even-even nuclei as function of $\chi'$. Lower left: The heavy cluster decay formation probabilities $\log |RF(R)|^{-2}$
		as a function of the charge number of the mother nucleus
		$Z$. Lower right: The formation probabilities $\log |RF(R)|^{-2}$
		for both $\alpha$ and heavy cluster decays as a function of
		$\rho'$. Taken from Ref. \cite{Qi2009}.\label{formation2}}
	\end{figure}

There have been a number of papers dealing with generalizations of the 
Geiger-Nuttall law  which includes cluster decay. 
One advantage of these generalizations is  that if reliable values of decay 
$Q$ values can be obtained, it is easy to extrapolate to all kinds of cluster decays
throughout the nuclear chart.
Systematic calculations of the decays of clusters heavier than $^{4}$He were 
done in Refs. \cite{Qi2009,0954-3899-39-1-015105,QI2009a,PhysRevC.70.034304}. In particular, as mentioned before, the UDL works well for all proton, alpha and heavy cluster decays. This is because the cluster formation probability, as shown in Fig. \ref{formation2}, follows roughly a linear behaviour as a function of $\rho'$. Therefore, all those decays can be described consistently within the same framework.
Such calculations were extrapolated to the decays of even heavier clusters from 
superheavy nuclei to daughter nuclei around $^{208}$Pb in 
Ref. \cite{PhysRevLett.107.062503} and later in 
Refs. \cite{PhysRevC.85.034615,PhysRevC.89.067301}. Although these efforts
have been very fruitful there are uncertainties behind the extrapolation
which requires further analysis. This is an open task.

There are a plethora of effective calculations of cluster decay performed within the 
framework of the Gamow theory. We will here mention a few, although the
interested reader can also consult the
references therein, corresponding to the latest years. In Ref.  
\cite{PhysRevC.70.034304} the width is evaluated by using Eq. (\ref{gamgam})
with $F_{{\rm eff}}=P_cF_c$, where $P_c$ is the preformation factor of
the cluster $c$ in the mother nucleus and $F_c$ is obtained through the
so-called normalization factor, leading to Eq. (14) of that reference. This
is actually based in Eq. (15) of Ref. \cite{gurvitz}.    
Essentially what is done is to use an effective  cluster-nucleus interaction which can be obtained using the double folding model. For example, in Ref. \cite{renIjmp}, it
is obtained from the double folding integral of the renormalized $M3Y$ nucleon-nucleon 
interaction and of the
density distributions of the $\alpha$- particle and daughter nucleus. 
The model reproduces fairly well the half lives corresponding to the decays
of clusters in the region $(^{14}C - ^{34}Si$).

Another similar but much more simpler  example is to take $F_{{\rm eff}}=\nu P_c$, where $\nu$ is the Gamow  "assault frequency"
and $P_c$ is, as above, the preformation probability. This very traditional
and simple Gamow approach gives, by using suitable values of the parameters,
a reasonable account of the half lives of 26 experimentally known cluster
decay half lives.

We notice that a quite common practice is to extract or define the effective cluster preformation factor
empirically based on the discrepancy between the calculated and experimental 
half-lives of the cluster emitters.

In the same fashion as trans-lead nuclei decay into daughters around the 
doubly magic nucleus $^{208}$Pb, a  second island of cluster radioactivities 
is expected in trans-tin nuclei decaying into daughters close to $^{100}$Sn (see, for example, Ref. \cite{Qi2009}). Such studies can be easily done within the generalization of the Geiger-Nuttall law to cluster decays. In particular, there have been tremendous efforts trying to detect the possible $^{12}$C emission from $^{114}$Ba \cite{Oganessian1994,PhysRevC.56.R2912} and even $^{112}$Ba but so far there has been no success. That decay process from $^{114}$Ba was calculated in Ref. \cite{PhysRevC.51.594} with different models.

\section{Effective approaches}

The great success of the Gamow theory can be measured by its wide acceptance
even today. In this theory, and many other similar ones, the decaying
$\alpha$ particle is assumed to be a preformed particle 
which is trapped inside the mother nucleus by the nuclear, Coulomb
and centrifugal potential. It is
assumed that this particle has no intrinsic structure and that the decay
width $\Gamma $ is just proportional to the penetrability, i. e. 
\begin{equation}
\label{gamgam}
\Gamma =F_{{\rm eff}}\exp\left[-2\int^{R_2}_{R_1}k(r)dr \right],
\end{equation}
where $F_{{\rm eff}}$ is an unknown  quantity that is taken as a free parameter
to be fitted to experimental data. Different authors
give different interpretations to this quantity. In the first effective
theory of alpha-decay \cite{Gamow1928} Gamow introduced the concept of
hitting frequency. That is, he assumed that the $\alpha$-particle moves
freely inside the mother nucleus. It would escape the nuclear and 
Coulomb barriers when hitting the wall of the nucleus from inside. The
hitting frequency, i. e.  the coefficient $F_{{\rm eff}}$, is given by 
$F_{{\rm eff}}=v/R$, where $R$ is the nuclear radius and $v$ the velocity of 
the $\alpha$-particle.
This, and the wave number $k$ in the equation above, are given by
$v=\sqrt{2Q_{\alpha}/\mu }$ and $k(r)=\sqrt{2\mu|Q_{\alpha} - V(r)|}/\hbar$ 
where $\mu$ is the effective mass and $V(r)$ the effective potential 
between the cluster and the daughter nucleus, including the Coulomb field. 
The radii $R_1$ and $R_2$ are turning points, i. e. the points $r$ that 
satisfies $V (r) = Q_{\alpha}$.

A different kind of effective model was proposed in Ref. \cite{Buck90},
where the $\alpha$-cluster is assumed to be an inert particle moving in
orbits with large values of a global quantum number. This alpha-cluster
model was actually proposed long before \cite{Buck1994.212}. In this paper
it was found that an $\alpha$-cluster model with parameters values which
optimize the fit to $\alpha$-decay lifetimes for a wide range of nuclei,
results in good agreement with the $\alpha$-and $\gamma$-decay properties of
states belonging  to the ground state  band of $^{212}$Po. 
The reliability of this model was strengthened by the work in Ref. \cite{Buck90}, where the
$\alpha$-particle-core potential is taken to be of squared-well form, with
fixed depth, and radius given by the Bohr-Sommerfeld condition. This rather
simple model reproduces well all the $\alpha$-decay half-lives for heavy
even-even nuclei. 
This result can be considered  farther evidence that the 
wave functions dealing with alpha-decay has to include cluster components, as
discussed above.

In spite of the accurate results of this model its use is
difficult in comparison with the effective procedures based upon the
Gamow model, where the calculations can be performed by using a simple
pocket calculator and its predictive power is good enough to guide experimental
searches.
That is the reason why variations of the Gamow model are mostly used when
dealing with alpha-decay studies. 
One sees that in the simple and outstanding Gamow theory there is only one free
parameter, namely the nuclear radius $R$, which is well determined within
less than an order of magnitude. In radioactive decay studies this is a
rather small error. And even more impressive is that only the
inverse of $R$ (and not the exponential) enters in Eq. (\ref{gamgam}). 
Instead, the exponential is very
sensitive to $R_1$ and $R_2$, but that is the usual limitation of the WKB
approximation. The problem with this beautiful theory is that it is too
simple and that it violates crudely the Pauli principle, although when the
theory was presented this was
compensated by the great breakthrough introduction of the concept of
quantum penetrability,  

Similar empirical approaches, even for cluster decays, were developed through 
the years. Most of these applications consist of various versions of the
nuclear potential in Eq. (\ref{gamgam}). This is odd since one expects that
the value of the short range nuclear potential
should be negligible in the integration range, i. e. between $R_1$ and
$R_2$, in Eq. (\ref{gamgam}). In this context it is worthwhile to point out 
that this nearly independence upon the nuclear potential in $\alpha$-decay
effective theories is not shared by the exact treatment of the Thomas decay
width \cite{Thomas}. Here the internal wave function ${\cal F}_c(R)$ is
mainly determined by the nuclear interaction, although the Coulomb part plays 
also a role. 

In spite of this weak dependence upon the nuclear interaction, different 
effective theories provide
different values of the width. We will not go into these details for all the large number of effective
$\alpha$-decay models introduced in the past. Instead we will refer to some of the 
recent publications, and references therein, on this subject. Thus, in Refs. 
\cite{MahBhaGam15,0954-3899-26-8-305,PhysRevC.74.014312,Denisov2009815,birbi08,
	0954-3899-37-10-105107,PhysRevC.78.057302,kusashgu13} the width is evaluated 
according to Eq. (\ref{gamgam}) with different potentials and different values of
$F_{{\rm eff}}$.  Among these theories we will briefly describe, 
as a typical example, the one in Ref. \cite{Xu2006322} which is valid for
all sort of clusters, including $\alpha$-clusters. 

The nuclear-cluster potential is approximated by the relation
\be
V_N(r)=\lambda (N_cv_n(r) + Z_cv_p(r))
\ee
where $\lambda$ is a folding factor, $N_c$ and $Z_c$ are neutron and proton 
numbers in the cluster, $v_n(r)$ and $v_p (r)$ are single neutron and 
proton fields (excluding the Coulomb potential).
For the single-particle potentials a square well form was adopted. 
The folding factor $\lambda$ was determined by using the Bohr-Sommerfeld condition 
with the Wildermuth rule as,
\begin{equation}
\int_{0}^{R_1}\sqrt{\frac{2\mu}{\hbar^2}[Q_{\alpha}-V(r)]}\text{d}
r=(G+1)\frac{\pi}{2}=(2N+L+1)\frac{\pi}{2},
\end{equation}
where $G$ is the global quantum number and $N$ the number of nodes.
$R_1$, $R_2$ and $R_3$ below are turning points obtained by
$V(r)=Q_{\alpha}$.

There is no microscopic approach to define a unique value for the
quantum number $G$. Since it is expected that the $\alpha$ cluster
should be built upon valence nucleons occupying orbits just above
the Fermi surface, the value of quantum number $G$ is usually taken
from the Wildermuth rule as,
\begin{equation}
G=(2N+L)=\sum_{i=1}^4(2n_i+l_i)=\sum_{i=1}^4g_i.
\end{equation}
$g$, $n$ and $l$ are corresponding quantum numbers of the nucleons
in the $\alpha$ cluster.
The decay width is given as
\begin{equation}\label{eqngamma}
\Gamma =
F\frac{\hbar^2}{4\mu}\text{exp}\left[-2\int_{R_1}^{R_2}drk(r)\right],
\end{equation}
where $k(r)$ the wave number,
\begin{equation}
k(r)=\sqrt{\frac{2\mu}{\hbar^2}|V(r)-Q_{\alpha}|}.
\end{equation}
$F$ is the normalization factor determined by
\begin{equation}
F\int_{0}^{R_1}dr\frac{1}{k(r)}\cos^2\left[\int_{0}^{r}dr'k(r') -
\frac{\pi}{4}\right] = 1.
\end{equation}
The decay width is as in Eq. (\ref{gamgam}) with $F_{{\rm eff}}=PF$ where $P$
is the cluster preformation probability and the normalization factor F is 
determined by a normalization condition. The preformation probability was
assumed to be $P=1$. This  and
the normalization condition were also taken from \cite{Buck90}. Notably, this
simple method reproduces the widths of a large number of alpha-decay data
within a factor of about 2.

An effective $\alpha$-particle decay equation was derived for the $\alpha$ particle 
on top of the $^{208}$Pb core in Refs. \cite{PhysRevC.90.034304,	
2015arXiv151107584X}, where an attractive pocket-like potential appears 
around the nuclear surface. That is related to the sharp disappearance of the 
nucleon density in the Thomas-Fermi model employed in those works. Using the
separation point $r_{sep}$ between the $\alpha$-particle and the daughter nucleus at
the rather large distance of $r_{sep}$= 15 fm good agreement with experiment
is obtained.

An effective method which does not rely upon the Gamow theory was developed
in the 1970s. This formalism treats radioactive decay as an extension
of the one corresponding to nuclear fission \cite{aurel}. This is not surprising,
since decay and fission are similar processes which are determined by the
Coulomb and centrifugal barriers. This method had some success in predicting
alpha-decay processes, but its great breakthrough occurs when it predicted
cluster decays which were later confirmed experimentally, as seen in the
next Section.

In spite of its shortcomings effective theories have been very successful in 
evaluating with reasonable accuracy and in a very simple and fast fashion 
$\alpha$- and cluster decay widths. 

\subsection{The cluster formation probability versus the  preformation factor}
The cluster formation
probability $|F_c(R)|^2$ (or the formation
amplitude $F_c(R)$) can be extracted from the experimental
half-lives data and is a model-independent quantity. As shown above, in practice we also often use the quantity $|RF_c(R)|^{2}$ indiscriminately as the formation probability since it is easier to extract from data. On the other hand, the so-called preformation factor, given as the difference between the calculation and the experimental datum, is often introduced in many effective models. This preformation factor depends strongly on the shape of the effective potential employed \cite{Buck90}. Particularly, it is not a physical quantity that can be calculated microscopically. The reduced width introduced in Ref. \cite{Ras69} is also a similar effective quantity that depends on the effective optical potential. 
We suggest that it is the formation
amplitude or formation probability that should be calculated and compared with those extracted from experimental data. On the other hand, the barrier penetration probability has been well understood since 1920s and contains on information on the nuclear structure property. Another model independent quantity is the ratio $N_0\equiv RF(R)/H_0^+(\chi,\rho)$ which can also be extracted from the experimental data.

The classical
expression for the decay width $\Gamma_c$ (Eq. (1)) can be rewritten written as
\begin{equation}\label{Tho}
\Gamma_c=\hbar\nu\left|\frac{R{\cal F}_c(R)}{H_l^+(\chi,\rho)}\right|^2=2 P_l(R) \frac{\hbar^2}{2\mu R} |{\cal F}_c(R)|^2,
\end{equation}
where $l$ is the angular momentum carried by the outgoing $\alpha$-particle,
${ P}$ is the penetration probability. The combination of regular and irregular Coulomb functions $F$ and $G$
is proportional to the Hankel function $H^+$.
The quantities $\rho$ and $\chi$ are  $\rho=\mu\nu R/\hbar$ and
$\chi = 2Z_cZ_de^2/\hbar\nu$. $\mu$ is the reduced mass and $Z_d$ is the 
charge number of the daughter nucleus. 
$R$ should be chosen that the nuclear
potential has vanishing value and the cluster and the daughter
nucleus has already been fragmented. Although we have discussed it already, it is convenient for the presentation 
to emphasize that the quantity ${\cal F}_{c}(R)$ is the formation
amplitude of the cluster at distance $R$ where only the Coulomb
interaction is relevant. At this point the internal wave function 
${\cal F}_{c}(R)$ is matched with the
wave function of the outgoing particle. 
The decay width is independent upon $R$, which implies that 
the strongly R-dependent Hankel function is balanced by the equally strongly
dependent formation amplitude. 

Here we show how the two-step mechanism is manifested in effective models that the formation process is not explicitly taken into account. 
In the semiclassical approach, the decay
width is given as
\begin{equation}
\Gamma = S_{{\rm eff}}F_{{\rm
eff}}\exp\left[-2\int^{R_2}_{R_1}k(r)dr \right],
\end{equation}
where $S_{{\rm eff}}$ is the effective preformation factor, $F_{{\rm
eff}}$ a proper normalization factor~\cite{Buck90} and $R_1$
and $R_2$ the classical turning points. Since the radius $R$ should
satisfy the relation of $R_1<R<R_2$, we have
\begin{equation}
\Gamma = S_{{\rm eff}}F_{{\rm eff}}\exp\left[-2\int^{R}_{R_1}k(r)dr
\right]P(R),
\end{equation}
where $k(r)=\sqrt{2\mu|Q_c - V(r)|}/\hbar$ with $V(r)$ being the
effective potential between the cluster and the daughter nucleus. The integration from $R_1$ to $R$ mimics the cluster formation probability in an effective way.
For convenience we have defined a penetration factor $P$ that is given as
\begin{equation}
P(R) = \exp \left\{ -2\int^{R_2}_{R} \sqrt{\frac{2\mu}{\hbar^2}
|V_{{\rm C}}(r)-Q_{c}|} dr \right\},
\end{equation}
where $V_{{\rm C}}(r)=Z_dZ_ce^2/r$ is the Coulomb potential. Above
equation can be integrated exactly, giving,
\begin{eqnarray}
\nonumber P &=& \exp  \left\{ -2Z_cZ_d e^2\sqrt{\frac{2\mu}{Q_c\hbar^2}}\right.
\left.\left[\arccos \sqrt{\frac{R}{R_2}} -
\sqrt{\frac{R}{R_2}} \sqrt{1-\frac{R}{R_2}} \right]\right\}.
\end{eqnarray}
Inserting $\chi=Z_cZ_d e^2\hbar\sqrt{{2\mu}/{Q_c}}$,
$R_2=Z_{c}Z_de^2/Q_{c}$ and
\begin{equation}
\frac{R}{R_2} = \frac{\rho}{\chi}=\frac{Q_c}{V_{{\rm C}}(R)},
\end{equation}
one immediately recognized that the penetration factor in the
effective approach is related to the Coulomb function as,
\begin{equation}\label{pn}
P=\frac{[H^+_0(\chi,\rho)]^{-2}}{\tan \beta} =
\exp\left[-2\chi(\beta-\sin\beta\cos\beta)\right].
\end{equation}

Similarly, the decay width in the fission model can be given
as~\cite{Poe02},
\begin{eqnarray}
\nonumber\Gamma&=&F_{{\rm f}}\exp\left[\frac
{-2}{\hbar}\int^{R_2}_{R_f}\sqrt{2B(r)E(r)}dr \right]\\
&=&F_{{\rm f}}\exp\left[\frac
{-2}{\hbar}\int^{R}_{R_f}\sqrt{2B(r)E(r)}dr \right]P(R),
\end{eqnarray}
where $F_{{\rm f}}$ is the frequency of assaults, $B(r)$ the nuclear
inertia, $E(r)$ the deformation energy from which the $Q$ value has
been subtracted. The relation above holds because one should have
$B(r)=\mu$ and $E(r)=V_{{\rm C}}(r)-Q_c$ beyond the radius $R$. Again, the inner side of the integration, which itself has no physical meaning, describes effective the cluster formation process.

To conclude this section, we would like to point out that it is more meaning even for effective approaches to calculate the cluster formation probability instead of the final half life value in relation to the fact that the half life is so sensitive to the barrier penetration which, however, is very well understood and does not contain any nuclear structure information.

\section{Superheavy nuclei}

In the 1960's emerged a keen interest in as yet undetected 
nuclei with large nucleon numbers. All elements with an atomic
number above Z=92 (U isotopes) are unstable, as can be seen from Fig. 1. It is those isotopes beyond uranium,  also known as transuranium elements, which are
called ``superheavy". It was thus predicted that stable nuclei 
would exist in the vicinity of the double magic nucleus $Z=110$ and $N=184$ (see, also, Ref. \cite{Nazarewicz2018} for a recent overview). This 
region of the nuclear
table of isotopes was named  "island of stability" \cite{SGNils}. The
method used to reach this conclusion consisted of evaluating the minima of
nuclear potential energy surfaces as a function of deformations by using the 
liquid drop model and the Strutinsky procedure.
This method was used to investigate the existence of superheavy nuclei
under the name macroscopic-microscopic (macro-micro) approach. In
particular, in Ref.\cite{adam} the half lives of nuclei in the vicinity 
of Z=108, N=162, also expected to form an island of stability, was 
properly predicted using the macro-micro method. 
However, no island of stability was ever found in spite of intense
efforts \cite{Ogane}. Fundings to reach islands of stability were abundant
in the 1960's and even afterwards. In the United States this was partly due 
to the Project Plowshare, which  was a program for the development
of techniques to use nuclear explosives for peaceful purposes. A review on the experimental techniques and alpha decay properties of superheavy nuclei is presented recently in Ref. \cite{asai_nuclear_2015}.

 The production of superheavy nuclei is attained through 
heavy-ion reactions
leading to fused superheavy nuclei \cite{0954-3899-34-4-R01,PT.3.2880,0954-3899-42-11-114001,MUNZENBERG20153,0034-4885-78-3-036301,1402-4896-92-8-083002}. A recent example of this is the
fusion reactions of $^{48}$Ca on $^{238}$U-$^{249}$Cf targets \cite{Ogane}.
In this reference one can also find a qualified review of the production cross 
sections and a summary of the decay properties, including the results of 
experiments performed in a number of facilities worldwide. Even a
comparison with theoretical calculations, performed within the framework of
the macro-micro approach, is presented. 

Fission and $\alpha$ decay are the dominant decay modes for available superheavy nuclei which are all on the proton-rich side. Those nuclei may also undergo $\beta$-decay (electron capture) as well as heavy cluster emissions which have not been observed so far. The detection of emitted $\alpha$ particles has been the principal method of identifying superheavy nuclei as well as their excited states (see, e.g., Ref. \cite{PhysRevLett.111.112502}).

The stability of the
superheavy elements generally decreases with rising atomic numbers. Yet, there 
are some of them that live a very long time. The most
striking example is $^{209}$Bi with a half life of $2\times 10^{19}$ years,
i. e. $10^9$ times the age of the Universe!. The heaviest known element, 
Oganesson, has a proton number Z=118 and $^{294}$Og has a half life of 0.69
milliseconds. 

Seaborgium (Z=116, with $^{269}$Sg half life of 3.1 minutes) was the first 
element ever to have been officially named 
after a living person (G. T. Seaborg). The second isotope to be so named
occurred in 2016 when Oganesson was discovered. This name was chosen to honor 
Yuri Oganessian, who is one of the authors of Ref. \cite{Ogane}.

An extensive account of the research done on production and properties of 
superheavy nuclei during the last decades was published in a special
issue of Nuclear Physics A \cite{dullman}. Here various articles written
by active researchers in the field are presented. In particular Ref. 
\cite{Ogane} belongs to this issue. One can also find here an
account of recent research on the production  of transuranium elements by the 
r-process in nucleosynthesis \cite{gabriel}. More to the aim of this review
is the discussion and comparison between macro-micro and microscopic 
models performed in another article of that issue \cite{heenen}. The
microscopic models include self-consistent mean-field approaches that use 
realistic effective nuclear interactions or energy density functional. 
It is concluded in this paper that fission is not the dominant decay mode,
but rather $\alpha$-decay.

Within the microscopic models one can investigate details of the wave
function which illuminates processes that are determined by special
wave function components. One such processes in alpha-decay is 
the hindrance factor corresponding to an excited state in the mother
nucleus. The value of the hindrance factor is given by the ground 
state to ground state formation probability divided by the excited state to
ground state  state formation probability.
This quantity was studied within a microscopic framework in Ref. \cite{DLW}.
It was found that the hindrance factor from a two-quasiparticle state (2qp) 
in the mother nucleus is very sensitive to the deformation, 
and thus can be a powerful
tool to determine the structure of superheavy nuclei. They are in all cases
very large but they can be divided into two sets. For 2qp
states that form an aligned configuration, the hindrance factors  are between 
10 and 100.
For nonaligned configurations they are larger than 100. This indicates that 
the decay of the nucleus may be strongly
hindered. If in the decay path the nucleus ends in such a
2qp state then it will not decay by $\alpha$-particle emission to the ground 
state of the daughter nucleus but rather to an excited
state having a structure similar to that of the parent 2qp state. In this
case the $\alpha$-decay chain will not proceed through gs to gs channels but 
rather from excited state to excited state. However, if such daughter state 
does not exist then the parent nucleus will remain in the 2qp state a long 
time, becoming an isomer.

Although these studies are important to understand decay processes occurring
in superheavy nuclei, the most important task at present is to find a
reliable formalism which would predict with acceptable accuracy the widths of 
the different decay channels. Since the widths are strongly dependent on the
Q-values, as discussed above in relation to Eq. (\ref{penet}), 
the first quantity which has to be reliable is the binding energy
corresponding to each  of the
nuclei involved in the decay process. This, which in fact is not related to
the formalism one uses to evaluate the decay width, is a task which has been
pursued in nuclear physics since a long time. The problem is more involved
at present because old formulas and methods that provided accurate binding 
energies in the past did not consider the very unstable isotopes that one
encounter when evaluating decay widths of superheavy nuclei. This maybe the
reason why different calculations predict quite different results. We will
present this contradiction  starting with a calculation already mentioned in the
previous Chapter. That is the evaluation of decay widths corresponding to
different clusters, including $\alpha$-clusters, by applying a formalism
which is as an extension
of the one corresponding to nuclear fission \cite{aurel}. The first time
that this formalism was used to perform a systematic search of cluster decay
probability was in Ref. \cite{PhysRevLett.107.062503}. The fission
probability is also extremely dependent upon the Q-values. In this reference
one compares experimental binding energies which are available for large
N-values with the corresponding ones predicted using the procedure of Ref. 
\cite{Koura} (called KTUY05 in \cite{PhysRevLett.107.062503}). It is found
that the agreement between theory and experiment is good. One then proceeds
to evaluate the branching ratios $B_\alpha$=$T_\alpha /T_c$, where $T_\alpha$
($T_c$) is the half live of the $\alpha$ (cluster) decay channel. One sees
that if $B_\alpha$ is large then the nucleus will decay by emitting a
cluster, since the cluster-decay half life would be  much smaller than the one
corresponding to $\alpha$-decay. It was found in 
\cite{PhysRevLett.107.062503} that for nuclei with
experimentally known binding energies the value of $B_\alpha$ is very small
(see Fig. 4 of Ref. \cite{PhysRevLett.107.062503}). But then the
extrapolation to heavier isotopes produces a notable result. For values of
the mother neutron number N lying between N=190 and N=200 the superheavy
nucleus with Z=124 would decay by emitting a heavy cluster with a branching
ratio $B_\alpha\approx 10^{10}$. This result implies that the
superheavy nucleus can be experimentally confirmed by just measuring the 
presence of the emitted cluster. 
This feature went against what one had expected at that time, namely that 
$\alpha$-decay would be the dominant form of decay of superheavy nuclei. Moreover, one did not consider the significantly large uncertainties in the binding energy predictions for super-heavy nuclei \cite{PhysRevC.92.024306,0954-3899-42-4-045104}.

There were a very large number of publications since then claiming opposite
views on this subject. We will not go into details of all these
investigations, which would add to the general confusion which already
exists. Instead we will refer to two recent publications in order to 
illustrate the nature of the problem. Thus, in Ref. \cite{manjsow} 
different decay
modes such as spontaneous fission, ternary fission and cluster decay of
super heavy nuclei are studied. The authors conclude that (sic) 
"The comparison of half lives for
different decay modes reveals that alpha decay is having smaller half lives
than the other studied decay modes. A detailed study of branching ratio of
alpha decay with respect to other decay modes also confirms that alpha decay
is most dominant decay mode for the super heavy nuclei $^{318}$126, 
$^{319}$126, $^{320}$126 and $^{323-326}$126 and hence these nuclei can be 
detected through the alpha decay mode only.".
Which is just the opposite conclusion reached in
\cite{PhysRevLett.107.062503}. The confusion about the decay modes of
superheavy nuclei can perhaps best be illustrated by the conclusions reached
in Ref. \cite{santosh}. It is stated that "From the entire study of odd
superheavy elements, it is seen that among 1051 nuclei, 233 nuclei exhibit
proton emission and 18 nuclei exhibit neutron emission. 56 nuclei are stable
against alpha decay with negative value for the decay. 92 nuclei show alpha
decay followed by spontaneous fission and 9 nuclei show alpha decay followed
by proton emission. 39 nuclei decay through full alpha chain and 595 nuclei
decay through spontaneous fission.".

It thus seems that in the decay process of super heavy nuclei a survey of
all available procedures to evaluate binding energies as well as models used
to calculate half lives are needed. One would thus hope to find a combination
of the best of these two schemes in order to predict with confidence the
main decay channels. 

The stability of superheavy nuclei may be enhanced from the high-spin K-isomerism phenomenon \cite{Xu04}. The alpha decay from such states can also be expected to be significantly hindered \cite{PhysRevC.98.014320,HEENEN2015415,PhysRevC.97.024333}. In Ref. \cite{PhysRevC.97.024333}, the pairing gap had to be reduced in order to reproduce the long alpha decay of the high-K isomer ($10^{-1}$) in $^{270}$Ds for which the alpha decay half life is 20 times longer than that of the ground state.
The reduction of the pairing parameter in that calculation dramatically reduced
the formation probability of the $\alpha$ particle by
about two orders of magnitude.

\section{Summary and outlook}

Understanding how nuclear many-body systems can self-organize in 
simple and regular patterns is a long-standing challenge in modern 
physics. The first case where this was realized is the Geiger-Nuttall law in 
$\alpha$ radioactivity which shows striking linear correlations between the 
logarithm of the decay half-life and the kinetic energy of the outgoing 
particle. Since that law was formulated the study of particle radioactive 
decay has leaped forward and other forms of decay were found, including 
proton, two-proton, cluster, as well as $\alpha$-decay from rare nuclei.
The origin and properties of these transitions have been investigated by many 
authors during the years, a process that still continues actively at
present. In this review we have gone through the developments that took
place in this subject during the last couple of decades. We have shown that 
in all cases a proper explanation of the decay process requires the 
microscopic description which started to be formulated just at the beginning
of quantum mechanics.

Proton decay is also an excellent 
tool to investigate and understand the intrinsic 
structure of deformed single- (or quasi-) particle orbitals. This 
development resulted in the presentation  of a simple formula relating 
the half-life of a proton decaying mother nucleus with the proton Q-value.
This formula enables the precise assignment of spin and parity of the decaying 
state, a property which is of great value in experimental searches. 
Similar developments, including the formulation of Geiger-Nuttal type laws,
were found in cluster decay. But in this case the formation of the decaying 
cluster within the mother nucleus 
has been exhaustively studied within the framework of macroscopic formalisms.
However, only for the case of alpha-decay these microscopic studies has
provided fruitful results.  
The shell model has been very successful in this endeavour. Thus, its
application in its quasiparticle representation showed that in an 
$\alpha$-decaying isotope chain the $\alpha$-formation probability is
proportional to the corresponding pairing gap (see Figure \ref{fap}).

It was also found within the shell model that 
the clusterization of the protons and neutrons in the $\alpha$-particle is 
induced by high-lying single particle configurations, i. e. the continuum
part of the spectrum. This is needed in order to 
properly describe the influence of the strong pairing correlation upon the
clusterization process. 
Yet, even including very  high-lying configurations the evaluated 
decay width is too small by about one order of magnitude. 
Even the explicitly introduction of the proper continuum through the
Berggren (complex) representation did not substantially affected the results. 
This, and other decay features (related to both particle as well as 
electromagnetic decays) showed that the representation has to contain cluster 
components.

In nuclei  with $N\sim Z$, where neutrons 
and proton move in the same shells and, therefore, the influence of the 
neutron-proton (np) correlation is expected to be strongest, particle 
radioactive decay may illuminate the role played by np pairing degrees of 
freedom in nuclear spectroscopy. This includes isoscalar
pairing modes and, specially important in this review, nuclear clusterization.
Experiments in this direction are being performed at present and also are
planned to be carried out in the near future. 

Systematics of 
the $\alpha$ formation probabilities revealed interesting local fluctuations 
which can provide invaluable information on the pairing correlation as well
as on the structure of the shells determining the clustering. Thus,
the reduction of the pairing gap at $Z=82$ and $N=126$ and the changes in the 
nuclear shapes in neutron-deficient nuclei around $Z=82$ induce suppression of 
the $\alpha$ clusterization. 

The emission of heavy clusters has provided an excellent tool in the
understanding of isotope formation in nucleosynthesis. But the thrilling
subject which is being pursued at present in relation to heavy cluster decay
from superheavy nuclei is whether such decay is relevant. Usually the
overwhelming form of decay from superheavy nuclei is $\alpha$-decay. But is
was found that in some cases heavy-cluster decay is more likely, by many
orders of magnitude, than $\alpha$-decay. Therefore by detecting the 
decaying cluster one can identify the otherwise elusive emitting superheavy
nucleus.

It will be possible in the near future to upgrade radioactive been facilities
to reach rare nuclei and their emitting particles. This will provide
invaluable information on neutron deficient nuclei as well as on nuclear  shell
structures far from the stability line. One may even hope that if this
experimental breakthrough is achieved then the old aim
of reaching a nuclear island of stability may be fulfilled.

We have included in this review an investigation related to $\alpha$-decay 
which may lead  to outstanding practical applications. Namely that under the 
influence of
strong electromagnetic fields, as those to be reached in coming laser facilities,
the $\alpha$-decay half life may be shortened by many orders of magnitude. One
may thus treat nuclear wastes in dedicated facilities by exposing them
to strong laser fields.

We also commented on the application of the effective theories or phenomenological Gamow models to $\alpha$ and cluster decays as well as proton decays. The success of those models in reproducing the decay half-lives is simply related to the fact that the decay process is dominated by the well understood penetration process which sensitively depends on the $Q$ value but does not carry any nuclear structure information. On the other hand, we recommend that the decay formation probability should be calculated and compared with those extracted from experimental data. The  formation probability removes the sensitivity to decay $Q$ value and is solely determined by nuclear structure effects.

A generalized fully microscopic description of nuclear particle radioactivity 
(including e. g. clusters) is still lacking and more progress on the
theoretical side is required to fully describe the fascinating process of 
particle decaying atomic nuclei.

Finally, it is important to underline that we have carefully included here a 
large number of references in order to guide interested readers in the search 
of specific subjects.

\section*{Acknowledgments}
This work was supported by the Swedish Research Council (VR) under grant 
Nos. 621-2012-3805, 621-2013-4323 and the G\"oran Gustafsson foundation.
We are grateful to the Swedish National Infrastructure for Computing (SNIC) at NSC 
in Link\"oping and PDC at KTH, Stockholm for computational support.

\begin{singlespace}

\setlength{\itemsep}{0pt}

\providecommand{\newblock}{}

\end{singlespace}

\end{document}